\documentclass[12pt]{article}
\pdfoutput=1

\usepackage{multirow}
\usepackage[pdftex]{graphicx}
\usepackage{amssymb}
\usepackage{amsmath}
\usepackage{cite}
\usepackage{lscape}

\usepackage{longtable}
\usepackage{booktabs}
\usepackage{array}
\usepackage{ulem,slashed}

\usepackage{xcolor}
\definecolor{tit}{rgb}{0.1,0.2,0.4}

\usepackage{subfig}

\newcommand{\eq}[1]{\begin{equation} #1 \end{equation}}

\newcommand{\eqa}[1]{\begin{eqnarray} #1 \end{eqnarray}}

\newcommand{\eps}{\varepsilon}

\newcommand{\GeV}{\,{\rm GeV}}

\newcommand{\Leff}{\mathcal{L}_{\rm eff}}

\newcommand{\Eq}[1]{Eq.~(\ref{#1})}

\newcommand{\Sec}[1]{Section~\ref{#1}}

\newcommand{\Reff}[1]{Ref.~\cite{#1}}

\newcommand{\Fig}[1]{Figure~\ref{#1}}

\newcommand{\A}{{\cal A}}
\newcommand{\F}{{\cal F}}

\begin{document}

\allowdisplaybreaks

$\ $
\vspace{-2mm}
\begin{center}
\fontsize{25}{30}\selectfont
\bf 
Dark Matter Direct Detection\\[2mm]
in $t$-channel mediator models
\end{center}

\vspace{2mm}

\begin{center}
{\rm Giorgio Arcadi{$^{\, a,b}$}, David Cabo Almeida{$^{\, a,b,c,d}$}, Federico Mescia{$^{\, c,d}$} and
Javier Virto{$^{\, c,d}$}}\\[5mm]
{\it\small
{$^{\, a}$} 
Dipartimento di Scienze Matematiche e Informatiche, Scienze Fisiche e Scienze della Terra\\
Universita degli Studi di Messina, 
Via Ferdinando Stagno d'Alcontres 31, I-98166 Messina, Italy\\[2mm]
{$^{\, b}$} 
INFN Sezione di Catania Via Santa Sofia 64, I-95123 Catania, Italy\\[2mm]
{$^{\, c}$}
Departament de Física Quàntica i Astrofísica, Universitat de Barcelona,\\
Martí Franquès 1, E08028 Barcelona, Catalunya\\[2mm]
{$^{\, d}$}
Institut de Ciències del Cosmos (ICCUB), Universitat de Barcelona,\\
Martí Franquès 1, E08028 Barcelona, Catalunya
}
\end{center}

\vspace{1mm}
\begin{abstract}\noindent
\vspace{-5mm}

We perform a comprehensive study of the Direct Detection phenomenology of singlet Dark Matter $t$-channel portal models.
For that purpose, we present a complete one-loop matching onto a Heavy Dark-Matter Effective Field Theory, leading to a complete computation of the loop induced direct detection cross-section
for both scalar and fermionic Dark Matter candidates.
The results are compared with current and future bounds from Direct Detection experiments, as well as with the requirement of the correct Dark Matter relic density.

\end{abstract}

\newpage

\setcounter{tocdepth}{2}
\tableofcontents

\newpage

\section{Introduction}
\label{sec:intro}

The introduction of a new cosmologically-stable particle to the Standard Model (SM) is one of the most popular solutions to the Dark Matter (DM) puzzle. Among the broad variety of the proposed particle physics frameworks, Weakly Interacting Massive Particles (WIMPs) have gathered most of the attention from the particle physics community in light of their testability. In recent years, Dark Matter Direct Detection (DD) experiments have made a significant progress in probing many particle physics models (see e.g. \cite{Arcadi:2017kky} for a review). In this regard, an interesting question is whether the current experimental sensitivity or the one expected in the next generation of detectors will allow to probe models in which the interactions relevant for Direct Detection emerge at the loop level. 

A notable example of this kind of scenarios is given by the class of DM simplified models known as ``$t$-channel portals''. In these models, the DM features a Yukawa-like interaction with a SM fermion and a new particle with non trivial quantum numbers under the SM gauge group. The interactions of the new particles are dictated by gauge invariance, and fixed by their representations under the SM gauge group.
Moreover, $t$-channel portals reproduce the features of several theoretically motivated models such as flavored Dark Matter \cite{Kile:2011mn,Agrawal:2011ze,Kile:2013ola,Agrawal:2014aoa,Kile:2014jea,Bhattacharya:2015xha} or SUSY DM models. 

In this work we review the direct detection prospects within $t$-channel portals. We provide the most-up-to-date evaluation of the DM scattering cross-section and scattering rate over nucleons for several variants of this model. We consider both scalar (real and complex) and fermionic (Dirac and Majorana) DM candidates. While extensive computations have been already performed for fermionic DM, see e.g.~\cite{Ibarra:2015fqa,Hisano:2010ct,Hisano:2011cs}, the case of scalar DM has been somehow overlooked (see nevertheless~\cite{Agrawal:2011ze,Hisano:2015bma,Arakawa:2021vih}).We present for the first time, to the best of our knowledge, a complete computation of the scattering cross-section of a real or complex scalar DM coupled with a $t$-channel fermionic mediator. One of the interesting outcomes, in this regard, is the fact that a theoretically consistent computation requires the presence, at least at some energy scale, of a portal coupling between a pair of DM candidates and the SM Higgs boson.

We then compare the theoretical predictions for the DM scattering cross-sections with the most up-to-date constraints from Direct Detection experiments as well as with the projected sensitivity of the next generation of detectors. The outcome of such analysis is in turn confronted with the requirement of the correct DM relic density, assuming the standard freeze-out paradigm.

It is also the case that $t$-channel portals are a very interesting benchmark models for collider searches. However we have not explicitly included such constraints in order to keep the focus on the DM side. For this we refer, for example, to the recent reference~\cite{Arina:2023msd} (see also \cite{Mohan:2019zrk}), which is hence strongly complementary to our work.

The paper is structured as follows.
In~\Sec{sec:lagrangian} we illustrate, from a theoretical perspective, the different realizations of the $t$-channel portal model. 
\Sec{sec:matching} is devoted to direct detection; for each model shown in~\Sec{sec:lagrangian}, we illustrate the corresponding effective Lagrangian relevant for direct detection and then compute the DM scattering cross-section and/or scattering rate. \Sec{sec:results} is devoted to a first series of results illustrating the direct detection prospects of the models under consideration without considering other possible constraints.
In~\Sec{sec:relic} a brief overview of complementary constraints, especially relic density, is presented. 
\Sec{sec:conclusions} contains a summary of all the considered constraints and the conclusions of our work.

\section{Model Lagrangian}
\label{sec:lagrangian}

We consider both the cases of scalar (real or complex) and fermionic (Dirac or Majorana) Dark Matter.
In the former case the relevant interactions for DM phenomenology are described by the following interaction Lagrangian:
\eqa{
\mathcal{L}_\text{scalar}
&=&{\Gamma_L^{f_i}\bar f_i }{P_R}{\Psi_{f_i}}{\Phi_{\rm DM}}+  {\Gamma_R^{f_i}\bar f_i }{P_L}{\Psi_{f_i}}{\Phi_{\rm DM}} +
{\rm{h}}{\rm{.c.}}
\nonumber\\
&&
+ \lambda_{1H\Phi}  (\Phi_{\rm DM}^\dagger \Phi_{\rm DM})  (H^\dagger H) + \lambda_{2H\Phi}  (\Phi_{\rm DM}^\dagger T^a_\Phi \Phi_{\rm DM} )  (H^\dagger \dfrac{\sigma^a}{2} H)\ ,
\label{eq:scalar_lagrangian}
}
where the scalar DM field has been labelled $\Phi_{\rm DM}$ while $\Psi_{f_i}$ represent new vector like fermions. 
In order to maintain the notation as general as possible, we have introduced generic couplings $\Gamma_L^{f_i},\Gamma_R^{f_i}$ of the new fermions with both left-handed and right-handed SM fermions.
The Lagrangian is written under the assumption of the existence of a global $Z_2/U(1)$ for real/complex DM representations under which the NP states are odd while the SM fields are even. This forbids the existence of interactions of a single NP state with two SM fermions, thus ensuring the stability of the lightest BSM state (the DM particle). Notice also that the global symmetry forbids operators such as $\bar f_i P_R H \Psi_{f_i}$ which would be responsible of a mixing between the SM and the BSM fermions. The symmetry cannot, however, forbid the existence of four-field interaction terms, shown in the second line of~\Eq{eq:scalar_lagrangian},  between the DM and the Higgs doublet~$H$. The term proportional to $\lambda_{2H}$ is present only if $\Phi_{\rm DM}$ belongs to a $SU(2)$ multiplet. As will be explained in the following these terms cannot be neglected since they are necessary for a theoretically consistent treatment of DM scattering on nucleons. We finally remark that, due to gauge invariance, only some specific couplings between the SM and BSM fermions can be simultaneously different from zero, depending on the gauge quantum numbers of the latter. One could, of course, introduce different NP fermions with different gauge quantum numbers in order to include DM interactions with all of the SM fermions.

In the case of fermionic DM (denoted by $\Psi_{\rm DM}$), the mediator is now a complex scalar field denoted by $\Phi_{f_i}$. The interaction Lagrangian is given by:
\eqa{
\mathcal{L}_\text{fermion}
&=&
{\Gamma_L^{f_i}\bar f_i }{P_R}{\Phi_{f_i}}{\Psi_{\rm DM}}+  {\Gamma_R^{f_i}\bar f_i }{P_L}{\Phi_{f_i}}{\Psi_{DM}} +
{\rm{h}}{\rm{.c.}}
\nonumber\\
&& + \lambda_{1H\Phi}  (\Phi_{f_i}^\dagger \Phi_{f_i})  (H^\dagger H) + \lambda_{2H\Phi}  (\Phi_{f_i}^\dagger T^a_\Phi \Phi_{f_i} )  (H^\dagger \dfrac{\sigma^a}{2} H) \ .
\label{eq:fermion_lagrangian}
}

The quantum numbers of the New Physics states are dictated by the requirement of gauge invariance of the Lagrangians above, with the additional condition that the lightest new state is electrically neutral. Unless differently stated, we work under the assumption that the DM is a pure SM singlet; in such a case the fields $\Phi_{f_i}, \Psi_{f_i}$ trasform, under the SM gauge group, as the corresponding SM fermion present in their Yukawa-like coupling with the DM. Consequently, in the minimal realization of the $t$-channel portal, namely the one containing just the (SM-singlet) DM candidate and a single mediator, the NP states will couple to a single fermion species. 
In this case last term in both Eqs.(\ref{eq:scalar_lagrangian}) and~(\ref{eq:fermion_lagrangian}) are missing, too.
We work in a basis in which the interactions are flavor diagonal.

\section{Dark Matter Direct Detection}
\label{sec:matching}

We now provide a detailed computation of the DM (SI) scattering cross-section on nucleons, considering both cases of scalar and fermionic DM. Since the energy transfer in DD experiments is of the order of $\sim 1\GeV\ll \Lambda_\text{EW}\lesssim \Lambda_\text{BSM}$, this requires the construction of an effective field theory (EFT). This EFT will contain as dynamical degrees of freedom the light quark flavors ($u$, $d$ and $s$), the gluons, and the slow-varying components of the DM field as a static source~\cite{Bishara:2016hek}.
Heavy quark flavors, the $t$-channel mediators~\footnote{
We assume the mediators to be heavy compared to the cut-off of the EFT.}
and the high-frequency components of the DM field are integrated out, their effects being contained in the Wilson coefficients (WCs) of the effective operators.
As already pointed out, the main focus of this work is represented by the case in which the effective operators for DM scattering are generated by interactions arising at the loop level. Thus the construction of the EFT requires a one-loop matching calculation in order to calculate the WCs in terms of the model couplings and heavy masses.
For a general discussion we refer, for example, to \cite{DelNobile:2013sia,Bishara:2016hek,Brod:2017bsw}.

\subsection{Complex Scalar DM}
\label{sec:ComplexScalar}

\begin{figure}[t!]
\begin{center}
\begin{tabular}{ccc}
\subfloat[]{\includegraphics[width=0.3\textwidth]{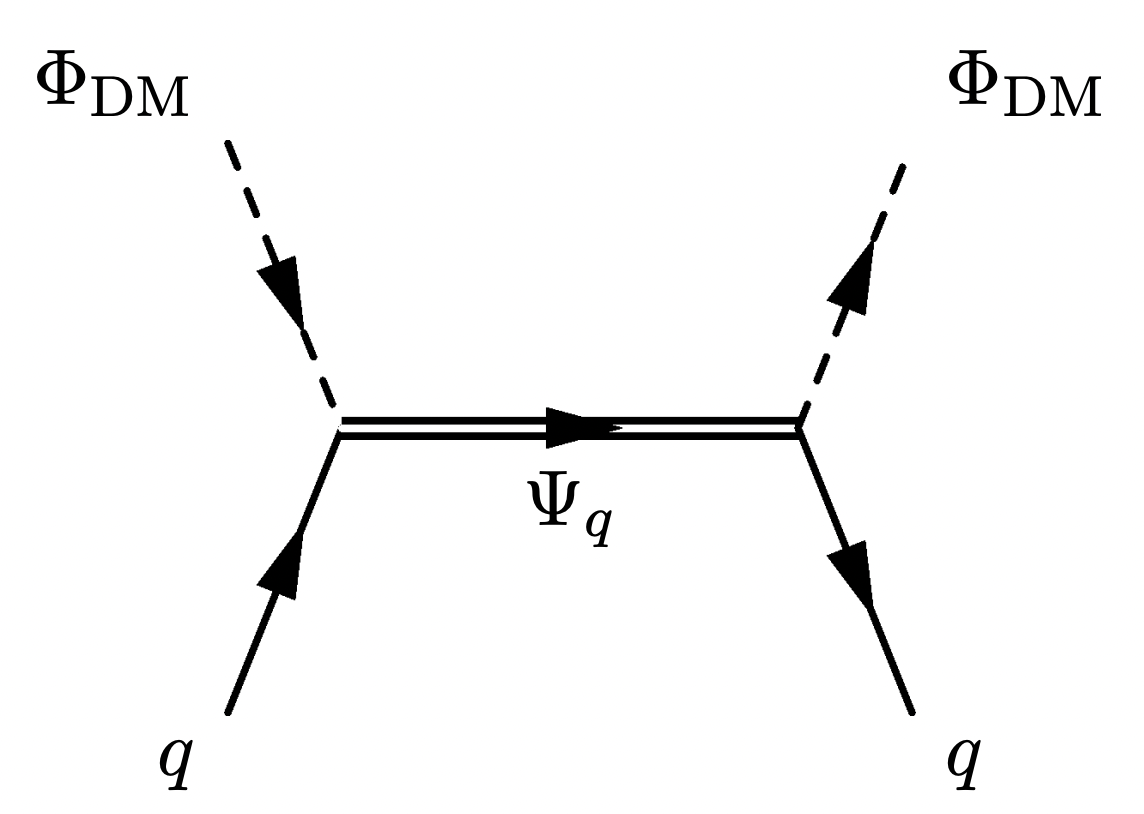}}
&
\subfloat[]{\includegraphics[width=0.3\textwidth]{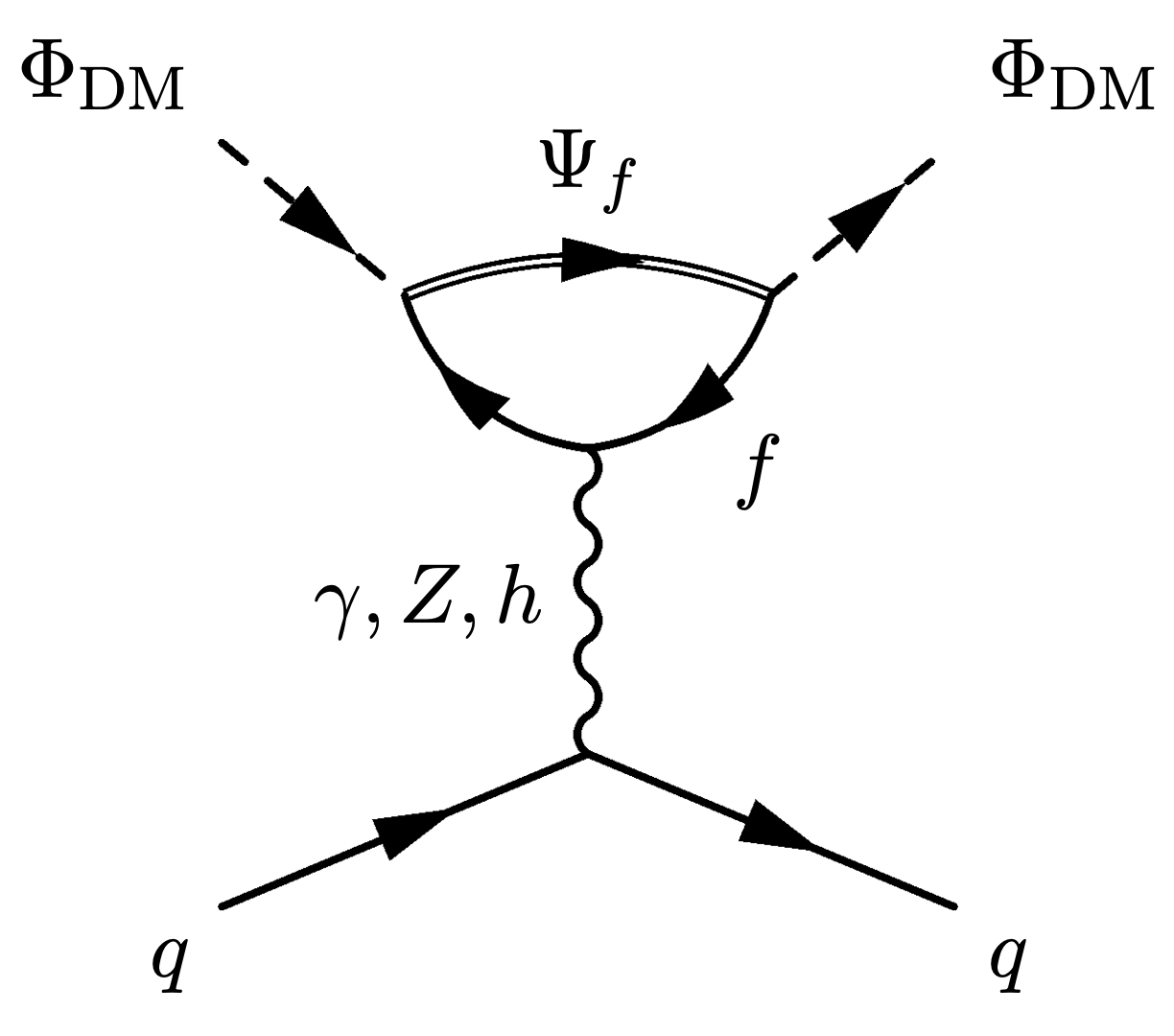}}
&
\subfloat[]{\includegraphics[width=0.3\textwidth]{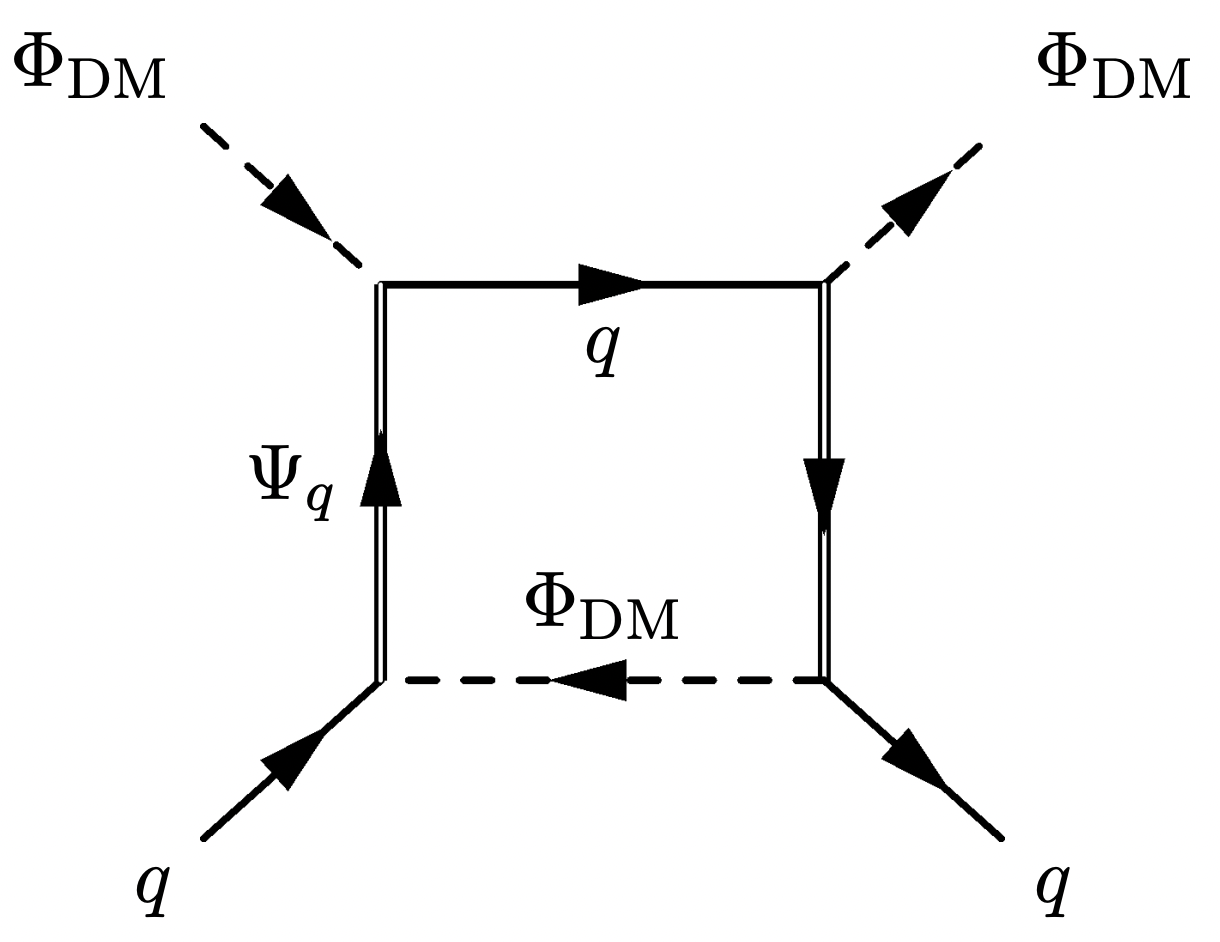}}
\\
\subfloat[]{\includegraphics[width=0.3\textwidth]{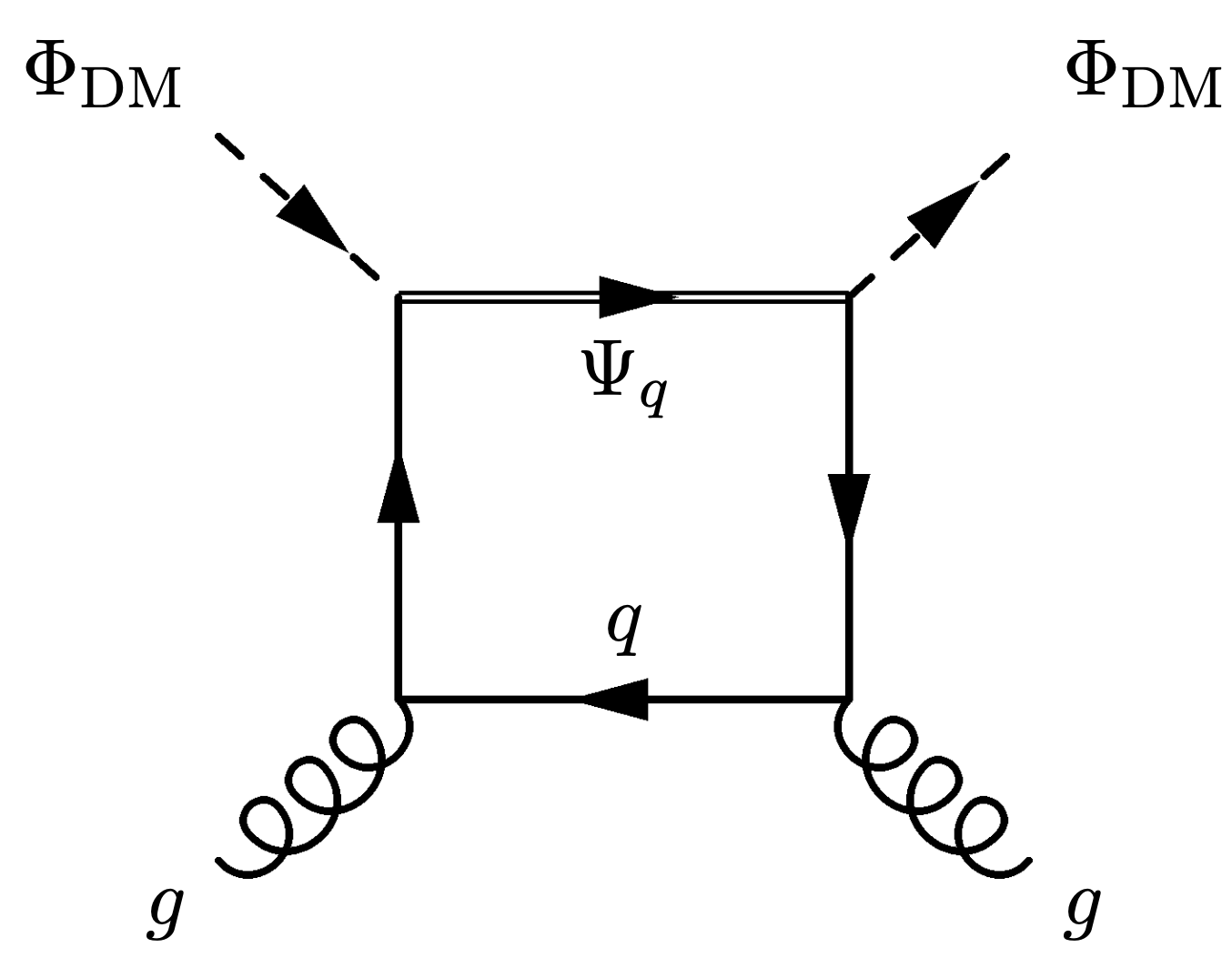}}
&
\subfloat[]{\includegraphics[width=0.3\textwidth]{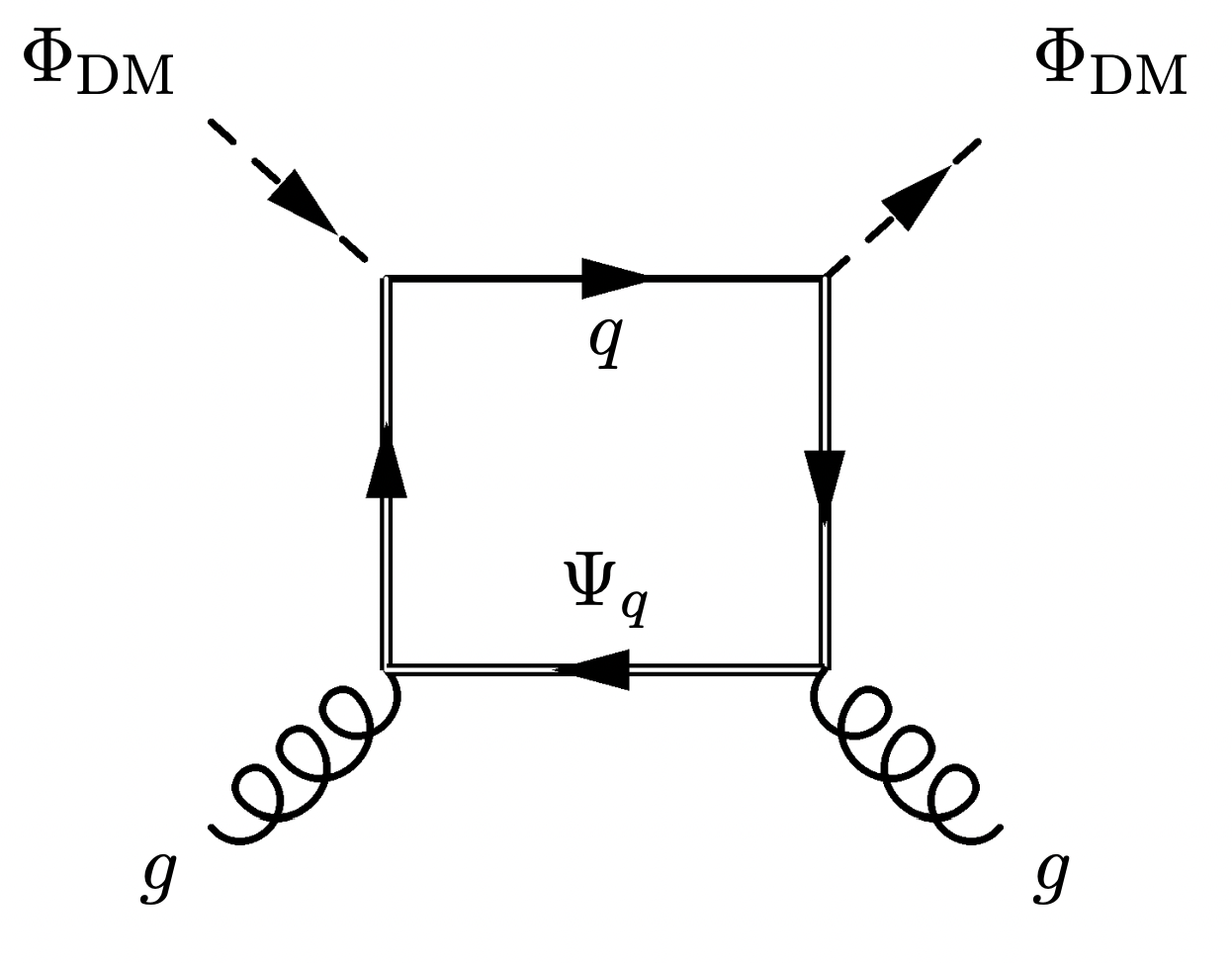}}
&
\subfloat[]{\includegraphics[width=0.3\textwidth]{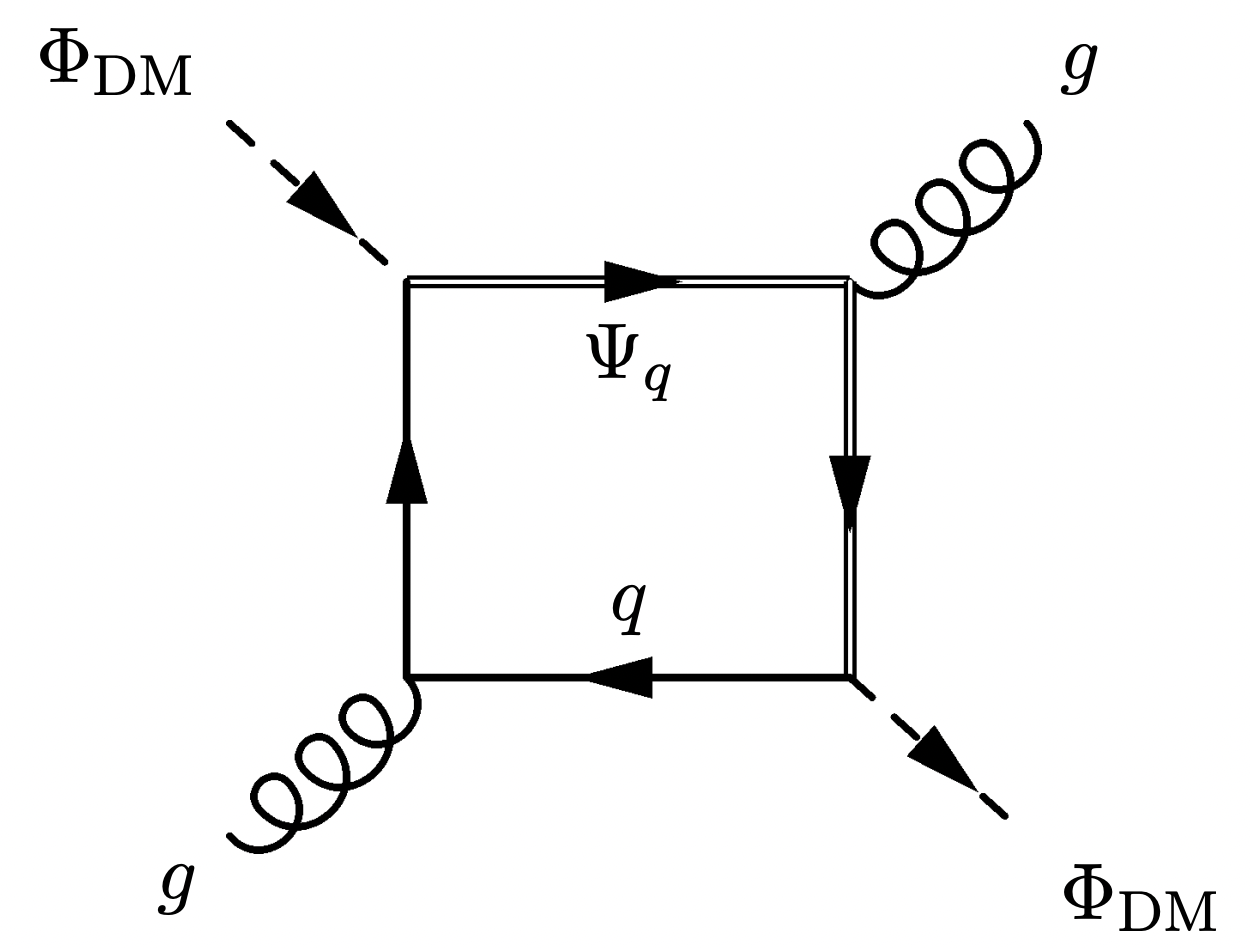}}
\\
\end{tabular}
\end{center}
\caption{\it Feynman diagrams contributing to the matching coefficients in the effective Lagrangian for DM Direct Detection, in the case of scalar DM. Diagram (b) has a partner (not shown) where the $\gamma,Z$ couple to $\Psi_f$ instead of $f$.}
\label{diags}
\end{figure}

We begin our discussion with the case of complex scalar Dark Matter. The corresponding EFT is given by the effective Lagrangian,
\eqa{    
\Leff^{{\rm Scalar},q}
&=&  \sum_{q=u,d} c^q
\left( \Phi_{\rm DM}^\dagger i\overset{\leftrightarrow}{\partial_\mu} \Phi_{\rm DM}\right) \bar q \gamma^\mu q + \sum_{q=u,d,s} d^q m_q \Phi_{\rm DM}^\dagger \Phi_{\rm DM}\, \bar q q+ d^g \frac{\alpha_s}{\pi}\Phi_{\rm DM}^\dagger \Phi_{\rm DM} \, G^{a\mu \nu}G^a_{\mu \nu}
\nonumber\\ 
&&
+ \sum_{q=u,d,s} \frac{g_1^q}{M_{\Phi_{\rm DM}}^2}
\Phi_{\rm DM}^\dagger (i \partial^\mu)(i \partial^\nu ) \Phi_{\rm DM}\, \mathcal{O}^{q}_{\mu \nu}
+ \frac{g_1^g}{M_{\Phi_{\rm DM}}^2}
\Phi_{\rm DM}^\dagger (i \partial^\mu)(i \partial^\nu) \Phi_{\rm DM}\, \mathcal{O}^{g}_{\mu \nu}
\ ,
\label{Scalar:leff}
}
where $\mathcal{O}^{q}_{\mu \nu}$ and $\mathcal{O}^{g}_{\mu \nu}$ are the twist-2 components:
\eq{
\mathcal{O}^q_{\mu \nu}=\bar q \left(\dfrac{iD_\mu \gamma_\nu+iD_\nu \gamma_\mu}{2}-\frac{1}{4}g_{\mu \nu}i\slashed{D}\right)q\ ,
\quad\quad
\mathcal{O}^g_{\mu \nu}=G_\mu^{a\rho} G^a_{\nu \rho}-\frac{1}{4}g_{\mu \nu}G^a_{\rho \sigma}G^{a\rho \sigma}\,.
} 
The various WCs can be decomposed into different contributions, depending on the Feynman diagrams involved (see~\Fig{diags}). 

We begin by considering the first operator in~\Eq{Scalar:leff}, in which the DM is coupled with the quark vector current $\bar q \gamma^\mu q$.
The quark current operator is associated to charge conservation, hence only the valence quarks $u,d$, will contribute. The associated Wilson coefficients $c^{u,d}$ can be decomposed as follows:
\eq{
\label{eq:Wilson_current}
c^{u,d} = c^{u,d}_{\text{tree}} +  c^{u,d}_{Z} + c^{u,d}_{\gamma}   + c^{u,d}_{\text{box}}\ .
}
The term $c^{u,d}_{\text{tree}}$ arises at leading order from the tree-level contribution shown in~\Fig{diags}(a).\footnote{
The terminology "$t$-channel mediator" for $\Psi_f$ refers to the fact that it mediates DM annihilation through $t$-channel exchange. In DM scattering off quarks (the present case), it acts as an $s$-channel mediator. 
}
It is given by~\cite{Kavanagh:2018xeh}
\eq{
c^{q}_{\rm tree}
= -\dfrac{|\Gamma_{L,R}^q|^2}{ 4(M_{\Phi_{\rm DM}}^2-M_{\Psi_f}^2)}\ ,
}
for $q=u,d$.
As already pointed out, the tree-level contribution to the Wilson coefficient is present only if the NP fields couple to first generation quarks. In case this does not occurs, the other contributions to~\Eq{eq:Wilson_current}, which originate at the loop level, are relevant.

The coefficients $c^{q}_{\gamma,Z}$ arise from photon and $Z$  penguins respectively (see~\Fig{diags}(b)). The coefficient associated to photon penguins can be written as:
\eq{
\label{eq:Sgamma}
c^{q}_{\gamma} = \sum_{f}\frac{\alpha |\Gamma_{L,R}^f|^2 N^f_c Q_f Q_q}{24 \pi M^2_{\Psi_f} } \mathcal{F}_\gamma
\Bigg[\frac{m_f^2}{M_{\Psi_f}^2},\frac{M_{\Phi_{DM}}^2}{M_{\Psi_f}^2}\Bigg]\ ,
}
where $\alpha=e^2/(4\pi)$ is the electromagnetic fine-structure constant. $Q_q$ and $Q_f$ are, respectively, the electric charges of the external quarks and of the fermion flowing inside the loop, and $N_c^q=3$, $N_c^\ell=1$ is the number of colors of SM quarks and leptons. The loop function $\F_\gamma$ is given by:
\eqa{
\F_\gamma (x_f,x_\phi) &=&
\frac{1}{x_\phi^2 }
\bigg\{
\frac{(1+x_{f}+2 x_{\phi})\log (x_{f})}{2}-\frac{(x_f-1)(1+x_f-x_\phi)x_\phi}{\Delta}
\nonumber\\
&&
+ \frac{(1-x_f)}{\Delta^{3/2}} \Big(x_{f}^{3}-x_{f}^{2}(1+x_{\phi})
+(1-x_{\phi})^{2}(1+x_{\phi})-x_{f}(1+10 x_{\phi}+x_{\phi}^{2})\Big)
\nonumber\\
&&\times \log\bigg[\frac{1+x_f-x_{\phi}+\sqrt{\Delta}}{2\sqrt{x_{f}}}\bigg] \bigg\}\ ,
}
where
\eq{
\Delta= x_{f}^{2}-2 x_{f}(1+x_{\phi})+(1-x_{\phi})^{2}\ .
\label{Delta}
}
This result is analytically equivalent to the result in~\Reff{Kawamura:2020qxo}, and agrees numerically with the result of~\Reff{Kavanagh:2018xeh} (see also~\cite{Bhattacharya:2015xha} for an earlier computation).
The index $f$ in~\Eq{eq:Sgamma} spans over all the SM fermions running in the loop. By looking at~\Fig{diags}(b), we can straightforwardly notice that a non-zero $c_\gamma^q$ would arise even if only an effective coupling between the DM, a suitable $t$-channel mediator and charged leptons was present. Indeed the NP loop is attached to the photon, which interacts with charged leptons, which in turn is coupled with the quarks inside the nucleon.

For the $Z$-penguin contribution, we find
\eq{
c_Z^q=\frac{G_F}{4\sqrt{2}}\sum_f \frac{T_3^f (T_3^q-2Q_q s_W^2)N^f_c| \Gamma_{L,R}^f|^2}{\pi^2}\,\frac{m_f^2}{M_{\Psi_f}^2}
\,\F_Z
\bigg[\frac{m_f^2}{M_{\Psi_f}^2},\frac{M_{\Phi_{DM}}^2}{M_{\Psi_f}^2}\bigg]\ ,
}
where
\eq{
\F_Z(x_f,x_\phi)=
\frac{1}{x_\phi}+\frac{1-x_f+x_\phi}{2 x_\phi^2}\log{x_f}
+\frac{1-2x_f + (x_f-x_\phi)^2}{x_\phi^2\sqrt{\Delta}}
\log{\bigg(\frac{1+x_f-x_\phi+\sqrt{\Delta}}{2\sqrt{x_f}}\bigg)}
\ ,
}
with $\Delta$ given in~\Eq{Delta}. Looking again at~\Fig{diags}(b) we can see that also a non zero $c_Z^q$ arises if the NP fields couple only to leptons. Contrary to the case of photon penguin, the $Z$-penguin contribution is proportional to $m_f^2$, with $f$ being the fermion running in the loop, and consequently strongly suppressed in this case. 

Finally, the coefficients $c^{q}_{\text{box}}$ arise from the box diagrams in~\Fig{diags}(c). We have,
\eq{
c^{q}_{\text{box}} =\sum_{f}\frac{1}{4}\dfrac{N_c^f |\Gamma_{L,R}^f|^2 |\Gamma_{L,R}^q|^2}{32 \pi^2 M_{\Phi_{\rm DM}}^2}\,
\mathcal{F}_\text{box}\Bigg[\frac{m_f^2}{M_{\Psi_f}^2},\frac{M_{\Phi_{\rm DM}}^2}{M_{\Psi_f}^2}\Bigg]\ ,
}
where
\eqa{
\mathcal{F}_\text{box}\left(x_f,x_\phi\right)
&=&
\frac{x_f-x_\phi}{x_\phi-1}+\frac{\beta_2 x_f^2(x_f-3 x_\phi)}{x_\phi(x_\phi-1)^2}\log \Bigg[\frac{\sqrt{x_f}(1+\beta_2)}{2\sqrt{x_\phi}}\Bigg]
\nonumber\\
&& \hspace{-12mm}
-\frac{x_f+x_\phi}{2x_\phi}\log{\left[x_f\right]}+\frac{(x_f^3-5x_f^2 x_\phi+4 x_f x_\phi^2+2x_\phi^3)}{2x_\phi(x_\phi-1)^2}\log\left[x_\phi\right]
\nonumber\\
&& \hspace{-12mm}
+\frac{-x_f^4+x_\phi(x_\phi-1)^3+x_f(x_\phi-1)^2(2x_\phi-1)+x_f^3(1+6x_\phi)+x_f^2(1-x_\phi(5+8x_\phi))}{ x_\phi \sqrt{\Delta}(x_\phi-1)^2}
\nonumber\\
&& \hspace{-12mm}
\times \log\Bigg[\frac{1+x_f-x_\phi+\sqrt{\Delta_\phi}}{2\sqrt{x_f}}\Bigg]\ ,
}
with
\eq{
\beta_2=\sqrt{1-4 \frac{x_\phi}{x_f}}\ .
}

\bigskip

We now consider the scalar operator, with Wilson coefficient $d^q$.
This Wilson coefficient receives two contributions. 
The first one comes from tree-level interactions between the DM, the fermionic mediator and the light SM quarks (\Fig{diags}(a)), and is given by
\eq{
d_{\rm QCD}^q=|\Gamma_{L,R}^q|^2\frac{2 M_{\Psi_q}^2-M_{\Phi_{\rm DM}}^2}{ 4 {(M_{\Psi_q}^2-M_{\Phi_{\rm DM}}^2)}^2}\ .
}
Here we have introduced the label QCD, which will be also used in the following, 
stemming from the fact that the coefficient is originated due to the presence of an explicit interaction vertex of a $t$-channel colored mediator with the quarks (and/or gluons) with the nucleon.

\begin{figure}
\begin{center}
\subfloat[]{\includegraphics[width=0.3\textwidth]{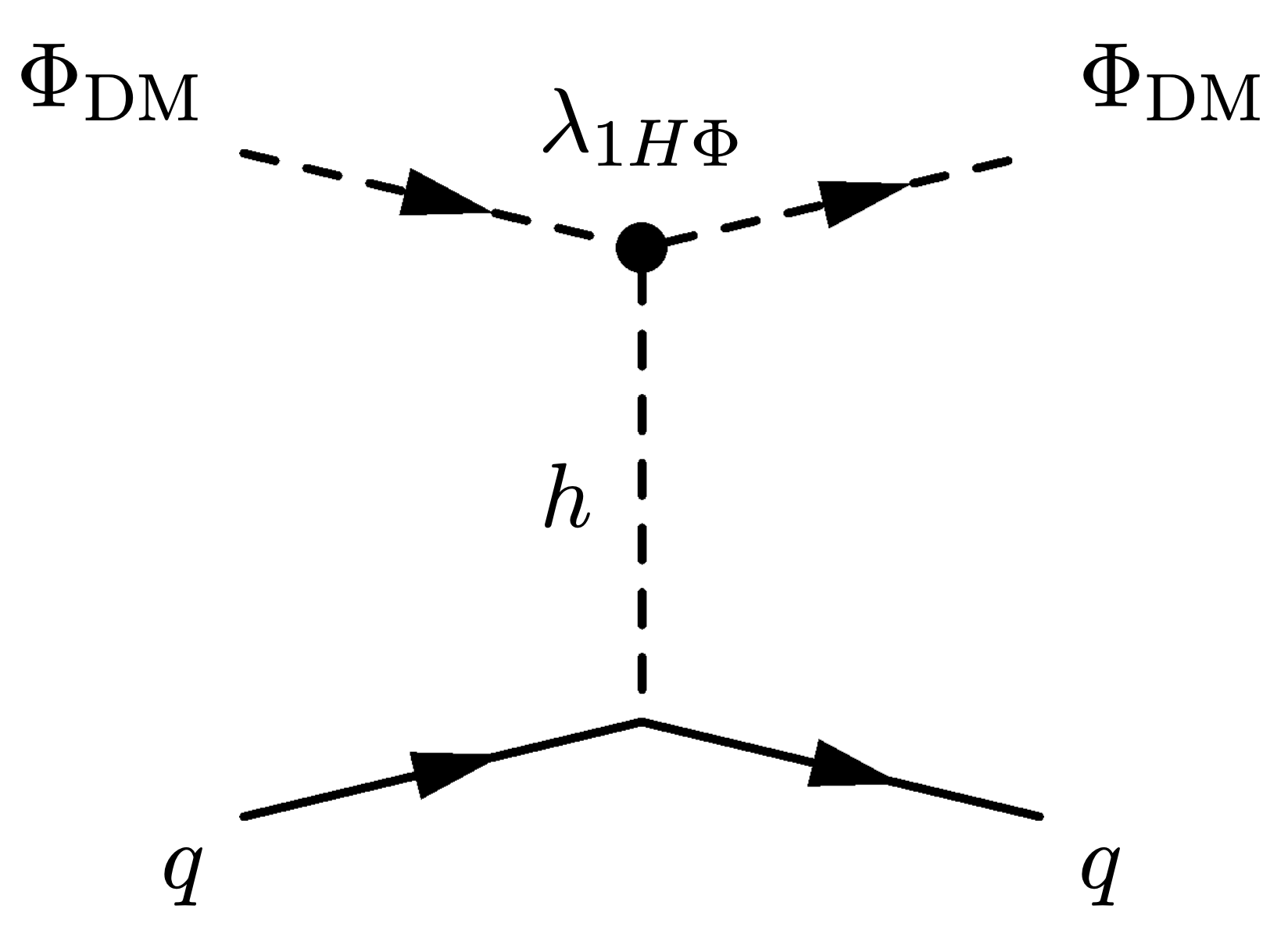}}
\quad
\subfloat[]{\includegraphics[width=0.3\textwidth]{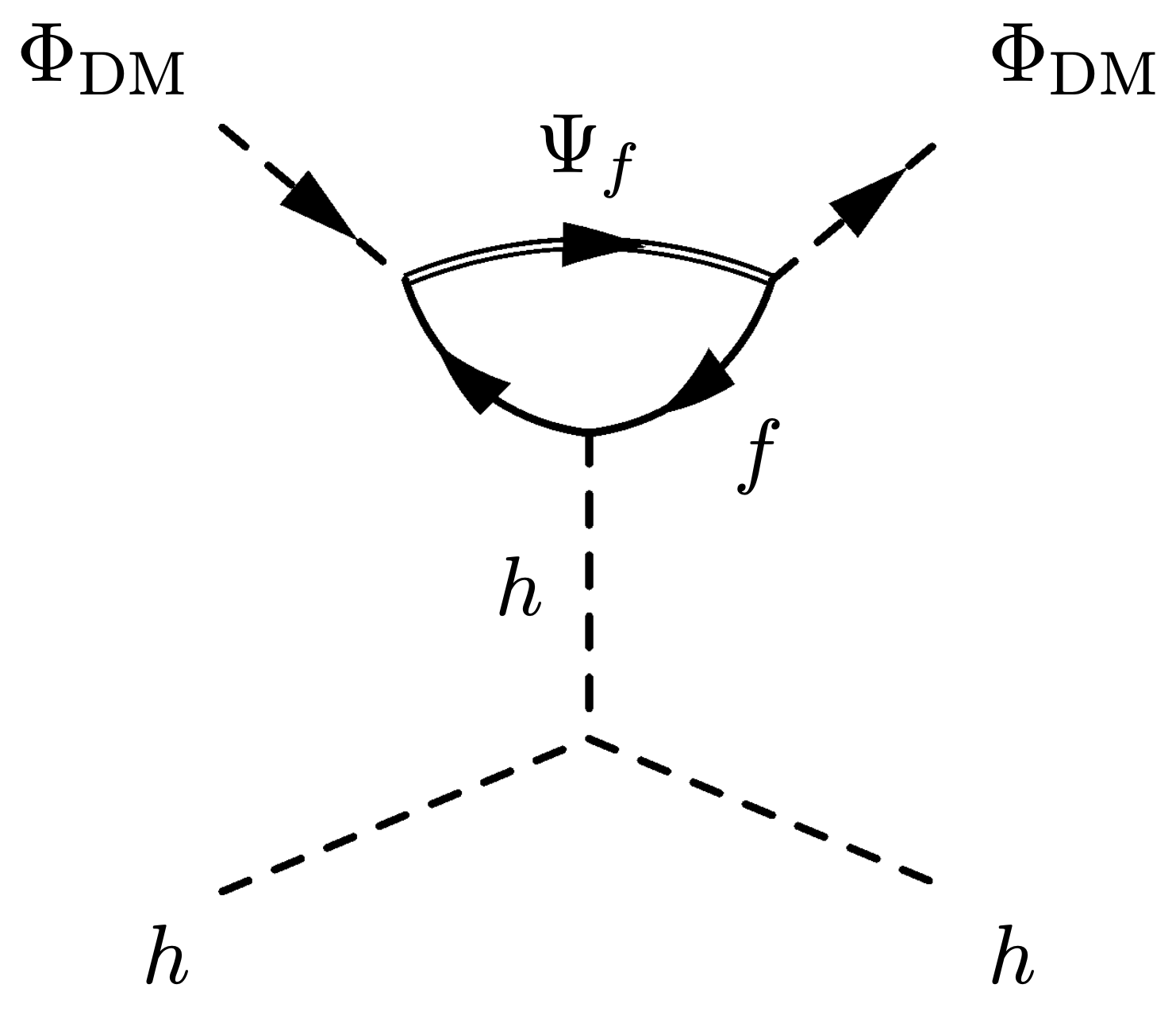}}
\quad
\subfloat[]{\includegraphics[width=0.3\textwidth]{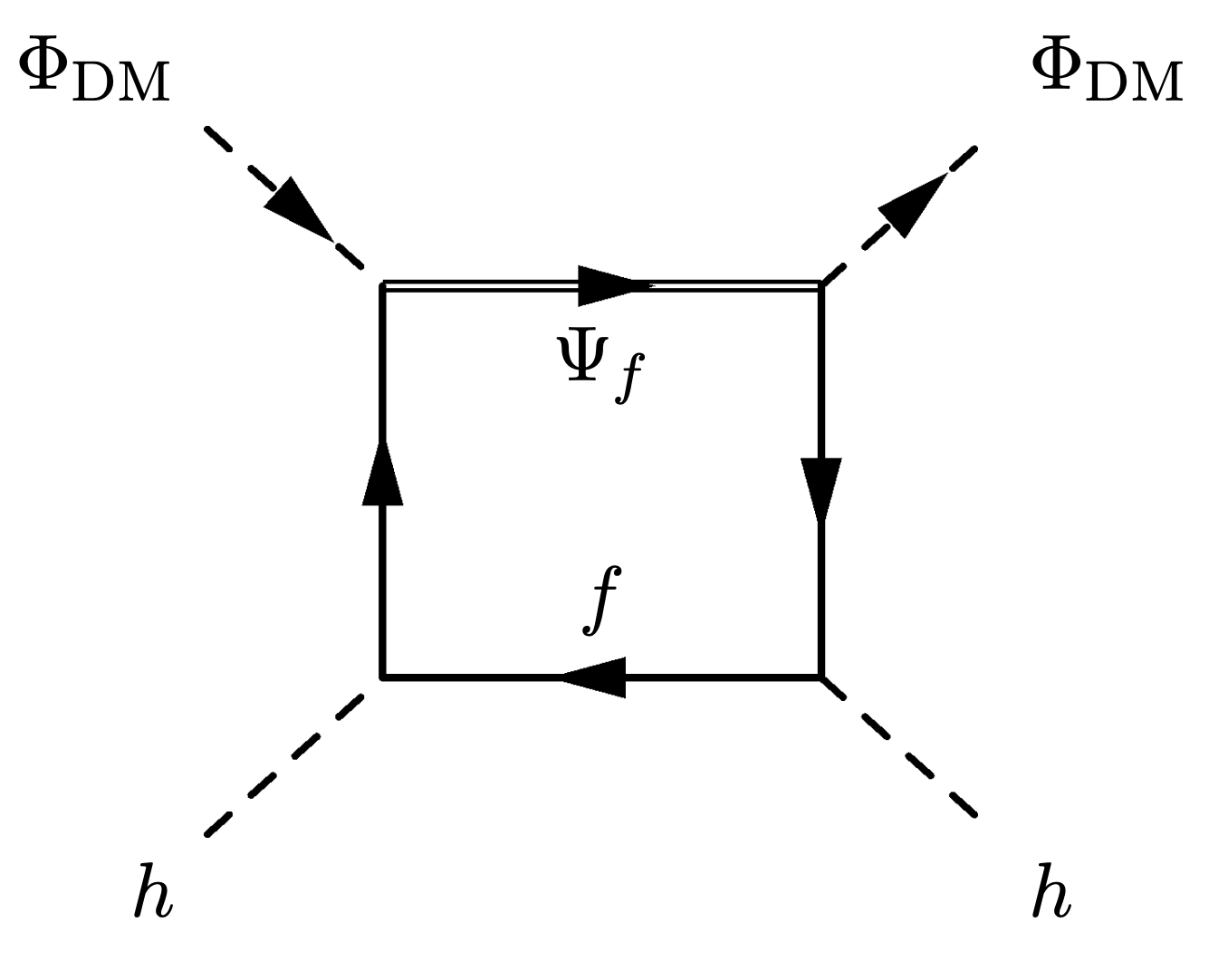}}
\end{center}
\caption{\it One-loop contributions to the process $\Phi\,H\to\Phi\,H$. Their divergences renormalize the $\Phi^2 H^2$ coupling non-multiplicatively. Thus this coupling can only be set to zero at a single scale.}
\label{diags2}
\end{figure}

The second contribution comes the from Higgs penguins in~\Fig{diags}(b),and leads to the Wilson coefficient
\eq{
d_H^{q}=\sum_f \frac{g^2 |\Gamma^f_{L,R}|^2 M_{\Psi_f}^2}{32\pi^2 m_H^2 m_W^2}\ \F_H\left(\frac{m_f^2}{M_{\Psi_f}^2},\frac{M_{\Phi_{\rm DM}}^2}{M_{\Psi_f}^2}\right) .
}
The Higgs penguin contribution by itself is UV divergent, giving the {\it unrenormalized} result for the loop function
\eqa{
\F_H^{(0)}(x_f,x_\phi)
&=&
2 x_f 
+ \frac{x_f(x_\phi-x_f)}{x_\phi}\log{x_f}
+2x_f \left(\frac{1}{\eps}+\log{\frac{\mu ^2}{m_f^2}}\right)
\nonumber\\
&&
+ 2 x_f\frac{x_f^2+(x_\phi-1)x_\phi-x_f(1+2x_\phi)}{x_\phi \sqrt{\Delta}}\log{\left[\frac{1+x_f-x_\phi+\sqrt{\Delta}}{2 \sqrt{x_f}}\right]}
\ ,
\label{FH0}
}
with $\epsilon=2-d/2$ and $\mu$ the $\overline{\text{MS}}$ UV regulator and scale, respectively, and $\Delta$ given in~\Eq{Delta}.
The counterterm needed to cancel this UV divergence comes from the $\Phi_\text{DM}^\dagger \Phi_\text{DM} H H^\dagger$ term in~\Eq{eq:scalar_lagrangian}.
This operator induces the tree-level contribution to $d^q$ shown in~\Fig{diags2}(a), and it is strongly constrained by phenomenology~\cite{Arcadi:2019lka}.
However, even if we assume $\lambda_{1H\Phi}$ to be negligible (at some renormalization scale), this operator is renormalized by terms proportional to the Yukawa couplings $\Gamma_{L,R}^f$
arising from the UV divergencies of the diagrams in~\Fig{diags2}(b,c).
The counterterm cancels the divergence in~\Eq{FH0}, but one is forced to include the tree level contribution in~\Fig{diags2}(a) with a radiatively generated $\lambda_{1H\Phi}$ which is of the same order in the couplings, i.e. of order $g^2 |\Gamma_{L,R}^f|^2$. 

A calculation of the diagrams in~\Fig{diags2}(b,c) gives the following unrenormalized one-loop contribution to the $\Phi_\text{DM} H\to \Phi_\text{DM} H$ amplitude
\eq{
\A(\Phi_\text{DM} H\to \Phi_\text{DM} H)
=-\sum_f \frac{g^2 m_f^2 |\Gamma_{L,R}^f|^2}{16 \pi^2 m_W^2}\left(1-3\right)\left(\frac{1}{\eps}+\log{\frac{\mu^2}{m_f^2}}\right)+ \text{counterterms} + \text{finite}\ ,
}
where in the term $(1-3)$ the `$1$' comes from the box diagram and the `$-3$' comes from the penguin diagram.
The divergence is cancelled by the same counterterm as before, by defining the bare $\Phi_\text{DM}^\dagger \Phi_\text{DM} H H^\dagger$ coupling
\eq{
\lambda_{1H\Phi}^{(0)}=\lambda_{1H\Phi}(\mu)
+ \sum_f
\frac{g^2 m_f^2 |\Gamma_{L,R}^f|^2}{16 \pi^2 m_W^2}\left(\frac{1}{\eps}+\log{\frac{\mu^2}{m_f^2}}\right)
\label{lambda0}
}
in terms of the renormalized $\overline{\text{MS}}$ coupling $\lambda_{1H\phi}(\mu)$ at the renormalization scale $\mu$. 
This fixes the scale dependence of the coupling $\lambda$ to order $g^2$ through the RG equation:
\eq{
\frac{d\lambda_{1H\Phi}(\mu)}{d\log{\mu^2}}  =- \sum_f\frac{g^2 m_f^2 |\Gamma_{L,R}^f|^2}{16 \pi^2 m_W^2}\ .
}
Without resummation, we have
\eq{
\lambda_{1H\Phi}(\mu)=\lambda_{1H\Phi}(M)- \log\frac{\mu^2}{M^2} \sum_f \frac{g^2 m_f^2 |\Gamma_{L,R}^f|^2}{16 m_W^2 \pi^2}\ ,
\label{lambda(mu)}
}
with $M$ being the scale at which the coupling is fixed (for example through experiment). We will impose $\lambda_{1H\Phi}(M)=0$ at the scale $M=M_{\Psi_f}$.

Taking into account these considerations and including the diagram in~\Fig{diags2}(a), one finds the renormalized version of~\Eq{FH0},
\eqa{
\F_H (x_f,x_\phi)
&=&
2 x_f +2x_f\frac{x_f^2+(x_\phi-1)x_\phi-x_f(1+2x_\phi)}{x_\phi \sqrt{\Delta}}\log{\left[\frac{1+x_f-x_\phi+\sqrt{\Delta}}{2 \sqrt{x_f}}\right]}
\nonumber\\
&& + \frac{x_f(x_\phi-x_f)}{x_\phi}\log{x_f}+\frac{32 \pi^2 m_W^2\lambda_{1H\Phi}^{(f)}(\mu)}{g^2 M_{\Psi_f}^2 |\Gamma^f_{L,R}|^2}
+ 2 x_f \log{\frac{\mu^2}{m_f^2}}\ .
\label{Fxfxphi}
}
Here $\Delta$ is again given by~\Eq{Delta}, and $\lambda_{1H\Phi}^{(f)}$ is the contribution to $\lambda_{1H\Phi}(\mu)$ from each fermion flavor (see~\Eq{lambda(mu)}).
This function is manifestly scale-independent.

In summary, this result can be explained by interpreting $d_H^q$ as a radiative contribution to the coupling of the Higgs portal operator $\Phi_\text{DM}^\dagger \Phi_\text{DM} H^\dagger H$.
In other words, a Higgs portal coupling is an unavoidable feature of the theory; even if it is set to zero at some scale $M$, it will be radiatively generated at a lower energy scale $\mu$. Throughout this work we will identify the scale $M$ with the mass of the $t$-channel mediator and take $\lambda_{1H\Phi}(M_{\Psi_f})=0$.

\bigskip

We now consider the third operator in~\Eq{Scalar:leff}, coupling the DM with gluons. The relevant contributions are shown in~\Fig{diags}(d-f). We find
\eq{
d_{\rm QCD}^g=\frac{1}{2}\frac{|\Gamma_{L,R}^q|^2 }{24  M_{\Psi_q}^2}\,
\F_{gg}^{(1)}\Bigg[\frac{m_q^2}{M_{\Psi_q}^2},\frac{M_{\Phi_{DM}}^2}{M_{\Psi_q}^2}\Bigg]
\ ,
}
with
\eqa{
\mathcal{F}_{gg}^{(1)}\left(x_f,x_\phi\right)&=&
\frac{12 \log \left[\frac{1+x_{f}-x_{\phi}+\sqrt{\left(x_\phi-1\right)^{2}-2\left(1+x_{\phi}\right) x_{f}+x_{f}^{2}}}{2 \sqrt{x_{f}}}\right] x_{f}x_\phi\left(1+x_{f}-x_{\phi}\right)}{\left(x_{f}^{2}+\left(-1+x_{\phi}\right)^{2}-2 x_{f}\left(1+x_{\phi}\right)\right)^{5 / 2}}
\nonumber\\
&&
-x_\phi \frac{x_{f}^{2}-2 x_{f}\left(-5+x_{\phi}\right)+\left(-1+x_{\phi}\right)^{2}}{\left(x_{f}^{2}+\left(-1+x_{\phi}\right)^{2}-2 x_{f}\left(1+x_{\phi}\right)\right)^{2}}\ .
}
Finally the matching coefficients for the two twist-2 operators are given by
\eqa{
g_{1,\rm QCD}^q &=&
|\Gamma_{L,R}^q|^2\frac{ M_{\Phi_{\rm DM}}^2}{{\left(M_{\Psi_q}^2-M_{\Phi_{\rm DM}}^2\right)}^2} \ ,
\nonumber\\
g_{1,\rm QCD}^g &=&
\frac{1}{2}\frac{|\Gamma_{L,R}^q|^2 \alpha_s M_{\Phi_{DM}}^2}{6\pi M_{\Psi_q}^4}\,
\mathcal{F}_{gg}^{(2)}\Bigg[\frac{m_q^2}{M_{\Psi_q}^2},\frac{M_{\Phi_{DM}}^2}{M_{\Psi_q}^2}\Bigg]\ ,
}
with
\begin{align}
& \mathcal{F}_{gg}^{(2)}\left(x_f,x_\phi\right)=\frac{3\left(x_{f}^{2}+\left(-1+x_{\phi}\right)^{2}-2 x_{f}\left(3+x_{\phi}\right)\right)}{\left(x_{f}^{2}+\left(-1+x_{\phi}\right)^{2}-2 x_{f}\left(1+x_{\phi}\right)\right)^{2}}\nonumber\\
&  -\frac{4 \log \left[\frac{1+x_{f}-x_{\phi}+\sqrt{\left(x_{f}-1\right)^{2}-2\left(1+x_{f}\right) x_{\phi}+x_{\phi}^{2}}}{2 \sqrt{x_{f}}}\right]\left(1+x_{f}-x_{\phi}\right)\left(x_{f}^{2}+\left(x_{\phi}-1\right)^{2}-x_{f}\left(5+2 x_{\phi}\right)\right)}{\left(x_{f}^{2}+\left(x_{\phi}-1\right)^{2}-2 x_{f}\left(1+x_{\phi}\right)\right)^{5 / 2}}\ .
\label{eq:Cs}
\end{align}

From the effective Lagrangian in~\Eq{Scalar:leff} one can calculate the following scattering cross-section for DM on nucleons (for illustration we focus on the case of the proton):
\eq{
\sigma_{\Phi_{\rm DM}}^{{\rm SI},\, p}= \frac{\mu_{\Phi_{DM}\,p}^2}{\pi}\, \frac{\left[Z f_p +(A-Z)f_n\right]^2}{A^2}\,,
}
where $\mu_{\Phi_{DM}\,p} = M_{\Phi_{\rm DM}} M_p/(M_{\Phi_{\rm DM}}+M_p)$ is the DM/proton reduced mass.  The extra factor depending on $A,Z$, being respectively the mass and atomic number of the detector material, allows a consistent comparison with experimental limits which assume equal coupling of the DM with protons and neutrons \cite{Feng:2013vod}.
$f_{p,n}$ represent, in fact, the effective coupling of the DM with protons and neutrons and read, in terms of the Wilson coefficients illustrated above:
\eqa{
f_p
&=&
c_{\rm tree}^p+c_Z^p+c_\gamma^p+c_{\rm box}^p+M_p\sum_{q=u,d,s}\left(f_q^p d_q  +\frac{3}{4}g_1^q \left(q(2)+\bar{q}(2)\right)\right)
\nonumber\\
&& +\frac{3}{4}M_p\sum_{q=c,b,t}g_1^g G(2) -\frac{8}{9}f_{TG}f_G\ , 
\label{eq:fpfn_cs}
\\
f_n
&=& c_{\rm tree}^n+c_Z^n+c_{\rm box}^n +\frac{3}{4}M_n\sum_{q=u,d,s}\left(q(2)+\bar{q}(2)\right)g_1^q
\nonumber\\
&& + \frac{3}{4}M_n\sum_{q=c,b,t}g_1^g G(2) -\frac{8}{9}f_{TG}f_G\ ,
}
where  $c_{i}^p=2 c_i^u+c_i^d$, $c_i^n=c_i^u+2 c_i^d$ and  $f_G=\sum_{q=c,b,t} d_q^g+d_H^q$.
The parameters $f_q^{N=n,p},\,f_{TG},\,q(2),$ and $G(2)$ are nucleon form factors defined as:
\eqa{
\langle N | m_q \bar q q|N \rangle
&=&
M_N f_q^N,\,\,\,\,\,f_{TG}=1-\sum_{q=u,d,s}f_q^N \ ,
\\
\langle N |\mathcal{O}^q_{\mu \nu} |N \rangle
&=&
\frac{1}{M_N}\left(p_\mu p_\nu-\frac{1}{4}M_N^2 g_{\mu \nu}\right) \left(\bar q(2)+q(2)\right)\ ,
\\
\langle N |\mathcal{O}^g_{\mu \nu} |N \rangle
&=&
\frac{1}{M_N}\left(p_\mu p_\nu-\frac{1}{4}M_N^2 g_{\mu \nu}\right) G(2)\ .
}
In the numerical analysis we will use the default values implemented into the micrOMEGAs package~\cite{Belanger:2015nma}.

\subsection{Real Scalar DM}

In the case of real scalar DM, the operator $\Phi_{\rm DM}^\dagger i\overset{\leftrightarrow}{\partial_\mu} \Phi_{\rm DM}$ identically vanishes. The effective Lagrangian is thus given by~\Eq{Scalar:leff} with~\cite{Hisano:2015bma}
\eq{
c^{q}_\text{Real\ Scalar} = 0\ ,
}
and with the Wilson coefficients $d^{q,g}$, $q_1^{q,g}$ identical to the ones given in~\Sec{sec:ComplexScalar}.
Due to phase-space symmetry, the scattering cross-section of the real scalar case will be multiplied by a factor 4 with respect to its complex scalar counterpart.

\subsection{Dirac DM}

We move now to the case of fermionic DM, denoted by $\Psi_{\rm DM}$.
The effective Lagrangian describing the effective interactions of a Dirac DM candidate with quarks and gluons is given by
\eqa{
\Leff^{\text{Dirac},q}
&=&
\sum_{q=u,d} c^q\, \bar\Psi_{\rm DM} \gamma_\mu \Psi_{\rm DM}\,\bar q \gamma^\mu q 
+\sum_{q=u,d,s}\tilde{c}^q\, \bar\Psi_{\rm DM} \gamma_\mu \gamma_5 \Psi_{\rm DM} \,\bar q \gamma^\mu \gamma_5 q
\nonumber\\
&&
+ \sum_{q=u,d,s} d^q\, m_q\bar \Psi_{\rm DM} \Psi_{\rm DM}\,\bar q q 
+\sum_{q=c,b,t} d_q^g\, \bar \Psi_{\rm DM}\,\Psi_{\rm DM} G^{a\mu \nu}G^a_{\mu \nu}
\nonumber\\
&&
+ \sum_{q=u,d,s} \left(g_{1}^{q} \frac{\bar \Psi_{\rm DM} i \partial^\mu \gamma^\nu \Psi_{\rm DM} \mathcal{O}^q_{\mu \nu} }{M_{\Psi_{\rm DM}}}+ g_{2}^{q} \frac{\bar \Psi_{\rm DM} \left( i  \partial^\mu \right)\left(i \partial^\nu \right) \Psi_{\rm DM} \mathcal{O}^q_{\mu \nu} }{M_{\Psi_{\rm DM}}^2}\right)
\nonumber\\
&&
+ \sum_{q=c,b,t}\left( g_{1}^{g,q} \frac{\bar \Psi_{\rm DM} i \partial^\mu \gamma^\nu \Psi_{\rm DM} \mathcal{O}^g_{\mu \nu} }{M_{\Psi_{\rm DM}}}+ g_{2}^{g,q} \frac{\bar \Psi_{\rm DM} \left( i  \partial^\mu \right)\left(i \partial^\nu \right) \Psi_{\rm DM} \mathcal{O}^g_{\mu \nu} }{M_{\Psi_{\rm DM}}^2}\right)\ .
\label{eq:Lfermion}
}
In addition to the Lagrangian above, DM DD is influenced as well by effective coupling of the DM with the photons, described by the following Lagrangian:
\eq{
\label{eq:Lfermionphoton}
\Leff^{\text{Dirac}, \gamma}=\frac{\tilde{b}_\Psi}2 \,\bar \Psi_{\rm DM} \sigma^{\mu \nu}\Psi_{\rm DM} F_{\mu \nu}+b_\Psi \bar \Psi_{\rm DM} \gamma^\mu \Psi_{\rm DM} \partial^\nu F_{\mu \nu}\,,
}
featuring two terms dubbed, respectively, magnetic dipole moment and charge radius operators. 
Notice that, even if interactions with photons were also present in the case of scalar DM, we did not need to write a separate Lagrangian since we could use the relation $\partial_\nu F_{\mu\nu} = -e\,Q_q \bar q\gamma^\mu q$ to reduce the operators $\big(i \Phi^\dagger \overset{\leftrightarrow}{\partial_\mu} \Phi\big) \partial_\nu F_{\mu\nu}$ and $\partial_\nu \Phi^\dagger  \partial_\mu  \Phi F_{\mu\nu}$ into $\big(i \Phi^\dagger \overset{\leftrightarrow}{\partial_\mu} \Phi\big) (\bar q_i \gamma^\mu q_i)$.
Similarly to the case of scalar DM, we will illustrate individually the coefficients associated to the different operators.

\begin{figure}[t!]
\begin{center}
\begin{tabular}{ccc}
\subfloat[]{\includegraphics[width=0.3\textwidth]{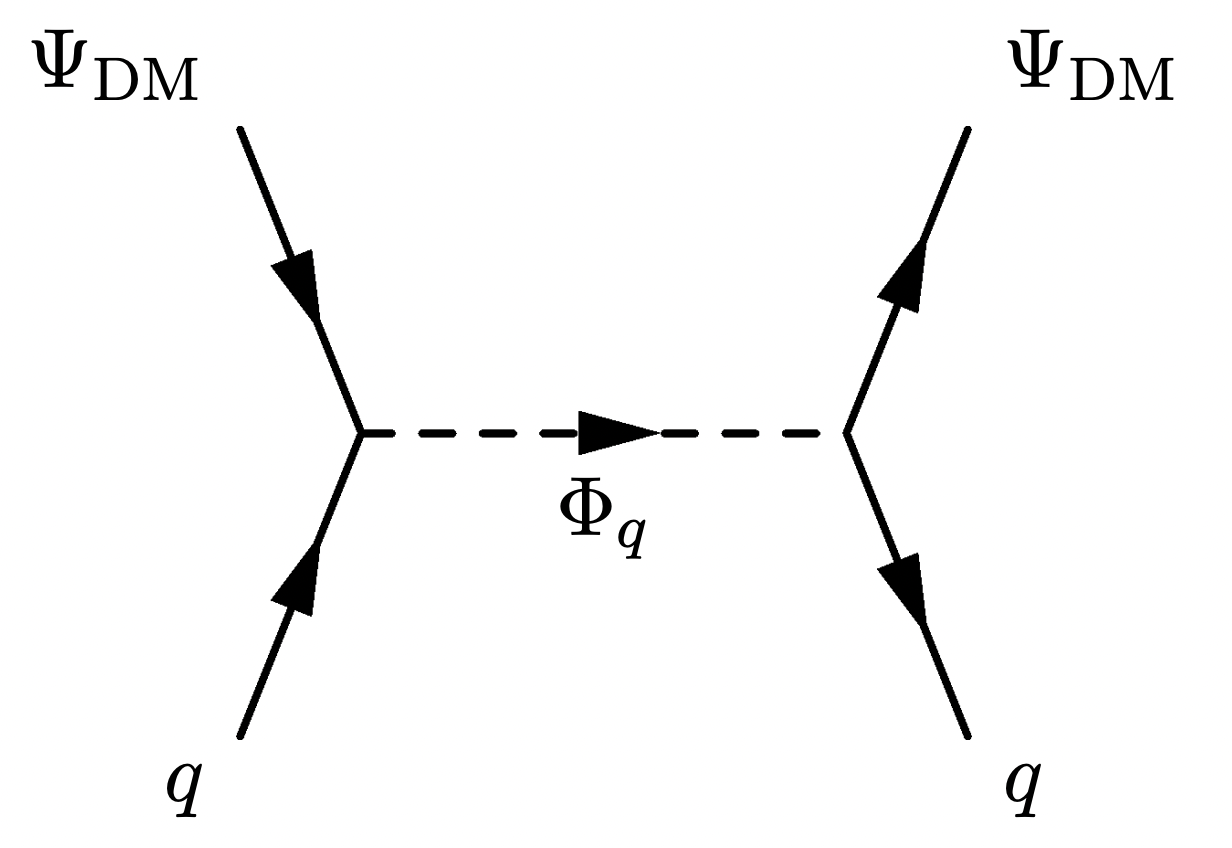}}
&
\subfloat[]{\includegraphics[width=0.3\textwidth]{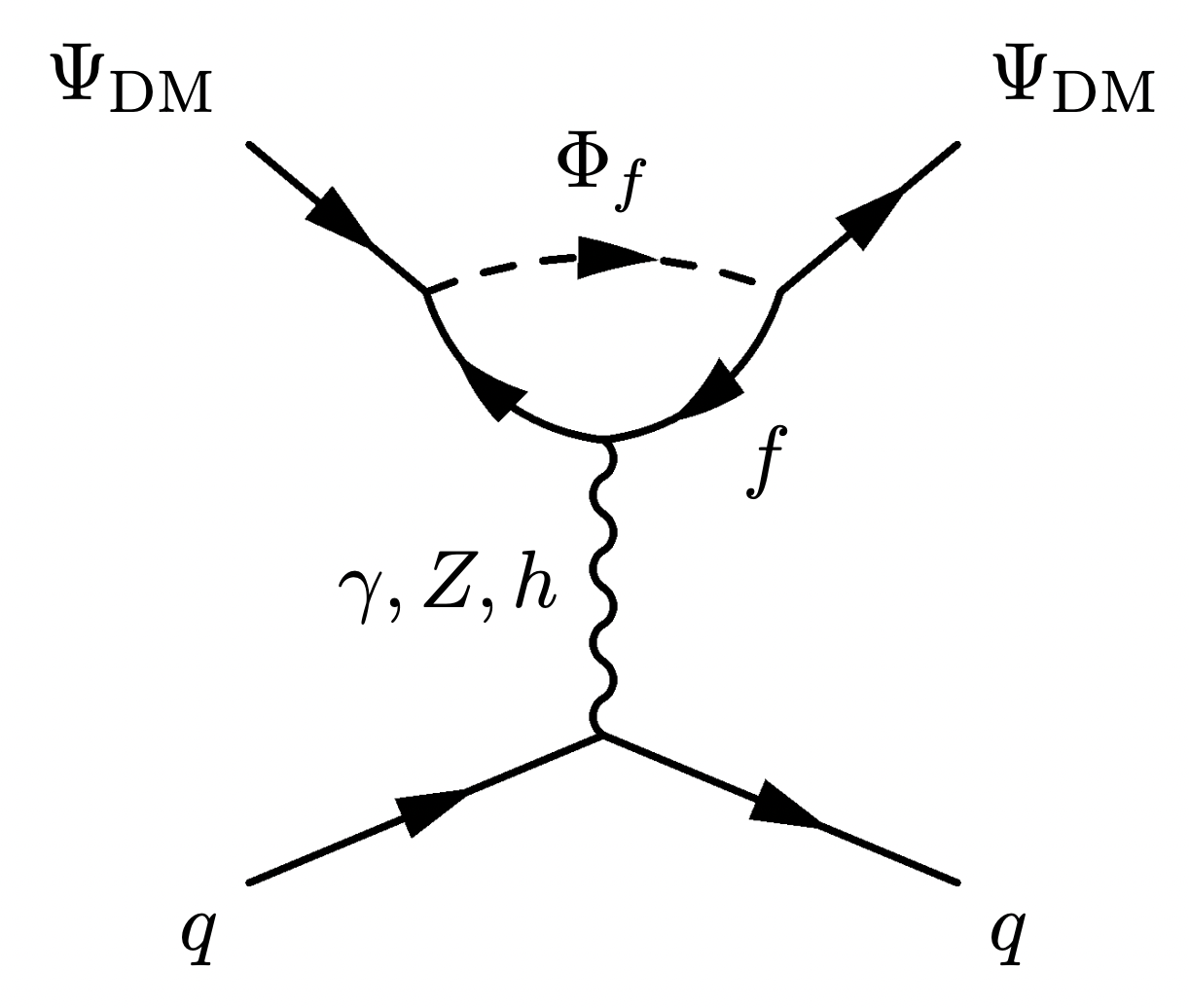}}
&
\subfloat[]{\includegraphics[width=0.3\textwidth]{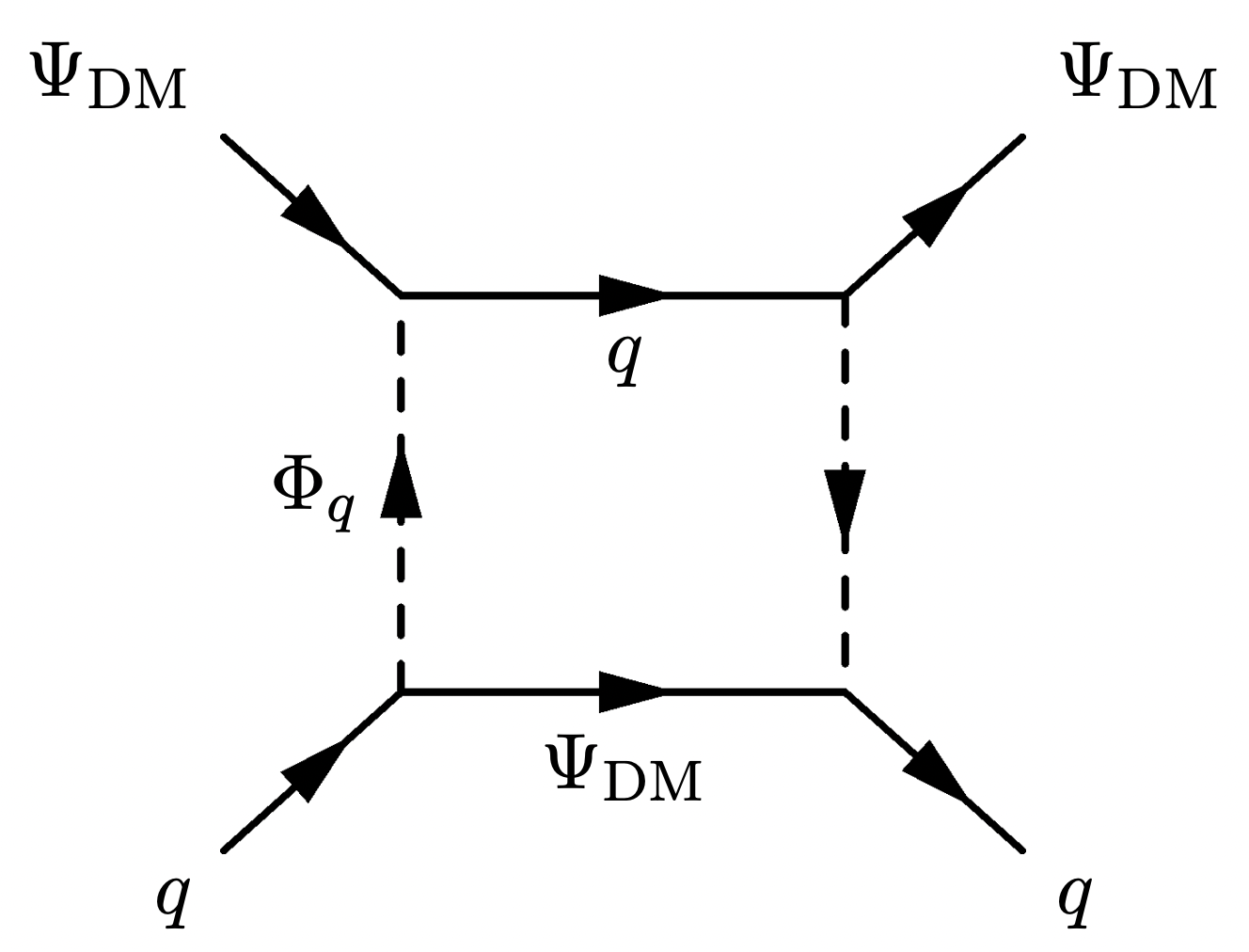}}
\\
\subfloat[]{\includegraphics[width=0.3\textwidth]{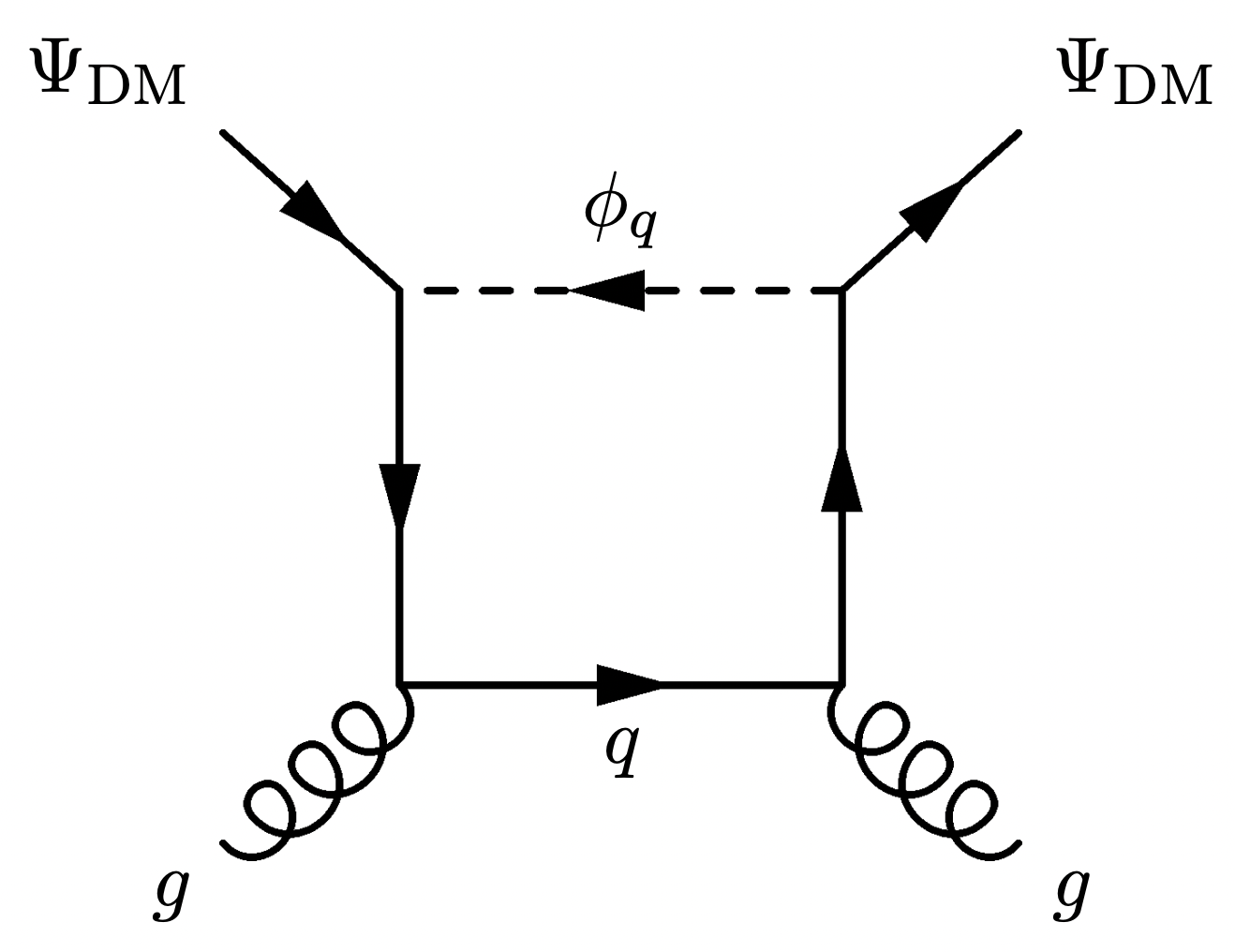}}
&
\subfloat[]{\includegraphics[width=0.3\textwidth]{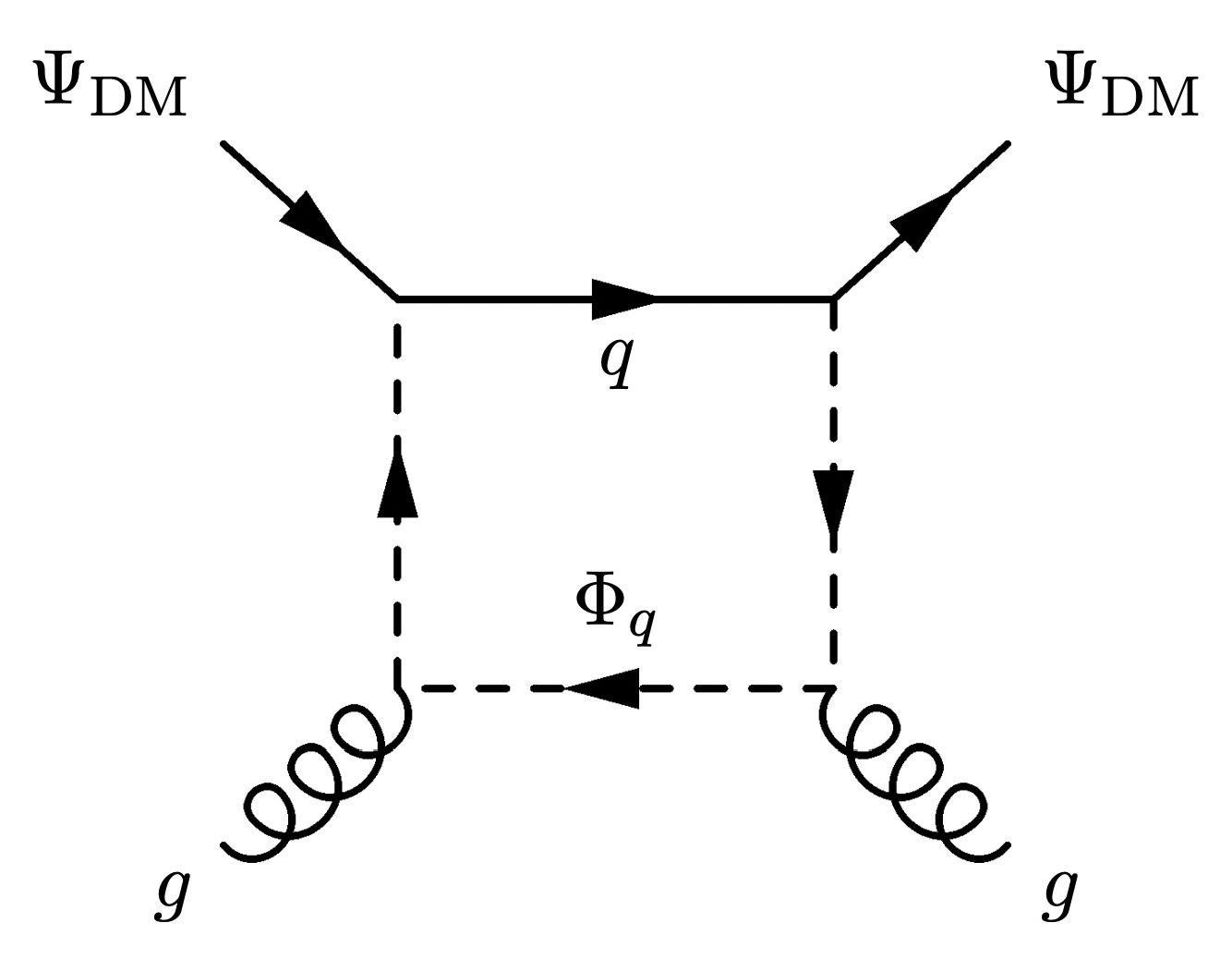}}
&
\subfloat[]{\includegraphics[width=0.3\textwidth]{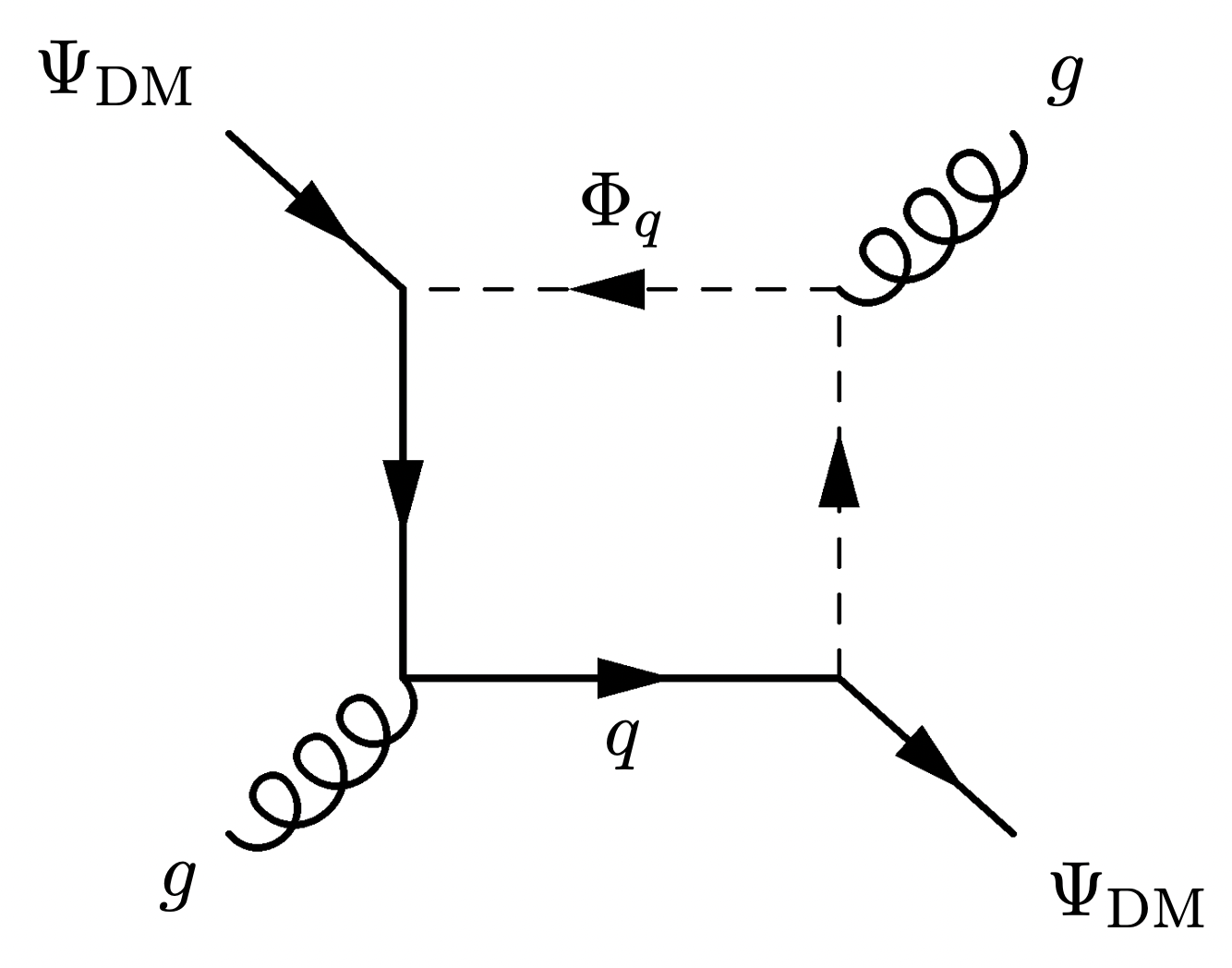}}
\\
&\subfloat[]{\includegraphics[width=0.3\textwidth]{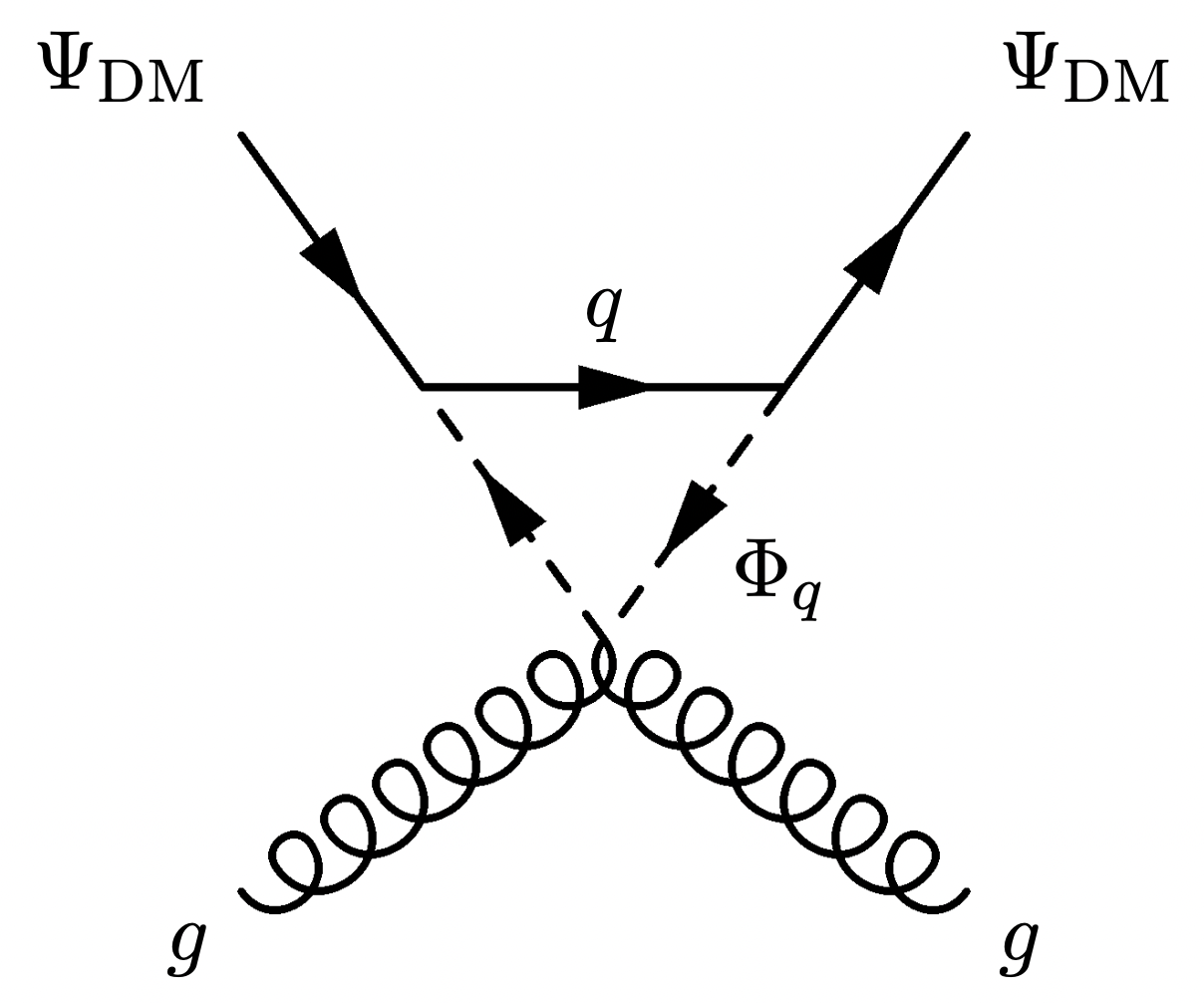}}
&
\\
\end{tabular}
\end{center}
\caption{\it Feynman diagrams contributing to the matching coefficients in the effective Lagrangian for DM Direct Detection, in the case of fernionic DM. Diagram (b) has a partner (not shown) where the $\gamma,Z$ couple to $\Phi_f$ instead of $f$.}
\label{diags3}
\end{figure}

We start again with the operator coupling the DM to the quark current. Its coefficient is the combination of three contributions:
\eq{
c^q=c_{\rm tree}^q+c_Z^q+c_{\rm box}^q\ ,
}
The first one comes from tree level interactions (\Fig{diags3}(a)) and it is different from zero only in case of coupling with first generation quark flavors. Its expression is totally analogous to the one in scalar DM model,
\eq{
c_{\rm tree}^q=\frac{|\Gamma_{L,R}^q|^2}{8 (M_{\Phi_q}^2-M_{\Psi_{\rm DM}}^2)}\ .
}
The second contribution arises from $Z$-penguin diagrams (\Fig{diags3}(b)) and is given by
\eq{
c_Z^{q,\,f_R}=\big[T^3_q-2Q_q \sin^2{(\theta_W)}\big] \sum_f \frac{G_F}{\sqrt{2}}\frac{N^f_c}{16 \pi^2}T^3_f  |\Gamma_{L,R}^f|^2 \F_Z\left(\frac{m_f^2}{M_{\Phi_f}^2},\frac{M_{\Psi_{\rm DM}}^2}{M_{\Phi_f}^2}\right)\ ,
}
\eq{
c_Z^{q}=\big[T^3_q-2Q_q \sin^2{(\theta_W)}\big]
\sum_f \frac{G_F}{\sqrt{2}}\frac{N^f_c}{16 \pi^2}\big(Q_f-\frac{1}{6}\big)  \big[\big(\Gamma_{L}^f\big)^2-\big(\Gamma_{R}^f\big)^2 \big]\F_Z\left(\frac{m_f^2}{M_{\Phi_f}^2},\frac{M_{\Psi_{\rm DM}}^2}{M_{\Phi_f}^2}\right)\ ,
}
with
\eqa{
\F_Z(x_f,x_\psi) &=&
x_f \log (x_f)-\frac{2 x_f \left(x_f-x_\psi-1\right)}{\sqrt{\Delta}}\log\left[\frac{x_f+1-x_\psi+\sqrt{\Delta}}{2 \sqrt{x_f}}\right]\ ,
\\
\Delta
&=&x_f^2-2 x_f \left(x_\psi+1\right)+(x_\psi-1)^2\ .
}
Here we show the term coupling with the left fermions, the relation with the right are the following, $c_Z^{(u_L,d_L)}=-c_Z^{(u_R)}-c_Z^{(d_R)}$ and $c_Z^{(\nu_l,e_L)}=-c_Z^{(e_R)}$
The last term arises from quark box diagrams and reads
\eq{
c_{\rm box}^q =
\sum_f \frac{|\Gamma_{L,R}^u|^2 |\Gamma_{L,R}^f|^2 }{16\pi^2 M_{\Phi_f}^2 } \F_{\rm box}\left(\frac{m_f^2}{M_{\Phi_f}^2},\frac{M_{\Psi_{\rm DM}}^2}{M_{\Phi_f}^2}\right)\ ,
}
with
\eq{
\F_{\rm box}(x_f,x_\psi) = \frac{ x_f^2(x_\psi-1)^2 \log x_f - x_\psi^2 (x_f-1)^2 \log x_\psi + (x_f-1)(x_\psi-1)(x_f-x_\psi)}{4(x_f-1)^2(x_\psi-1)^2(x_f-x_\psi)}\ .
}

\bigskip

Contrary to the case of scalar DM, the effective Lagrangian includes an operator coupling an axial-vector DM current to an axial-vector quark current. This operator emerges at tree level, but gives rise to a Spin-Dependent (SD) interaction. We will discuss this in more detail below, keeping for the moment the focus on Spin-Independent (SI) interactions.

We hence move the discussion to the third term in~\Eq{eq:Lfermion} i.e., the scalar operator.
In analogy to the case of scalar DM, its Wilson coefficient is broken up in two pieces,
\eq{
d^q=d_{\rm QCD}^q+d_H^q\ .
}
The first one is the contribution coming from the Yukawa coupling of the DM with a quark and a colored $t$-channel mediator (\Fig{diags3}(a)),
\eq{
d_{\rm QCD}^q=\frac{M_{\Psi_{\rm DM}}|\Gamma_{L,R}^q|^2}{16 {\big(M_{\Phi_q}^2-(m_q+M_{\Psi_{\rm DM}})^2\big)}^2}\ ,
}
while the second one is a loop-induced contribution from Higgs penguin diagrams (\Fig{diags3}(b)),
\eqa{
d_{\rm H}^q
&=&
\frac{m_q}{32 \pi^2 m_H^2}\sum_f N_c^f \Bigg[\frac{m_f^2}{M_{\Psi_{\rm DM}}v^2}|\Gamma_{L,R}^f|^2 \F_{H,1}\bigg(\frac{m_f^2}{M_{\Phi_f}^2},\frac{M_{\Psi_{\rm DM}}^2}{M_{\Phi_f}^2}\bigg)
\nonumber\\
&&
\hspace{3.3cm}
\frac{\lambda_{H \Phi_f \Phi_f}}{2\sqrt{2} M_{\rm \Psi_{\rm DM}}}|\Gamma_{L,R}^f|^2 \F_{H,2}\bigg(\frac{m_f^2}{M_{\Phi_f}^2},\frac{M_{\Psi_{\rm DM}}^2}{M_{\Phi_f}^2}\bigg) \Bigg]
\ ,
}
with
\eqa{
\F_{H,1}(x_f,x_\psi)
&=&
\frac{x_f-1}{x_\psi}\log{x_f} - 2
- 2 \frac{(x_f-1)^2-x_\psi-x_f x_\psi}{x_\psi \sqrt{\Delta}}\log\left[\frac{1+x_f-x_\psi+\sqrt{\Delta}}{2 \sqrt{x_f}}\right]\ ,
\\
\F_{H,2}(x_f,x_\psi)
&=& -\F_{H,1}(x_f,x_\psi) -\log{x_f}
+\frac{2 (x_\psi+x_f-1)}{\sqrt{\Delta}}\log\left[\frac{1+x_f-x_\psi+\sqrt{\Delta}}{2 \sqrt{x_f}}\right]\ .
}
In the expression above $\lambda_{H \Phi_f \Phi_f}$ represents the trilinear coupling between the SM Higgs and a pair of NP scalars (c.f.~\Eq{eq:fermion_lagrangian}).
In the present case of fermionic DM, and contrary to the scalar DM case discussed above, this operator is not renormalized at this order in the perturbative expansion, and thus we can set it to zero at any scale.

We now consider the operator describing the effective coupling of a DM scalar bilinear with gluons. The matching condition receives contributions from the diagrams in~\Fig{diags3}(d-g), and is given by
\eq{
d_{q,\rm QCD}^{g}
=
\frac{\alpha_s}{96\pi}\frac{M_{\Psi_{\rm DM}}}{M_{\Phi_q}^4} f_S^q
\ ,
}
with
\eqa{
f_S^q
&=&
\frac{\Delta_{\rm QCD} (x_\psi-1-x_q)-6 x_q \left(x_q-1-x_\psi\right)}{2 \Delta_{\rm QCD}^4}+\frac{3 x_q (x_q^2-1+x_\psi)}{\Delta_{\rm QCD}^2}L_{\rm QCD}\ ,
\\
\Delta_{\rm QCD} &=& 2 x_\psi (x_q+1)-x_\psi^2-(1-x_q)^2\ ,
\\
L_{\rm QCD}
&=&
\left \{
\begin{array}{cc}
\frac{2}{\sqrt{\Delta_{\rm QCD}}}\arctan\left[\frac{|\Delta_{\rm QCD}|}{x_q+1-x_\psi}+\theta(x_\psi-1-x_q)\right]     & \Delta_{\rm QCD} \geq 0 \\
\frac{1}{\sqrt{|\Delta_{\rm QCD}|}}\log \left[\frac{x_q+1-x_\psi+\sqrt{|\Delta_{\rm QCD}|}}{x_q+1-x_\psi-\sqrt{|\Delta_{\rm QCD}|}}\right]+2\pi i \theta(x_\psi-1-x_q)     & \Delta_{\rm QCD}<0 
\end{array}
\right.
\ .
}
Finally, the matching conditions for the twist-2 operators are given by
\eq{
g_1^{q,g}+g_2^{q,g}=\frac{1}{8} M_{\Psi_{\rm DM}} g_S^q\ ,
}
with
\eqa{
g_S^q
&=&
-\frac{\alpha_s\,\log{x_q}}{4\pi M_{\Psi}^4}
-\frac{\alpha_s}{3\pi M_{\Phi_q}^4}\Bigg[
\frac{3 x_q (x_q-1-x_\psi)}{\Delta_{\rm QCD}^2}
+\frac{2 x_q^2-x_q-1-4 x_q x_\psi-4 x_\psi+2 x_\psi^4}{2 \Delta_{\rm QCD}x_\psi}
+\frac{1}{x_\psi}
\nonumber\\
&&
\hspace{-10mm}
-L_{\rm QCD} \left(\frac{3 (x_q-1+x_\psi)}{4 x_\psi^2}+\frac{3 x_q^2-3 x_q x_\psi-2 x_\psi+2 x_\psi^4}{2 \Delta_{\rm QCD} x_\psi^2}+\frac{3 x_q (x_q^2-x_q-2 x_q x_\psi-x_\psi+x_\psi^2)}{\Delta^2 _{\rm QCD}}\right)\Bigg]
\ .
\nonumber\\
}

\bigskip

Concerning the matching conditions for dipole and charge radius operators in~\Eq{eq:Lfermionphoton}, we have~\cite{Ibarra:2015fqa,Hisano:2018bpz}
\eqa{
b_\Psi
&=&
\frac{\alpha}{8\pi M_{\Psi_{\rm DM}}^2}\sum_f N_c^f Q_f |\Gamma_{L,R}^f|^2 \F_\gamma \left(\frac{M_{\Psi_{\rm DM}}^2}{M_{\Phi_f}^2},\frac{m_f^2}{M_{\Phi_f}^2}\right)\ ,
\\
\tilde{b}_\Psi
&=&
\frac{\alpha}{8\pi M_{\Psi_{\rm DM}}}\sum_f N_c^f Q_f |\Gamma_{L,R}^f|^2 \widetilde{\F}_\gamma \left(\frac{M_{\Psi_{\rm DM}}^2}{M_{\Phi_f}^2},\frac{m_f^2}{M_{\Phi_f}^2}\right)\ ,
}
with
\eqa{
\F_\gamma (x_f,x_\psi)
&=&
\frac{1}{12}\left \{-\frac{8 (1-x_f)+x_\psi}{x_\psi}\log x_f-\frac{4}{\Delta} \left[4 \Delta+x_\psi (1+3 x_f)-x_\psi^2\right]\right.
\nonumber\\
&& \left.-\frac{1}{x_\psi \Delta}\left[8 \Delta^2+(9-5 x_\psi+7 x_f)x_\psi \Delta-4 x_f x_\psi^2 (3-x_\psi+x_f)\right]L_{\rm EW} \right \}\ ,
\\
\widetilde{\F}_\gamma (x_f,x_\psi)
&=&
1+\frac{1-x_f}{2 x_\psi}\log x_f+\frac{\Delta +x_\psi (1-x_\psi+x_f)}{2 x_f}L_{\rm EW}\ ,
}
and where $L_{\rm EW}(x_f,x_\psi)=L_{\rm QCD}(x_f,x_\psi)$.

Since the effective couplings of the DM with the photon cannot be reduced to the conventional SI Interactions, DD phenomenology is not fully captured by the scattering cross-section over nucleons but, on the contrary, one has to rely on the DM scattering rate over nuclei: 
\eq{
\label{eq:DM_scattering_rate}
\frac{d\sigma}{dE_R}=
\left(\frac{M_T}{2\pi v^2}|f^T|^2 + \alpha_{\rm em}\tilde{b}_\Psi^2 Z^2 \left(\frac{1}{E_R}-\frac{M_T}{2 \mu_{T }^2 v^2}\right)\right)|F_{\rm SI}(E_R)|^2 
+\tilde{b}_\Psi^2 \frac{\mu_T^2 M_T}{\pi v^2}\frac{J_T+1}{3 J_T}|F_{\rm D}(E_R)|^2\ ,
}
where
\eq{
f^T=Z f_p+(A-Z) f_n\ ,
}
and $\mu_T=M_{\Psi_{\rm DM}} M_T/(M_{\Psi_{\rm DM}}+M_T)$,  with $M_T$ and $J_T$ being the mass and spin of target nucleus. $F_{SI}$ is the conventional SI nuclear form factor~\cite{Lewin:1995rx}, while $F_{\rm D}$ is the form factor associated to dipole scattering~\cite{Banks:2010eh}. Experimental limits have been obtained, in this case, with the procedure illustrated in~\Reff{Hisano:2018bpz}. The coefficients $f_p$ and $f_n$ are given by
\eqa{
\label{eq:fpfn}
f_p &=&
c_{\rm tree}^p+c_Z^p+c_{\rm box}^p-e b_\Psi-\frac{e \tilde{b}_\Psi}{2 M_\Psi}
+M_p\sum_{q=u,d,s} \left( f_q^p d_q  +\frac{3}{4}\left(q(2)+\bar{q}(2)\right)\left(g_1^q+g_2^q\right)\right)
\nonumber\\
&&
+\frac{3}{4}M_p\sum_{q=c,b,t}G(2)\left(g_1^{g,q}+g_2^{g,q}\right)
-\frac{8}{9}f_{TG}f_G\ ,
\\
f_n
&=& 
c_{\rm tree}^n+c_Z^n+c_H^n+c_{\rm box}^n  
+\frac{3}{4}M_n\sum_{q=u,d,s}\left(q(2)+\bar{q}(2)\right)\left(g_1^q+g_2^q\right)
\nonumber\\
&&
+\frac{3}{4}M_n\sum_{q=c,b,t}G(2)\left(g_1^{g,q}+g_2^{g,q}\right)
-\frac{8}{9}f_{TG}f_G\ .
}

As already pointed out, contrary to the case of scalar DM, the effective Lagrangian contains also the operator proportional to $\bar \Psi_{\rm DM} \gamma^\mu \gamma_5 \Psi_{\rm DM}$, which is responsible for SD interactions. The corresponding cross-section is given by
\eq{
\label{eq:SDcross}
\sigma_{\Psi_{\rm DM}p}^{\rm SD}=\frac{3}{16\pi}\sum_{f=u,d,s}\frac{\mu_{\Psi_{\rm DM},p}^2}{{\left(M_{\Psi_{\rm DM}}^2-M_{\Phi_f}^2\right)}^2}{\left \vert (\Gamma_L^f-\Gamma_R^f)^2 \Delta_f^p\right \vert}^2
}
where the structure functions $\Delta_f^{N=p,n}$ account for the contribution of the light quark flavours to the nucleon's spin. Even if the SD cross-section arises at tree level, this does not guarantee a priori that the latter will give the most competitive constraints. Indeed, SI interactions are coherent, i.e. the contribution of the different nucleons add at the amplitude level, so that the scattering cross-section of the DM over a nucleus is enhanced by roughly a factor $A^2$ with respect to its counterpart at the nucleon level. In the case of heavy elements, like xenon, this enhancement can overcome an eventual loop suppression of the cross-section so that effective constraints are obtained. In the case of SD interactions, instead, the cross-sections at the nucleon and nuclear level are of the same order of magnitude; the reason is that the SD cross-section is essentially due to the contribution of an unpaired nucleon. We also further remark that the SD cross-section is not zero (at tree level) only if the the DM is coupled with at least one of the light quark flavors. SI interactions hence remain the only way to probe interactions of the DM with the heavy quark flavors (and possibly leptons).

\subsection{Majorana DM}

In the case of Majorana DM, the operators $\bar \Psi_{\rm DM} \gamma^\mu \Psi_{\rm DM}$ and $\bar \Psi_{\rm DM}\sigma^{\mu \nu}\Psi_{\rm DM}$ vanish identically. Hence, DD for Majorana DM is described by the following Lagrangian:
\eqa{
\label{eq:Lfermion_majo}
\Leff^{q,\text{Majorana}}
&=& \tilde{c}^q\, \bar \Psi_{\rm DM} \gamma^\mu \gamma_5 \Psi_{\rm DM} \bar q \gamma^\mu \gamma_5 q
\nonumber\\
&&
+ \sum_{q=u,d,s} d^q\, m_q\bar \Psi_{\rm DM} \Psi_{\rm DM} \bar q q 
+\sum_{q=c,b,t} d_q^g\, \bar \Psi_{\rm DM} \Psi_{\rm DM} G^{a\mu \nu}G^a_{\mu \nu}
\nonumber\\
&&
+ \sum_{q=u,d,s} \left(g_{1}^{q} \frac{\bar \Psi_{\rm DM} i \partial^\mu \gamma^\nu \Psi_{\rm DM} \mathcal{O}^q_{\mu \nu} }{M_{\Psi_{\rm DM}}}+ g_{2}^{q} \frac{\bar \Psi_{\rm DM} \left( i  \partial^\mu \right)\left(i \partial^\nu \right) \Psi_{\rm DM} \mathcal{O}^q_{\mu \nu} }{M_{\Psi_{\rm DM}}^2}\right)
\nonumber\\
&&
+ \sum_{q=c,b,t}\left( g_{1}^{g} \frac{\bar \Psi_{\rm DM} i \partial^\mu \gamma^\nu \Psi_{\rm DM} \mathcal{O}^g_{\mu \nu} }{M_{\Psi_{\rm DM}}}+ g_{2}^{g} \frac{\bar \Psi_{\rm DM} \left( i  \partial^\mu \right)\left(i \partial^\nu \right) \Psi_{\rm DM} \mathcal{O}^g_{\mu \nu} }{M_{\Psi_{\rm DM}}^2}\right)\ .
}
In absence of an effective coupling with photons, it is possible to rely again on the scattering cross-section of the DM on the nucleons to set constraints on the model.
The corresponding cross-section is given by:
\eq{
\sigma_{\Psi_{\rm DM}p}^{\rm SI}=4 \frac{\mu_{\Psi_{\rm DM}p}^2}{\pi}\frac{{\left \vert Z f_p+(A-Z)f_n\right \vert}^2}{A^2}\ ,
}
where $f_{p,n}$ can be easily obtained by adapting the expressions in the previous subsection. A SD scattering cross-section is present as well, whose expression substantially coincides with~\Eq{eq:SDcross} up to symmetry factors accounting for the fact that the DM belongs to a real representation.

\section{Results for Direct Detection limits}
\label{sec:results}

Having discussed in detail how to compute the DM scattering cross-section and recoil rate in the considered particle physics framework, we have all the elements needed to illustrate our numerical study.

We start our numerical study from the simplest scenarios, namely the case in which the DM is coupled with a single $t$-channel mediator. Neglecting for the moment possible complementary constraints we will only focus on the sensitivity to present and next future direct detection experiments showing plots in the $(M_{\Phi_\text{DM}},M_{\Psi_f})$ or $(M_{\Psi_\text{DM}},M_{\Phi_f})$ bidimensional planes while fixing to 1 the value of the corresponding couplings. As already pointed, all the results for scalar (both complex and real) DM are obtained by taking $\lambda_{1H\Phi}(M_{\Psi_f})=0$. On a similar footing we have taken $\lambda_{1H\Phi}=\lambda_{2H\Phi}=0$ for fermion DM models.

\subsection{Tree level Direct Detection for SM singlet DM}

Despite it not being the main focus of present paper, we report, for completeness, the scenarios in which DM direct detection is mostly accounted for interactions arising at tree level from the Lagrangians in~Eqs.~(\ref{eq:scalar_lagrangian}) and (\ref{eq:fermion_lagrangian}).
In the case of SM singlet DM, this can occur in the cases of complex scalar or dirac fermionic DM interacting with first generation quarks (in the case of SM singlet the Higgs portal operator for scalar DM can be safely neglected). In both cases, the low energy DM effective coupling is with a quark current $\bar q \gamma^\mu q$. Consequently a strong SI cross-section is expected. In the case of fermionic DM, SD interactions are present as well. Even if the corresponding limits are subdominant with respect to the SI case, we will nevertheless include them for completeness.

\begin{figure}[t]
    \centering
    \subfloat{\includegraphics[width=0.315\linewidth]{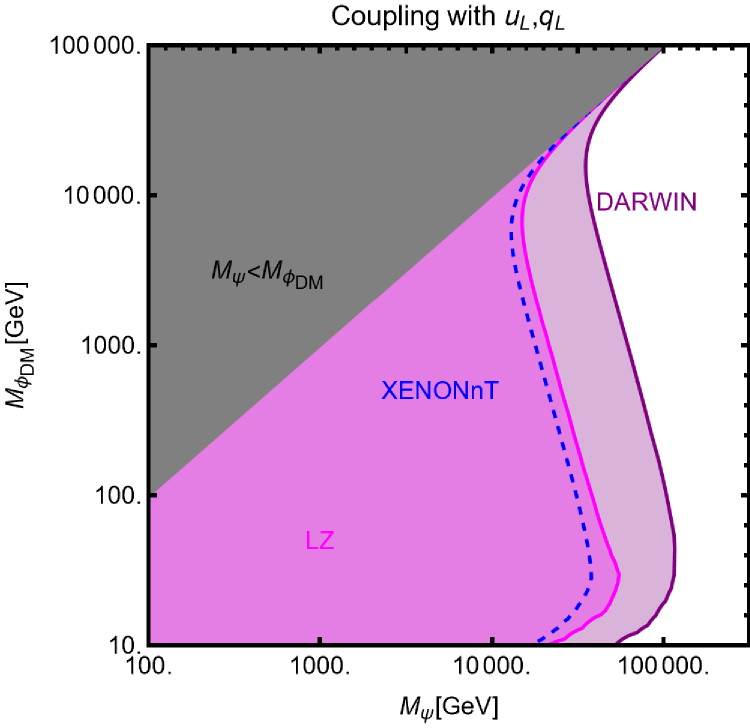}}
    \subfloat{\includegraphics[width=0.337\linewidth]{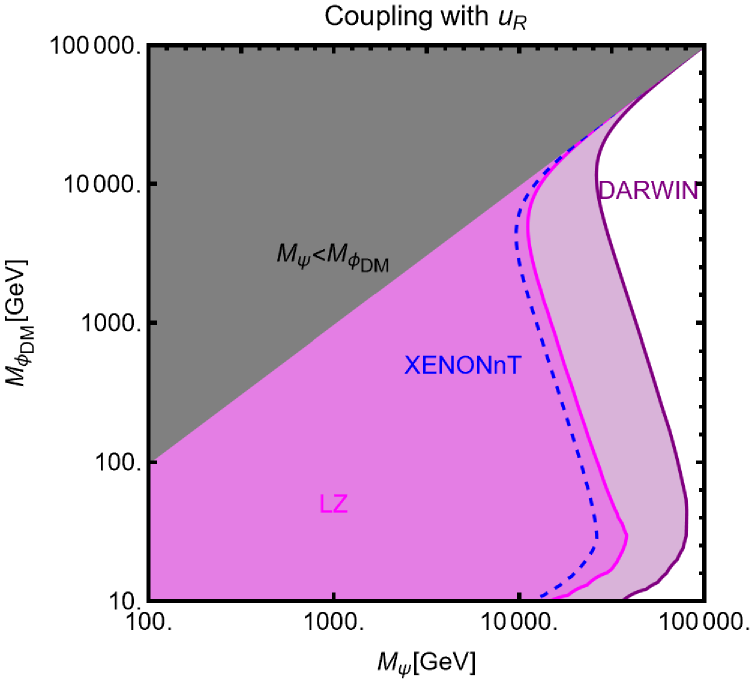}}
    \subfloat{\includegraphics[width=0.337\linewidth]{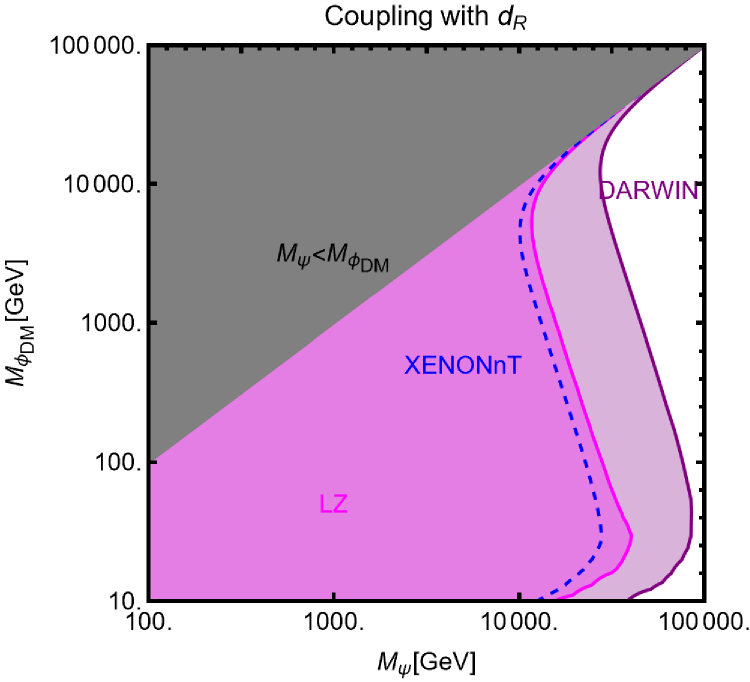}}
  \caption{\it Direct detection prospects for complex scalar DM, singlet under the SM gauge group, in the case of tree level couplings with SM quarks. The magenta region is excluded by the most recent bound from LZ. The purple region represents the expected sensitivity of the DARWIN experiment. The three panels of the figure represent, respectively, the cases of coupling with the first generation quark doublet, right-handed up- and down- quarks. In all cases such couplings have been set to 1. For reference, the bound from XENONnT, is shown as well as dashed blue contour. The gray region is excluded from the analysis since the DM is not cosmologically stable there.}
    \label{fig:complex_tree}
\end{figure}

\begin{figure}[t]
    \centering
    \subfloat{\includegraphics[width=0.328\linewidth]{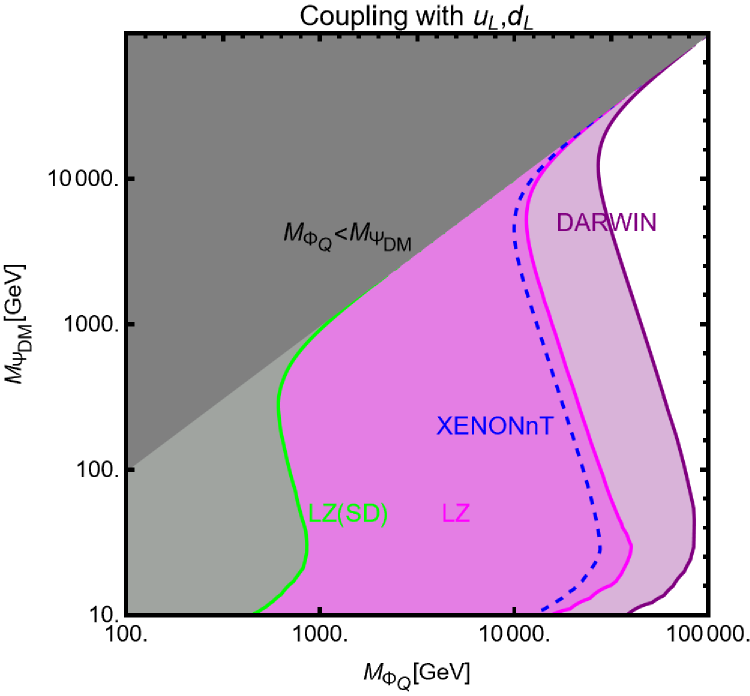}}
    \subfloat{\includegraphics[width=0.328\linewidth]{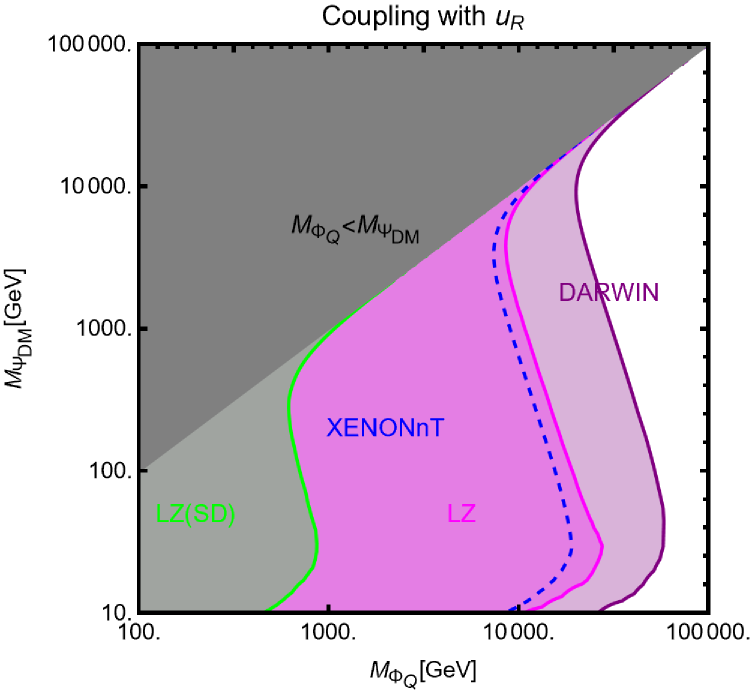}}
    \subfloat{\includegraphics[width=0.328\linewidth]{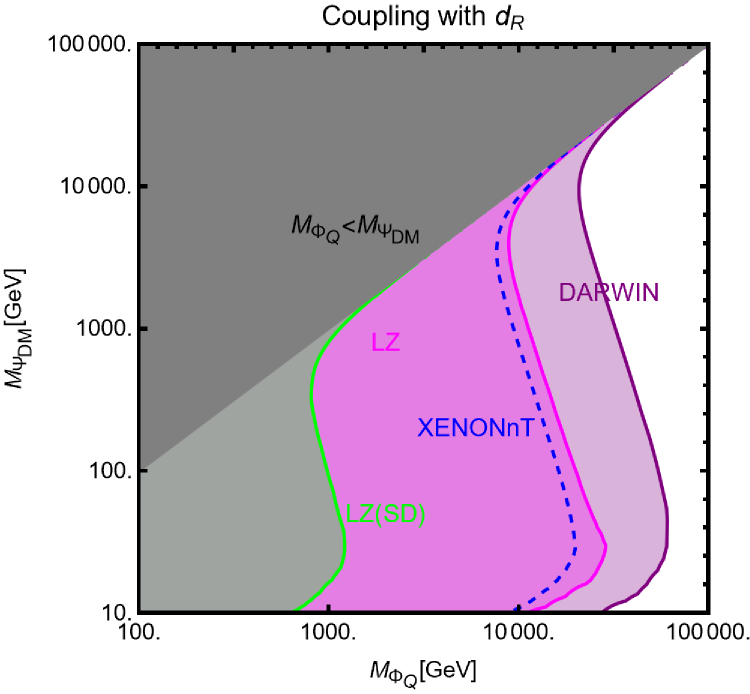}}
  \caption{\it Same as for~\Fig{fig:complex_tree} but in the case of dirac DM. Contrary to the case of scalar DM we have also reported the bound from SD interactions, always by LZ.}
    \label{fig:dirac_tree}
\end{figure}

According to the previous discussion, we show in Figures.~\ref{fig:complex_tree} and \ref{fig:dirac_tree}, the capability of present and next future Direct Detection experiments of probing models with complex scalar or dirac fermionic DM, coupled via $t$-channel mediators, to the left-handed first generation quark doublet $q_L=(u_L\,\,\, d_L)^T$, right-handed up quarks or right-handed down quarks. The figures shows the $(M_{\Phi_\text{DM}},M_{\Psi_q})$ and  $(M_{\Psi_\text{DM}},M_{\Phi_q})$ bisimensional planes while the coupling of the interaction of the DM with the mediator and the SM quark has been set to one. Each plot shows in magenta the current world leading limit on SI interactions as determined by the LZ collaboration \cite{LUX-ZEPLIN:2022qhg} (we have also reported as dashed blue line the very similar limit given by XENONnT \cite{XENON:2023sxq}\footnote{SI Interaction for WIMPs are also effectively probed by the PandaX experiment, see e.g. \cite{Liu:2022zgu}.}). Furthermore the sensitivity of the next generation of Direct Detection experiments, represented by DARWIN \cite{Aalbers:2016jon}, is shown as purple region. As evident, direct detection already excludes broad ranges of the DM and $t$-channel mediator masses. For the latter we have a limits as strong as 50 TeV in the case of coupling with the left-handed first generation quark doublet. Direct Detection experiments have substantially the same sensitivity to complex scalar and dirac fermion DM. This is due to the fact that the operators $\Phi_\text{DM}^{\dagger}\partial_\mu \Phi_\text{DM}$ and $\bar \Psi_\text{DM}\gamma_\mu \Psi_\text{DM}$ lead to the same scattering cross-section.  As already pointed, in the case of fermionic DM, SD interactions are present as well. As evidenced by~\Fig{fig:dirac_tree}, the corresponding limits are largely subdominant. Indeed, in this scenario, SI and SD interactions arise both a the tree-level; the corresponding experimental sensitivities are consequently largely different due to the coherent enhancement of SI interactions.

\subsection{One loop Direct Detection for SM singlet DM coupled with quarks}

In this subsection we illustrate the cases in which the DM is coupled with a single mediator and quark species but this time, the most relevant operators for Direct Detection emerge at the one-loop level.  

\begin{figure}[t]
    \centering
    \subfloat{\includegraphics[width=0.33\linewidth]{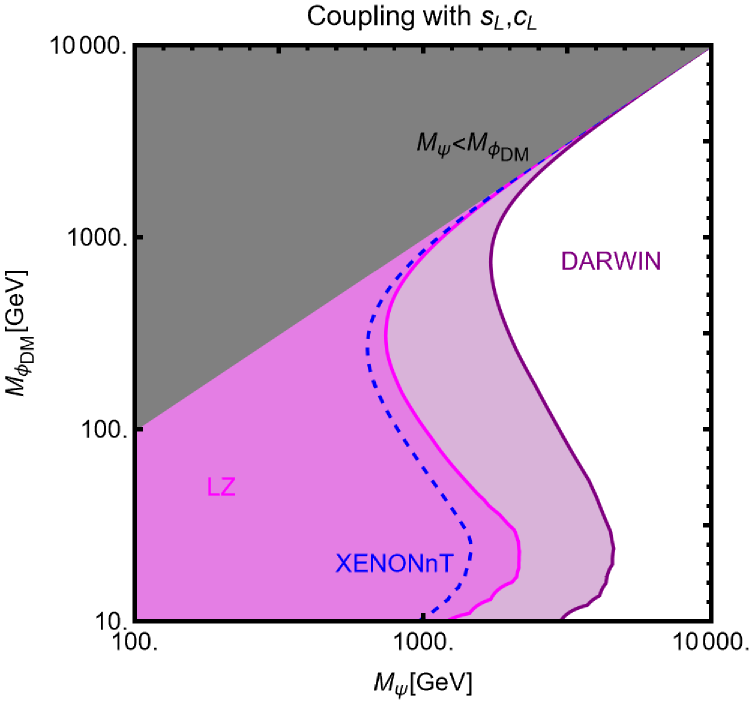}}
    \subfloat{\includegraphics[width=0.33\linewidth]{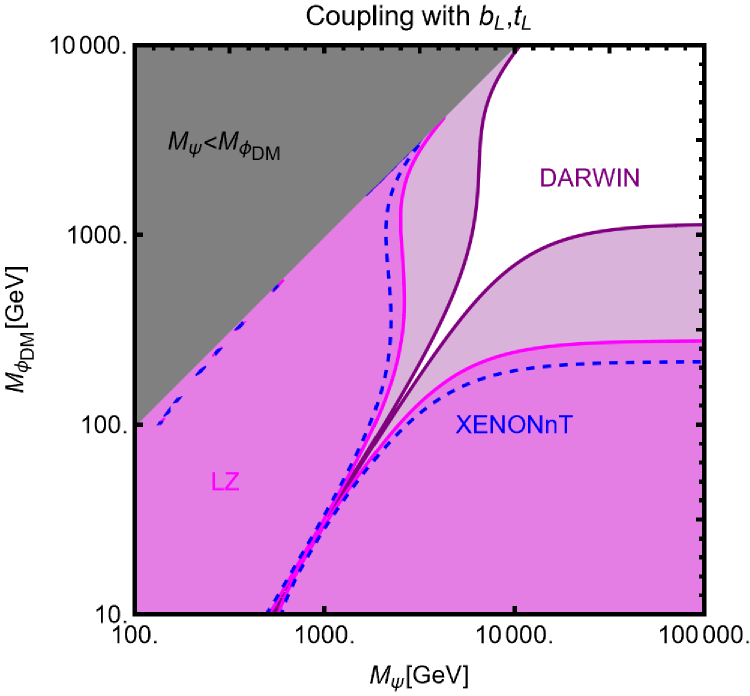}}
    \subfloat{\includegraphics[width=0.33\linewidth]{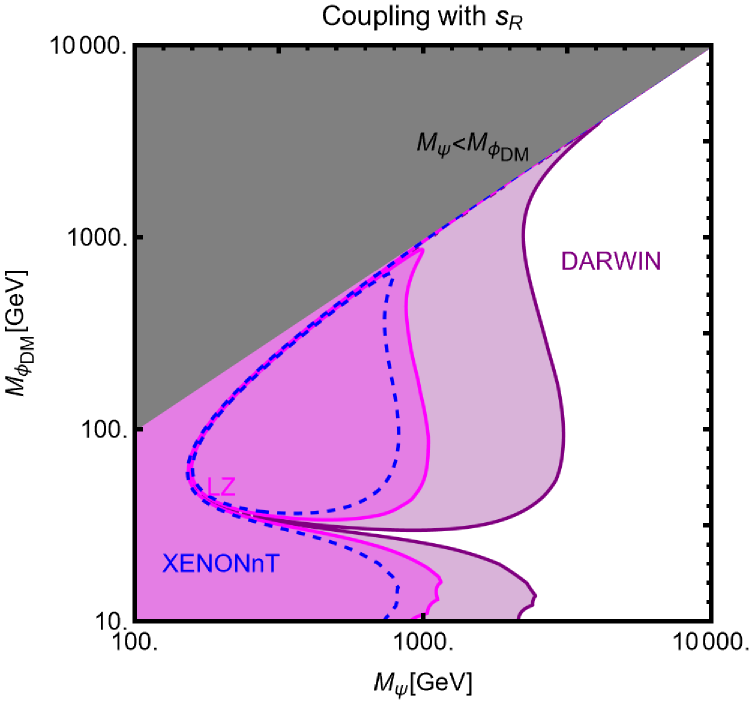}}\\
    \subfloat{\includegraphics[width=0.33\linewidth]{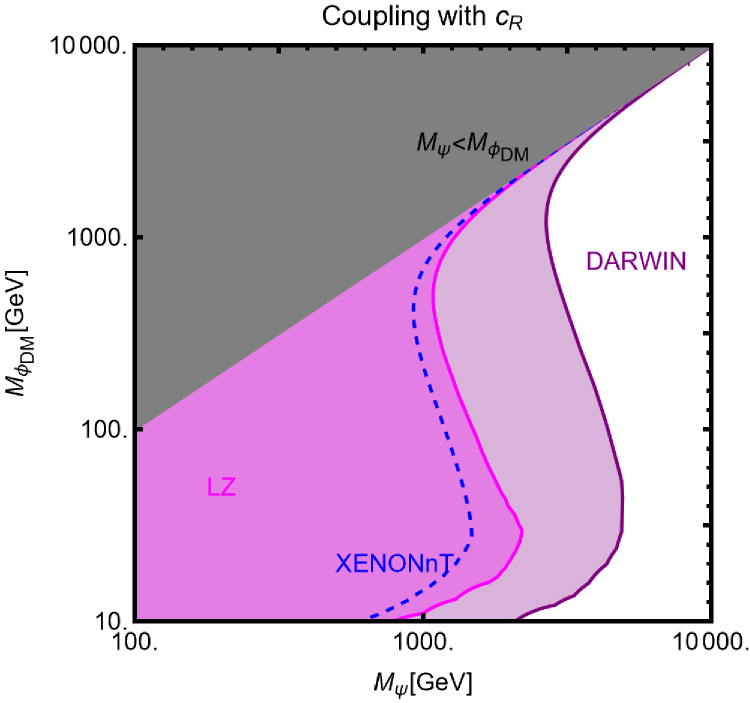}}
     \subfloat{\includegraphics[width=0.33\linewidth]{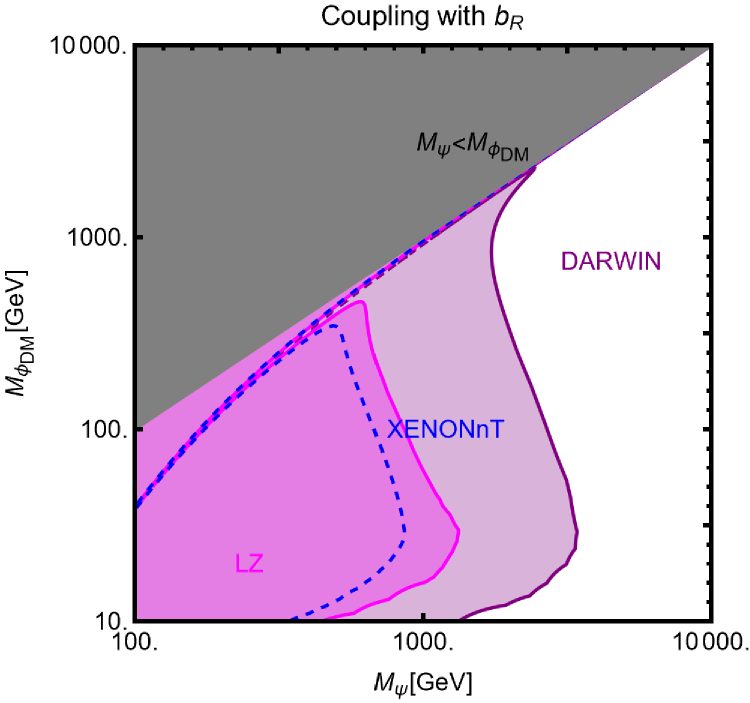}}
      \subfloat{\includegraphics[width=0.33\linewidth]{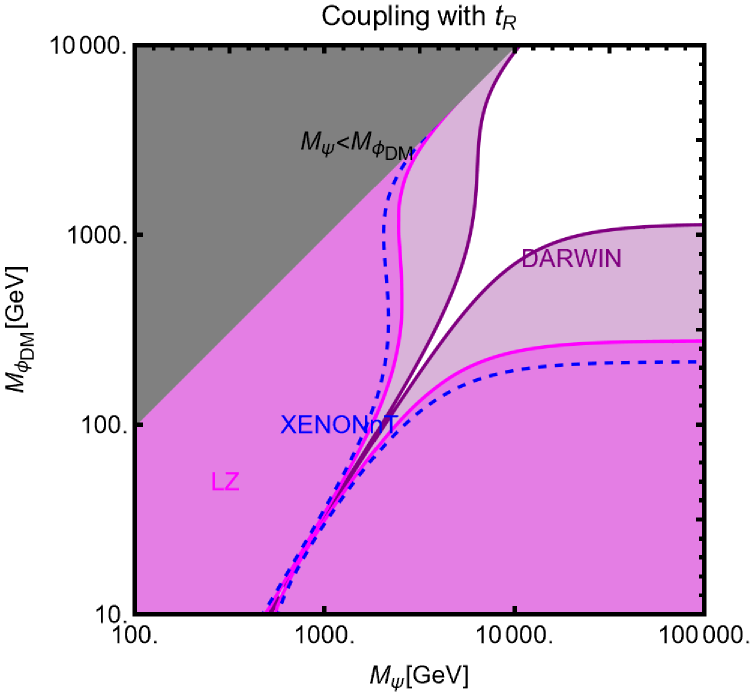}}
    \caption{\it Direct detection prospects for complex scalar DM, singlet under the SM gauge group, in the cases in which interactions relevant for Direct Detection arises at one loop. The magenta region is excluded by the most recent bound from LZ. The purple region represents the expected sensitivity of the DARWIN experiment. The different panels consider, individually, the different cases of couplings with left-handed or right-handed quarks. In all cases such couplings have been set to 1. For reference, the bound from XENON1T, is shown as well as dashed blue contour. The gray region is excluded from the analysis since the DM is not cosmologically stable there.}
    \label{fig:complex_singlet}
\end{figure}

\begin{figure}
    \centering
    \subfloat{\includegraphics[width=0.335\linewidth]{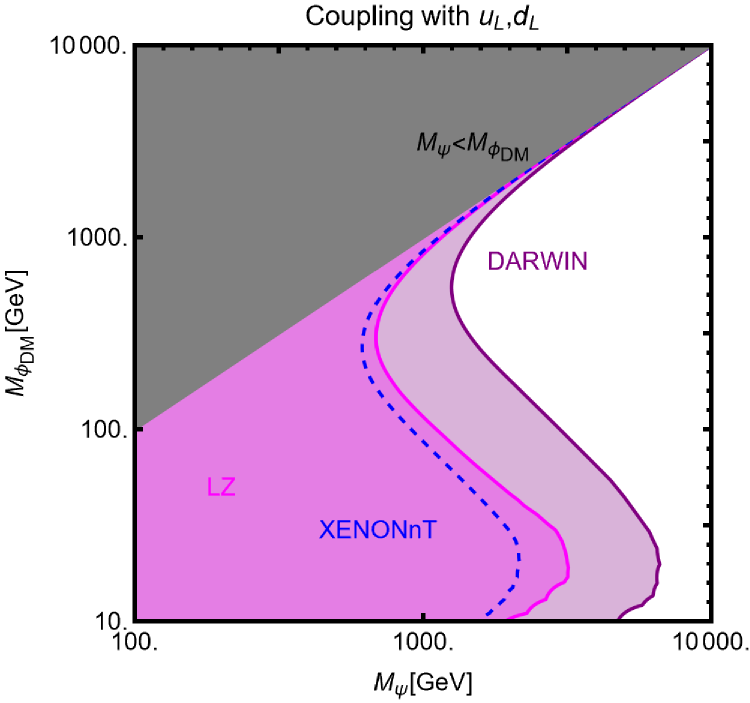}}
    \subfloat{\includegraphics[width=0.335\linewidth]{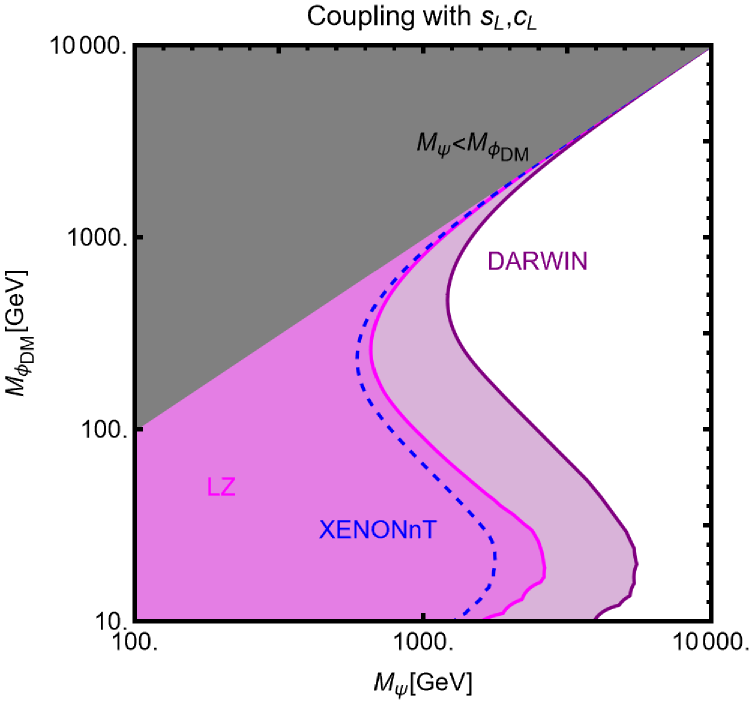}}
    \subfloat{\includegraphics[width=0.335\linewidth]{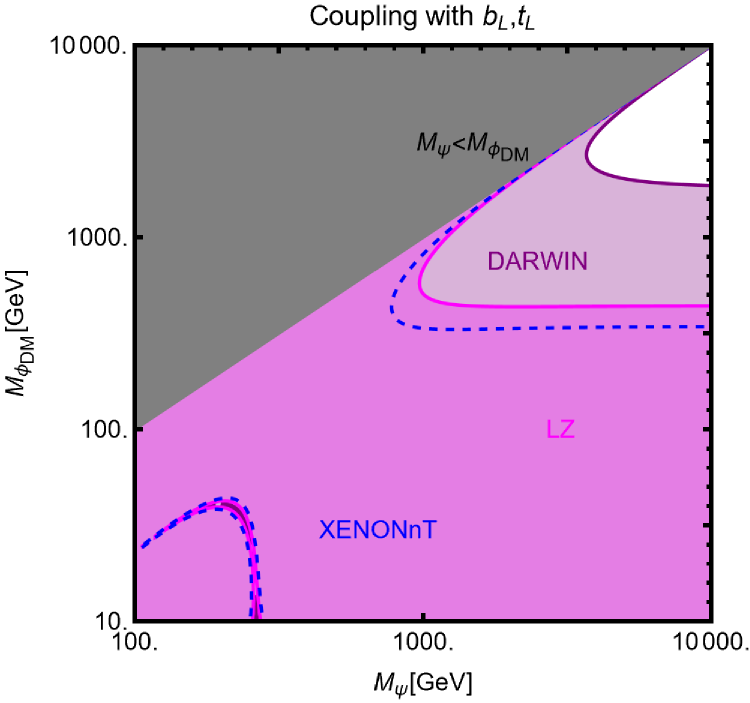}}\\
      \subfloat{\includegraphics[width=0.335\linewidth]{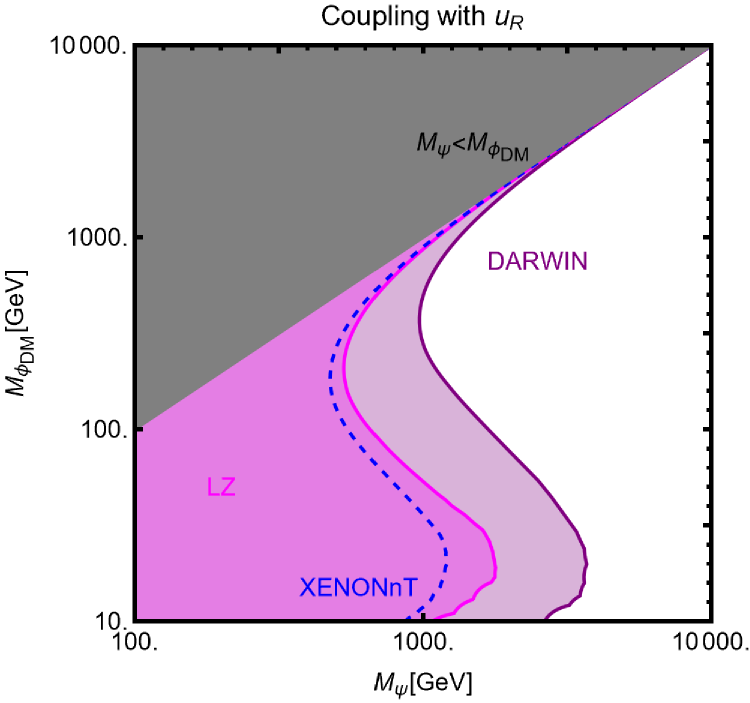}}
        \subfloat{\includegraphics[width=0.335\linewidth]{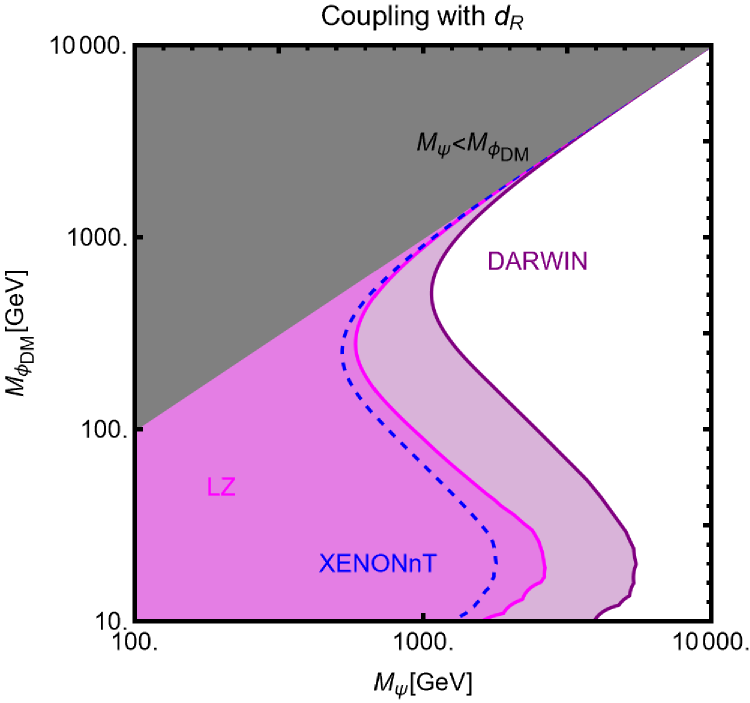}}
    \subfloat{\includegraphics[width=0.335\linewidth]{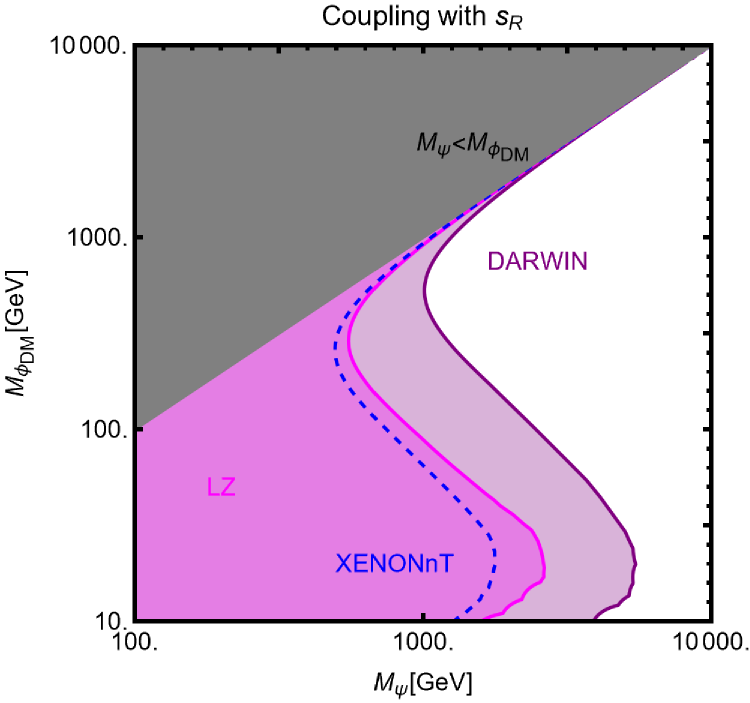}}\\
    \subfloat{\includegraphics[width=0.335\linewidth]{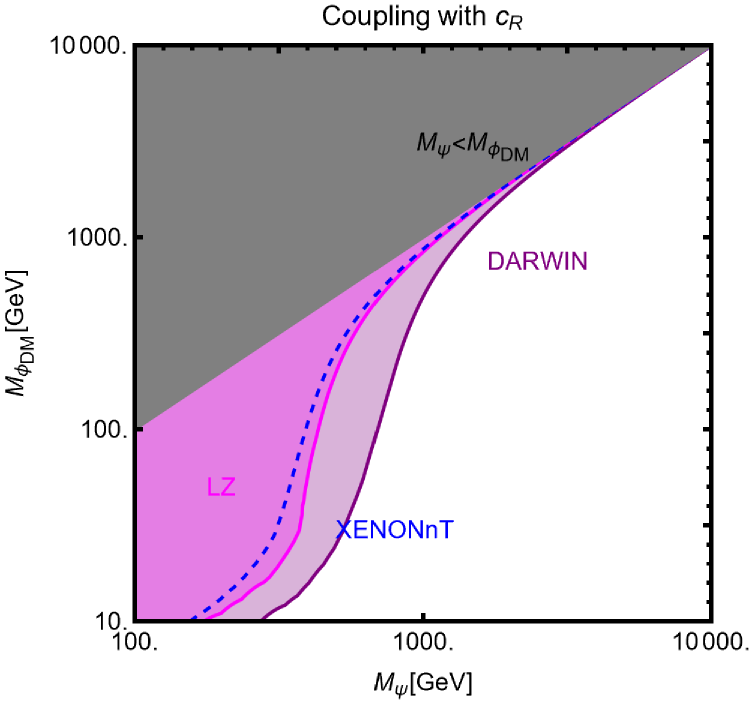}}
     \subfloat{\includegraphics[width=0.335\linewidth]{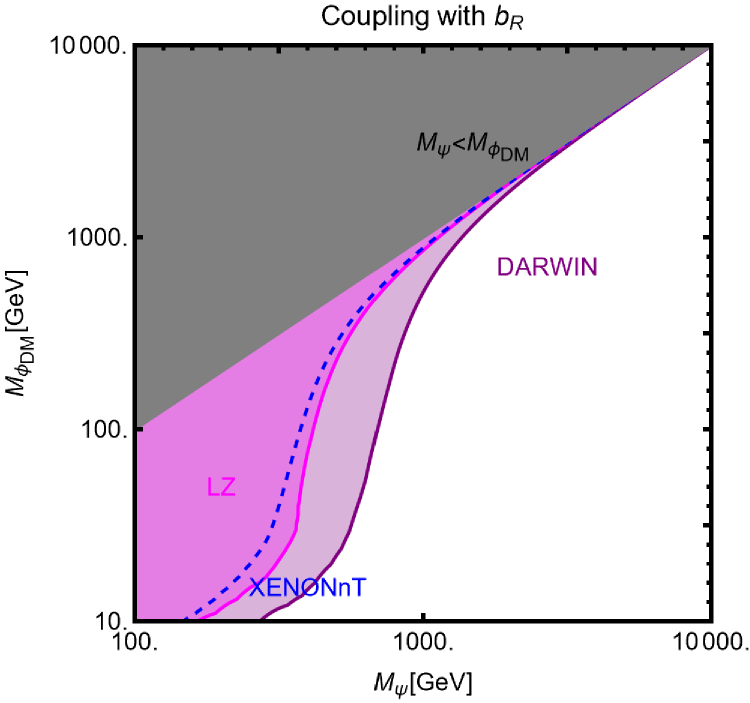}}
      \subfloat{\includegraphics[width=0.335\linewidth]{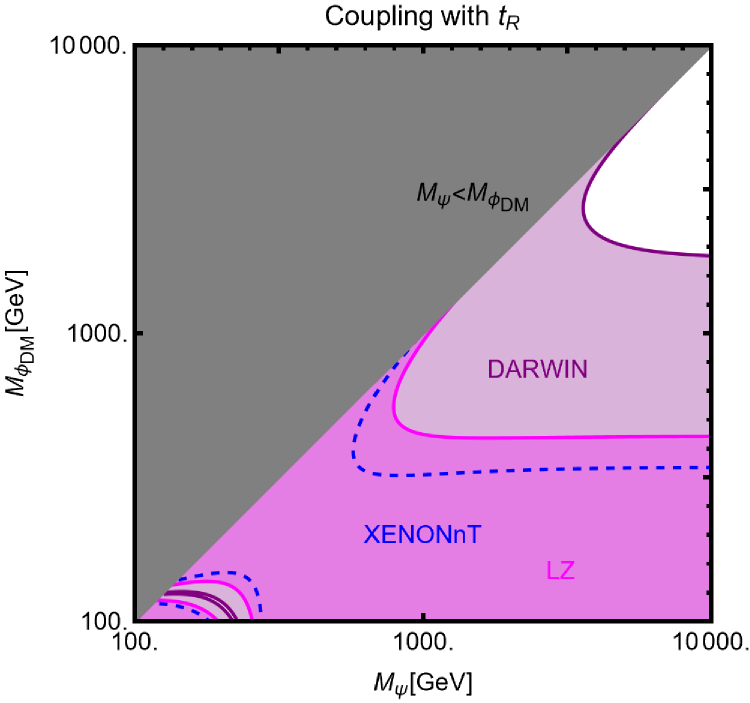}}
    \caption{\it Same as~\Fig{fig:complex_singlet} but for real scalar singlet.}
    \label{fig:real_singlet}
\end{figure}

\begin{figure}[t]
    \centering
    \subfloat{\includegraphics[width=0.33\linewidth]{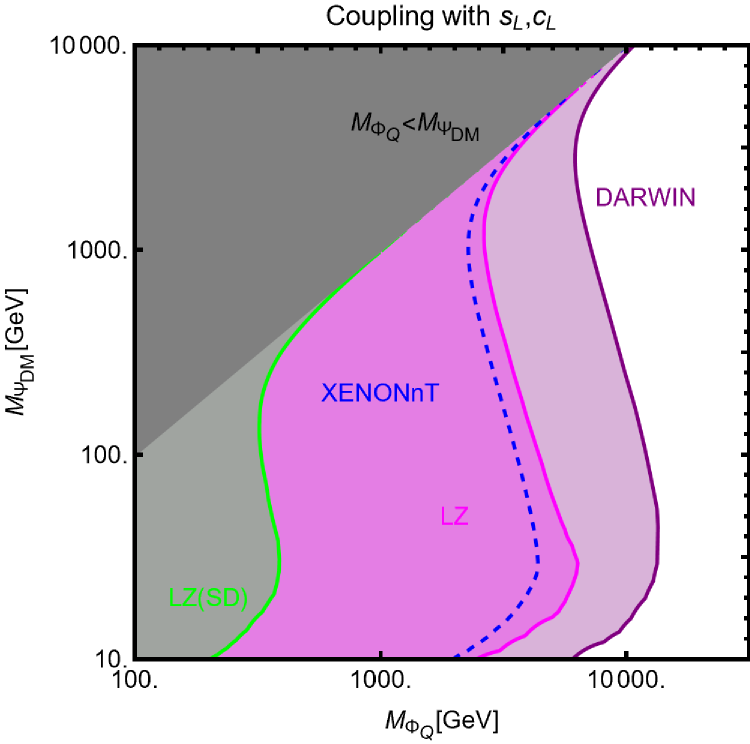}}
    \subfloat{\includegraphics[width=0.33\linewidth]{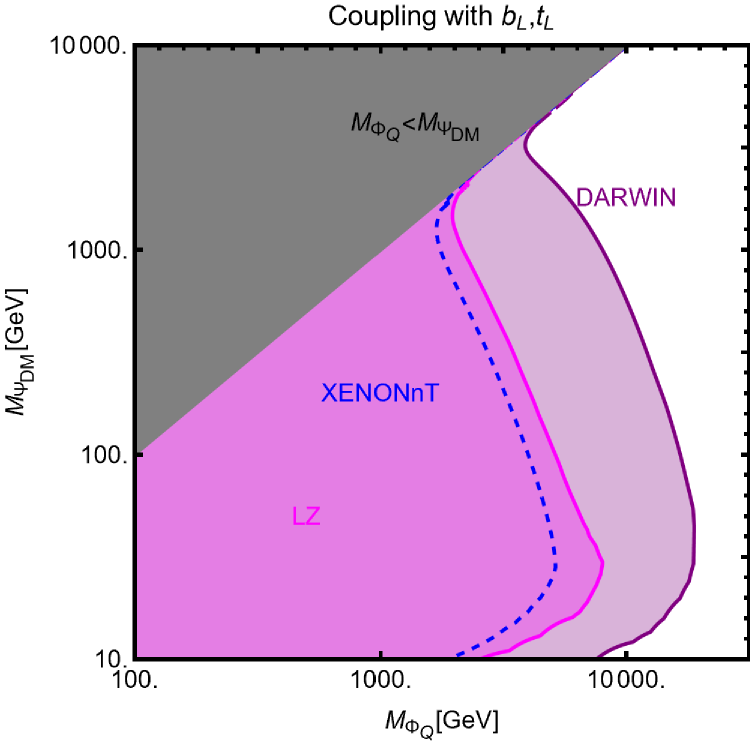}}
    \subfloat{\includegraphics[width=0.33\linewidth]{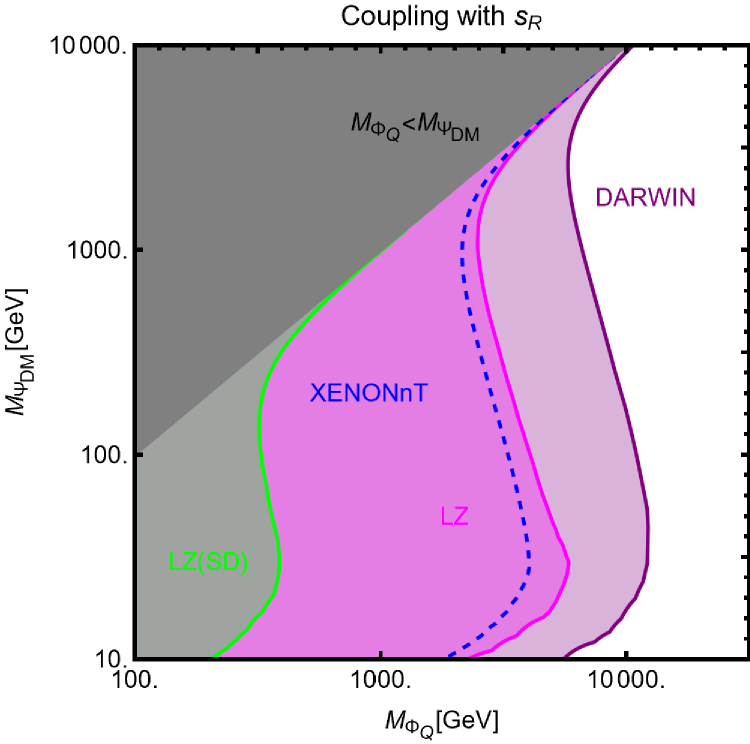}}\\
    \subfloat{\includegraphics[width=0.33\linewidth]{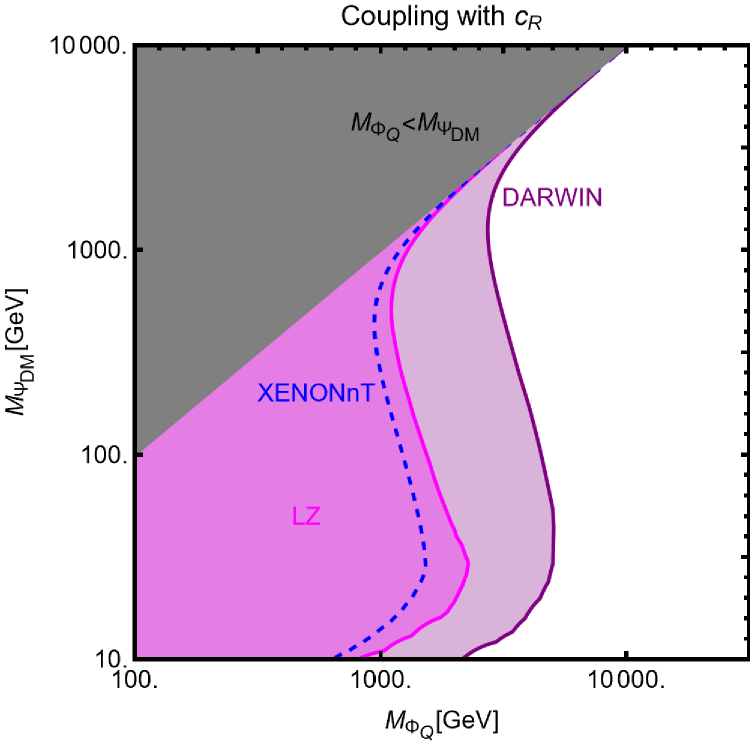}}
     \subfloat{\includegraphics[width=0.33\linewidth]{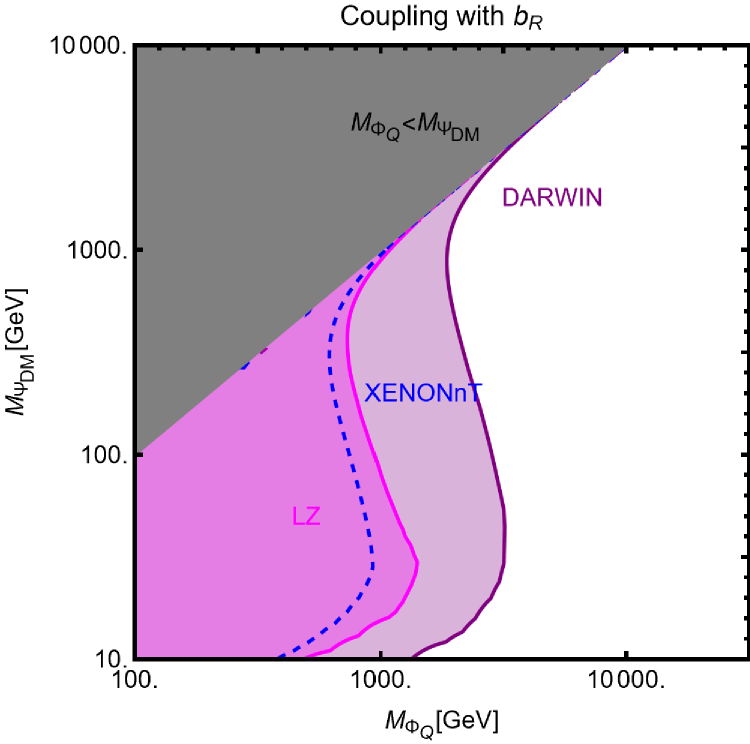}}
      \subfloat{\includegraphics[width=0.33\linewidth]{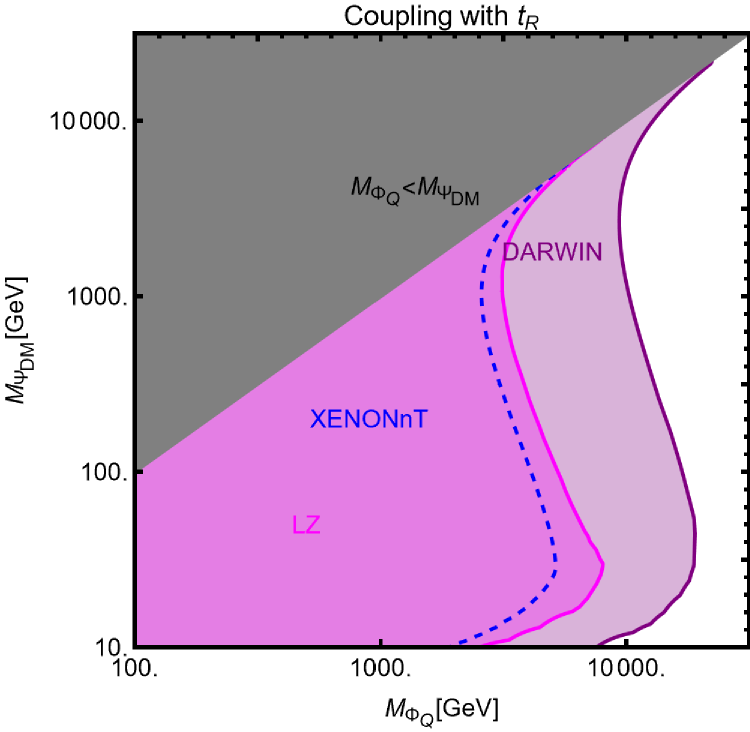}}
    \caption{\it Same as~\Fig{fig:complex_singlet} but for Dirac fermion singlet.}
    \label{fig:dirac_singlet}
\end{figure}

\begin{figure}
    \centering
     \subfloat{\includegraphics[width=0.335\linewidth]{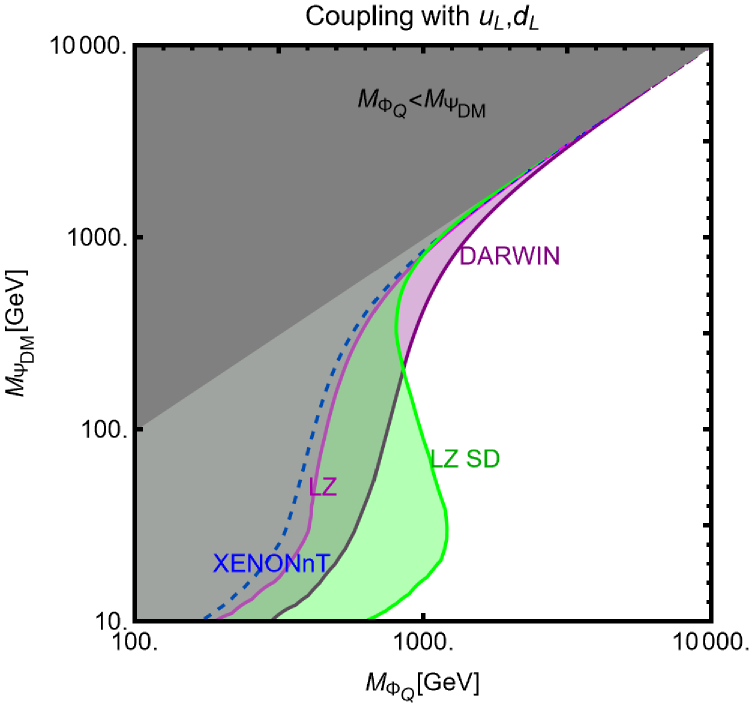}}
    \subfloat{\includegraphics[width=0.335\linewidth]{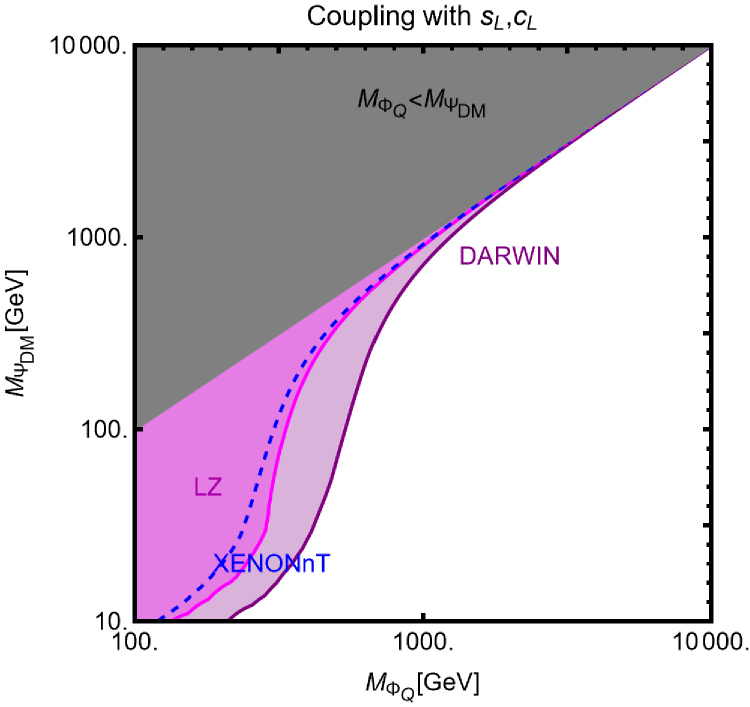}}
    \subfloat{\includegraphics[width=0.335\linewidth]{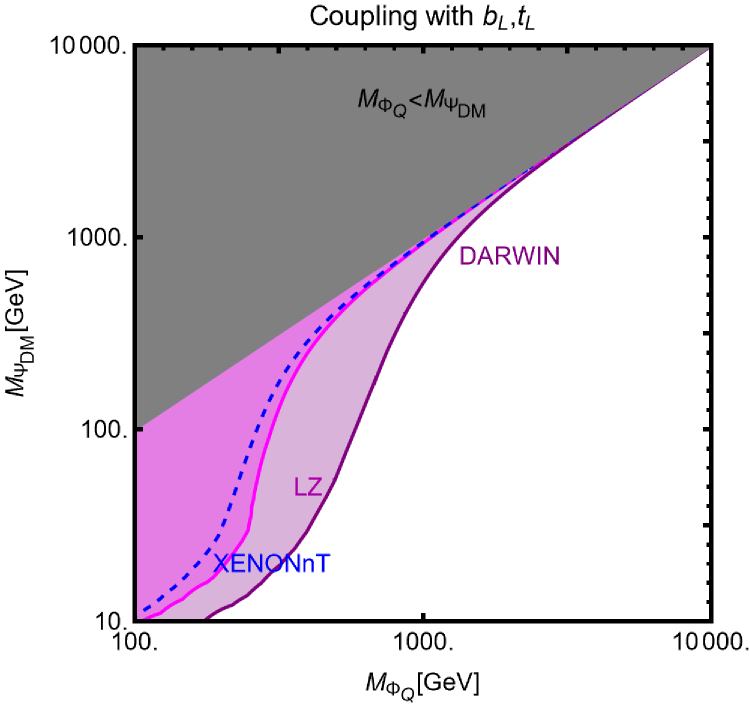}}\\
     \subfloat{\includegraphics[width=0.335\linewidth]{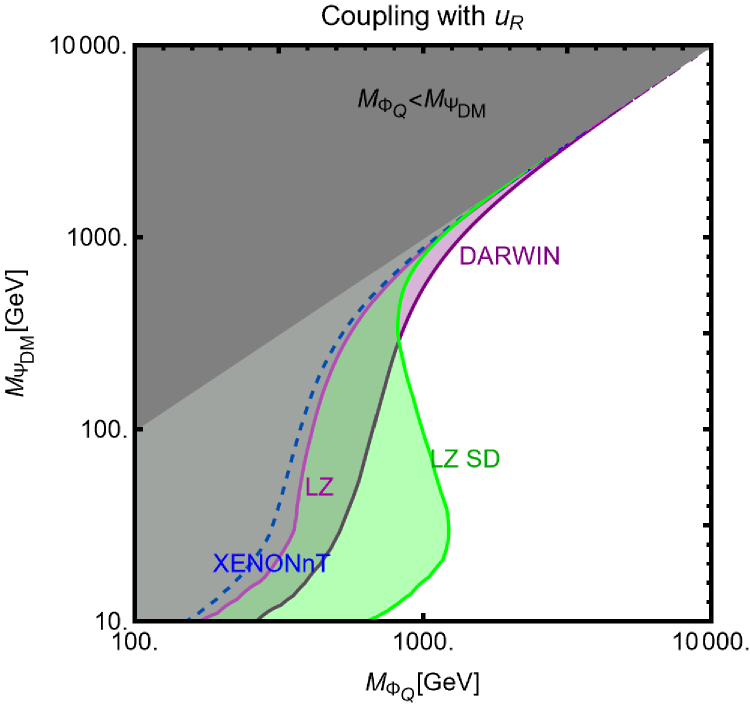}}
    \subfloat{\includegraphics[width=0.335\linewidth]{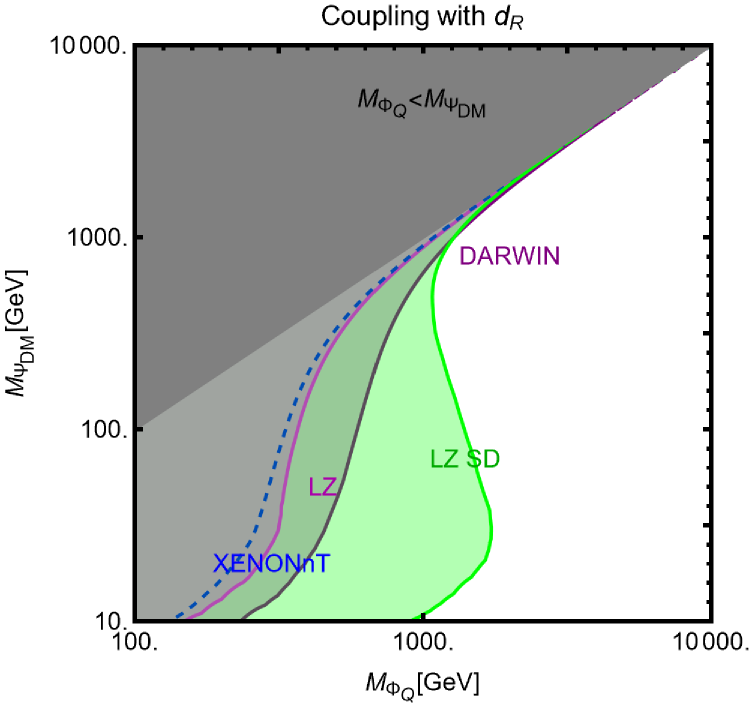}}
    \subfloat{\includegraphics[width=0.335\linewidth]{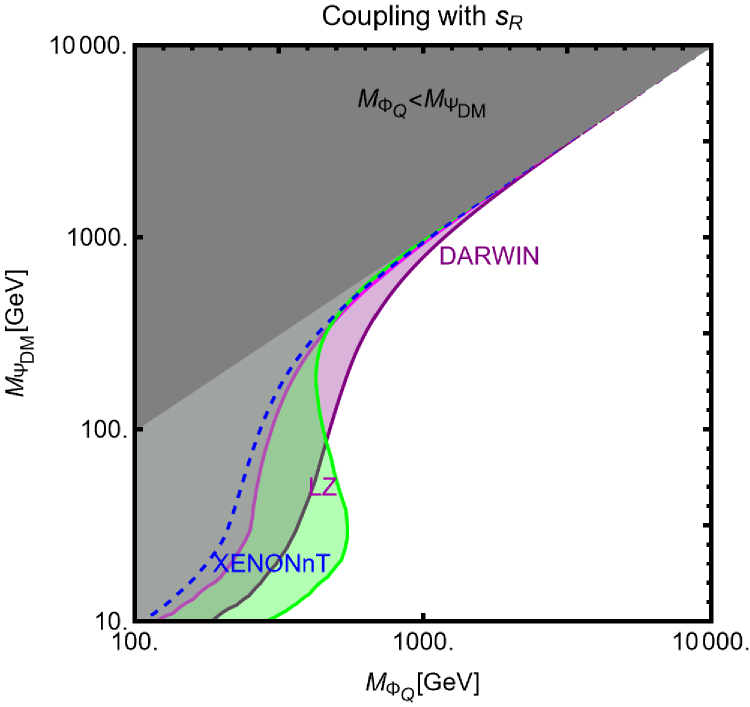}}\\
    \subfloat{\includegraphics[width=0.335\linewidth]{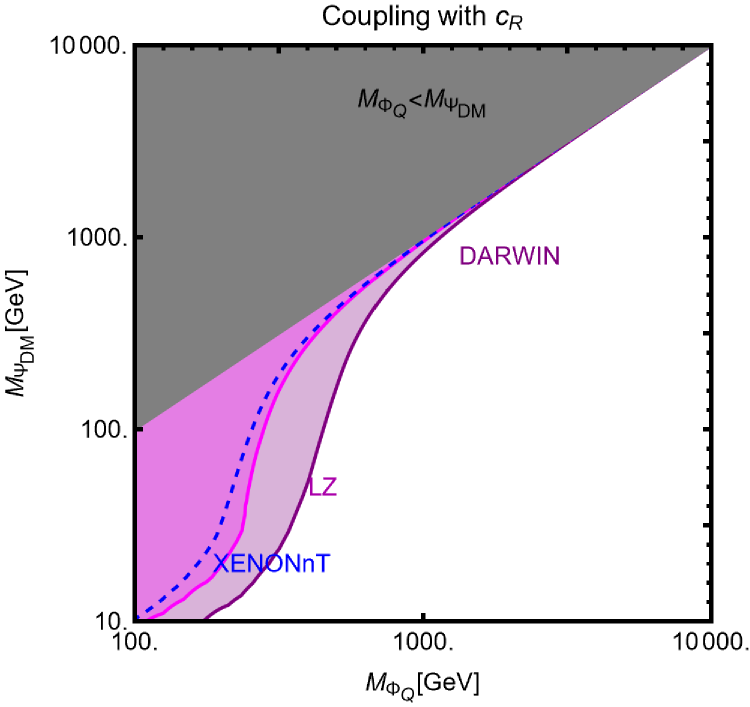}}
     \subfloat{\includegraphics[width=0.335\linewidth]{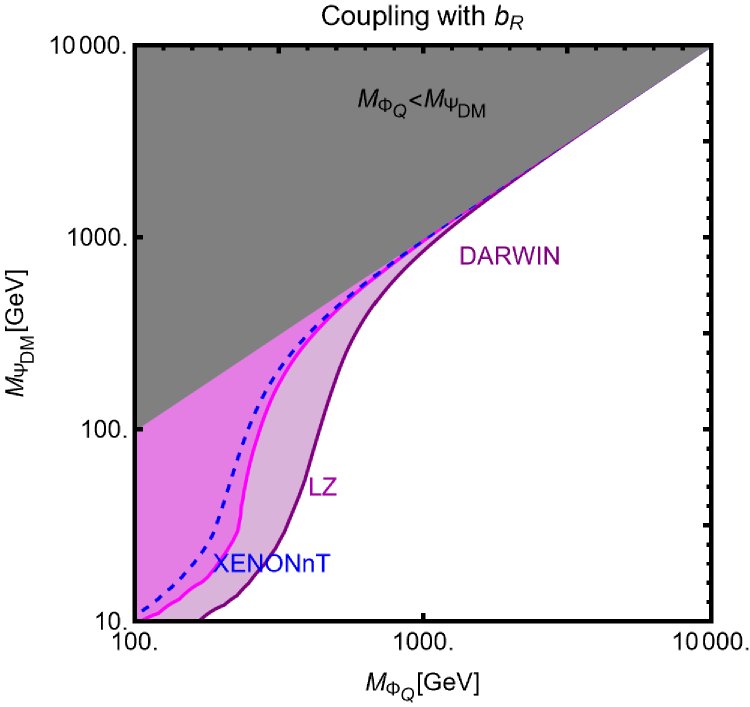}}
      \subfloat{\includegraphics[width=0.335\linewidth]{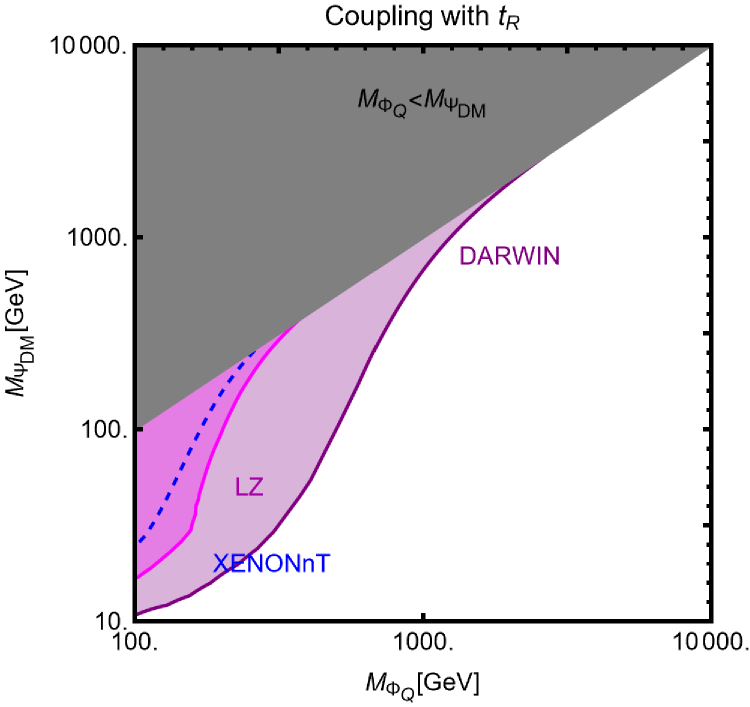}}
    \caption{\it Same as~\Fig{fig:complex_singlet} but for majorana fermion singlet.}
    \label{fig:majorana_singlet}
\end{figure}

Our results are shown in Figures~\ref{fig:complex_singlet}-\ref{fig:majorana_singlet} considering, in order, complex scalar, real scalar, dirac fermion and majorana fermion DM. The format and color code of the plots is the same as the plots illustrated in the previous subsection. 

Let's now discuss in more detail the different scenarios starting from the case of complex scalar DM. On general ground this scenario appear to be rather constrained regardless the quark species chosen to interact with the DM. As evident,~\Fig{fig:complex_singlet} shows only couplings with the second and third generation of quarks as in the case of coupling with $u,d$ quarks direct detection is dominated by tree-level induced couplings. In the case of coupling with second generation quarks or the right-handed bottom, the major contribution to the DM scattering cross-section comes from photon penguins. A sizable contribution comes also from DM-gluon operators, originated by QCD interaction of the mediator, especially for light values of the DM mass and in regions where the mass splitting between the DM and the $t$-channel mediator is small. A much larger region of parameter space is excluded if the DM features effective couplings with the top quark.  This is due to the radiatively induced Higgs portal coupling (see discussion in the previous section) which in the case of the aforementioned coupling with the top quark dominates the $d_H^q$ such that $d_H^q \propto \frac{m_t^2}{v^2 m_W^2}$, substantially independent on the mass $M_{\Psi_f}$ (notice that also the Wilson coefficient associated to $Z$-penguin diagrams is enhanced with the mass of the top. Its contribution is however subdominant for scalar DM). In summary, in the case of effective coupling of the DM with the top, we have a result more similar to the case of the Higgs portal, see e.g. \cite{Arcadi:2017kky,Arcadi:2021mag} (with respect to the latter we have a loop suppression of the coupling though).

Moving to the case of real scalar DM, we see that, in the case of coupling with the first two quark generations (we have included also the case of $u,d$ quarks since there is not tree level operator for real scalar DM) as well as $b_R$ masses of the DM above 100 GeV are less constrained with respect to the case of complex scalar DM. This because the operator responsible of the interaction of the DM with the photon vanishes if the DM belongs to a real representation. The excluded region again is drastically increases in the case of coupling with the top quark as consequence of the effective coupling with the Higgs, coinciding with the case of complex scalar DM.

Moving to fermionic DM we see, from~\Fig{fig:dirac_singlet}, that the exclusion regions are analogous to ones obtained in the case of complex scalar DM. Analogously to this case we have indeed that for coupling with the second generation and bottom quarks, DM scattering is mostly accounted for the effective coupling with photons and gluons. Again, an effective coupling of the DM with the top quark would translate into stronger limits. Contrary to the case of complex scalar DM, also the future expected limits from DARWIN can be evaded by considering a sufficiently heavy $t$-channel mediator. Indeed the strenghtening of the direct detection limits is due to the Z-penguin diagrams, whose Wilson coefficient is proportional to the mass of the SM fermion running in the loop. While also present, Higgs penguin diagrams give rise to finite amplitudes which do not induce the radiative generation of a Higgs portal coupling and are numerically subdominant with respect to $Z$-penguins \cite{Ibarra:2015fqa}.

Let's finally consider the case of Majorana DM. In such a case the DM scattering is essentially due to the operators, mostly scalar interaction with quarks and gluons, generated by the QCD interaction of the $t$-channel mediator (while present, the contribution from Higgs penguins is subdominant). As evident from the figure, in case of coupling with the $u,d,s$ quark, the strongest constraints come from the tree-level induced SD interactions so that the loop induced SI interactions are mostly relevant to probe the couplings with heavy quarks.

\subsection{DM coupled only with leptons}

Sizable SI interactions can arise in $t$-channel portals also in the case the tree level Lagrangian features only interactions between the DM and the leptons. This occurs only for complex scalar and dirac fermionic DM as due to the presence of $Z/h/\gamma$ penguin diagrams. Among them, the dominant contribution comes from $\gamma$ penguins, being the other Wilson coefficients strongly suppressed by the light fermion masses. Consequently, the direct detection sensitivity/prospects are substantially the same for coupling with left- and right-handed fermions.

\begin{figure}[t]
    \centering
     \subfloat{\includegraphics[width=0.336\linewidth]{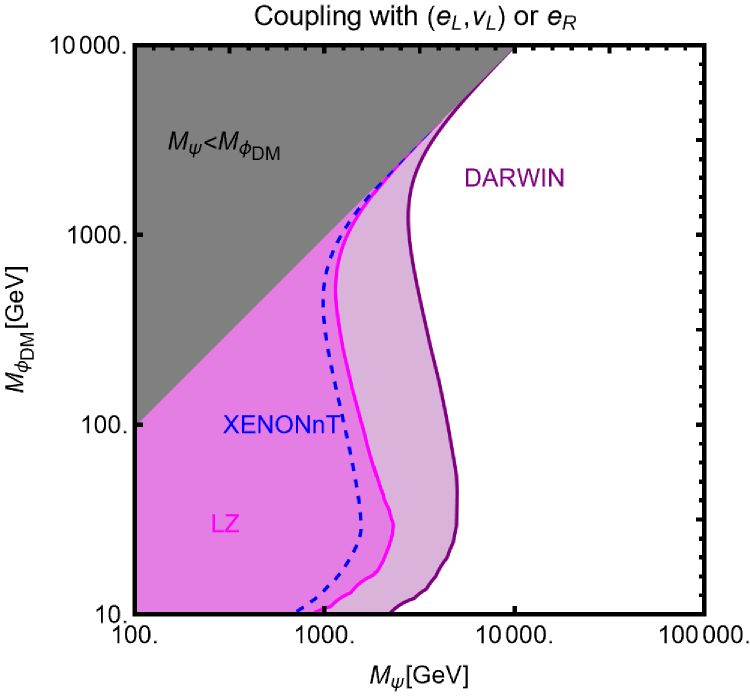}}
    \subfloat{\includegraphics[width=0.336\linewidth]{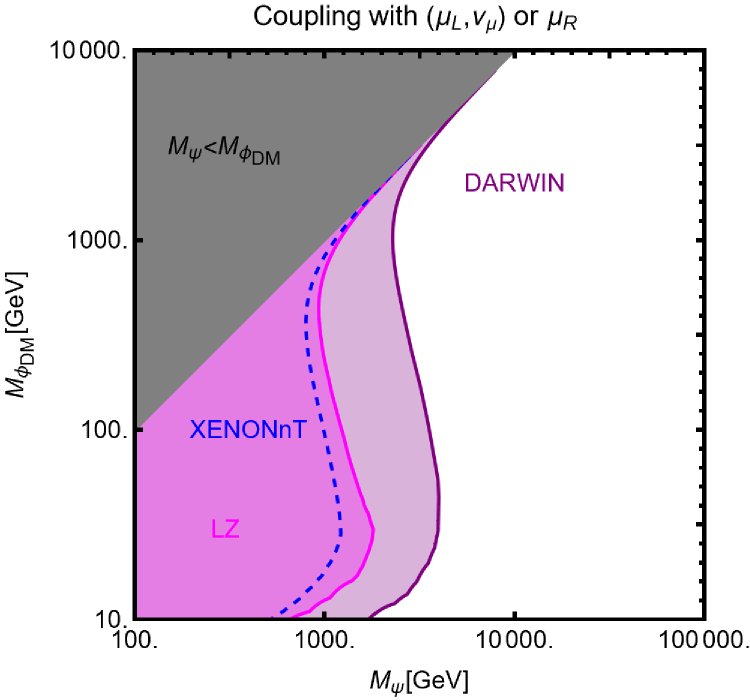}}
    \subfloat{\includegraphics[width=0.336\linewidth]{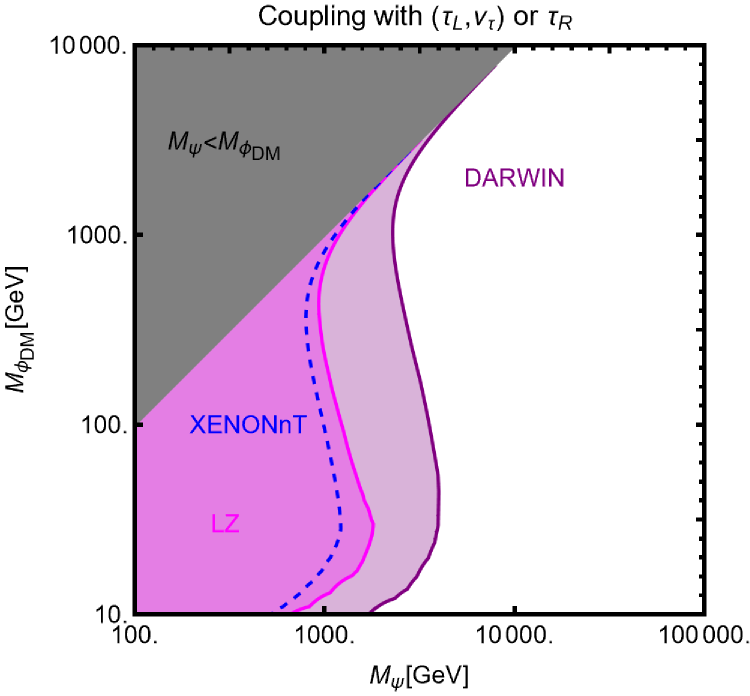}}
\caption{\it Collider limits/prospects for complex scalar DM coupled with SM leptons (as usual the different generations are considered individually) via a $t$-channel mediator.}
\label{fig:CSlepto}
\end{figure}

\begin{figure}[t]
    \centering
     \subfloat{\includegraphics[width=0.335\linewidth]{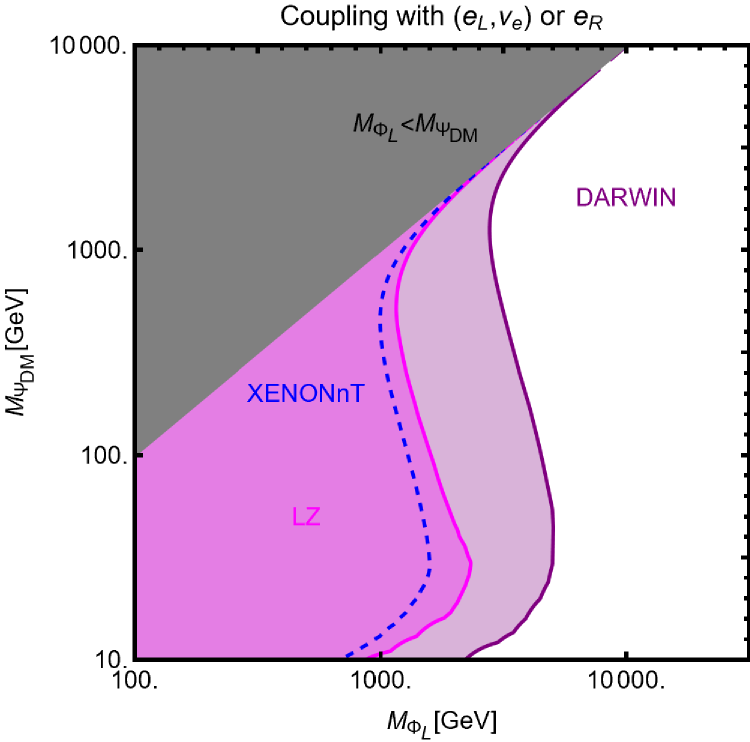}}
    \subfloat{\includegraphics[width=0.335\linewidth]{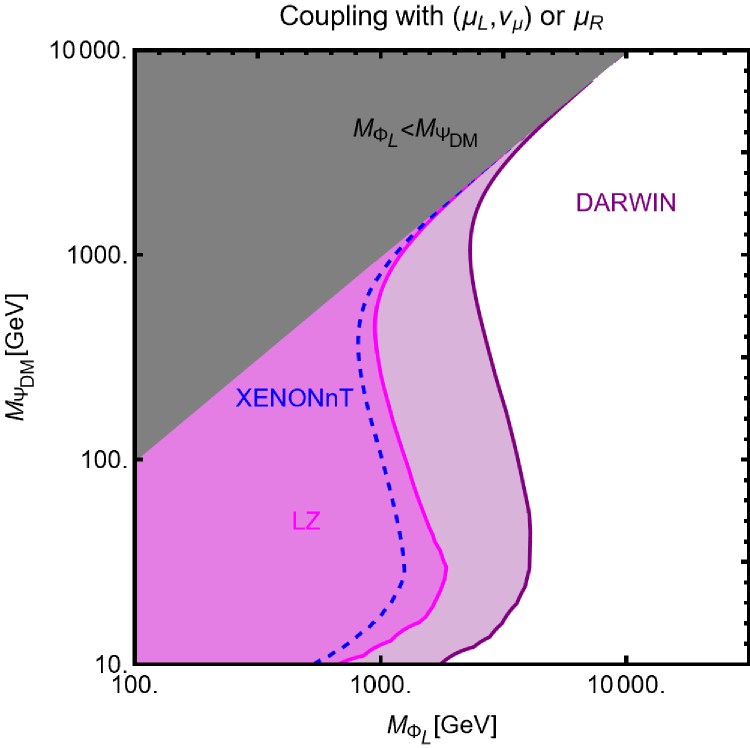}}
    \subfloat{\includegraphics[width=0.335\linewidth]{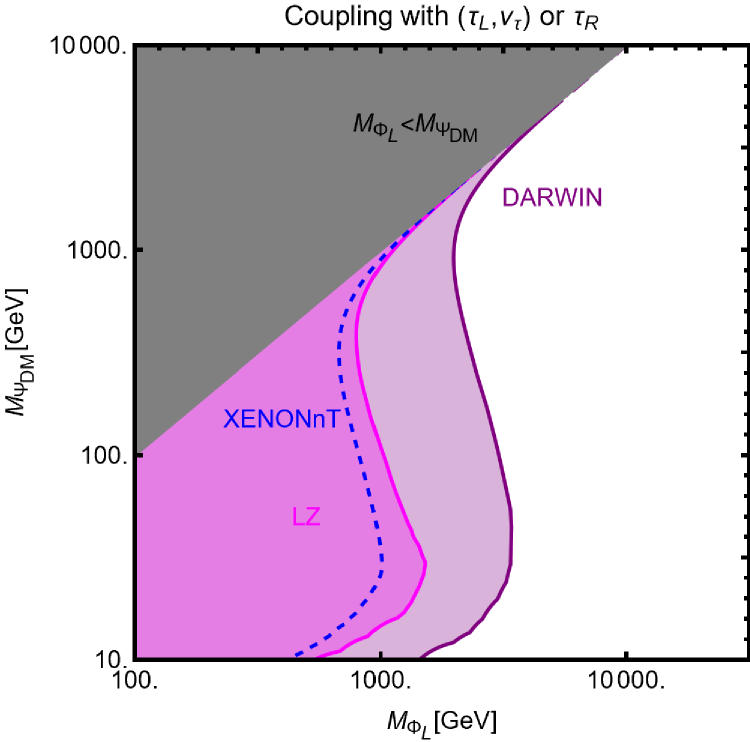}}
\caption{\it The same as~\Fig{fig:CSlepto} but for fermionic DM.}
\label{fig:DFlepto}
\end{figure}

The results of our numerical analysis are shown, with the usual color code, in Figures.~\ref{fig:CSlepto} and \ref{fig:DFlepto}, for complex scalar and dirac fermionic DM respectively. Similar to previously considered scenarios, the expected sensitivity to Direct Detection is substantially the same for both kinds of DM candidates. Current limits by LZ excluded masses for the DM and the $t$-channel mediators up to the TeV scale. An eventual negative signal also at the DARWIN detector would increase the lower bounds to around 5 TeV.

\section{DM relic density and Indirect Detection}
\label{sec:relic}

A strong complementary constraint on the parameter space discussed in the previous section comes from the DM relic density. The latter is indeed measured with great precision by the Planck experiment \cite{Planck:2018vyg}. The strongest correlation with Direct Detection is typically enforced in the case the thermal freeze-out paradigm.
In the latter case, the only particle physics input accounting for the DM relic density is the DM thermally averaged pair annihilation cross-section through the relation:
\cite{Gondolo:1990dk,Edsjo:1997bg}:
\begin{equation}
\label{eq:relic_thermal}
\Omega_{\rm DM}h^2 \approx 8.76 \times 10^{-11}\,{\mbox{GeV}}^{-2}{\left[\int_{T_{\rm f.o.}}^{T_0} g_{*}^{1/2} \langle \sigma v \rangle_{\rm eff}\frac{dT}{M_{\rm DM}}\right]}^{-1}\,,
\end{equation}
with $g_*$ being the effective number of relativistic
degrees of freedom while, $T_{\rm f.o.} \sim \frac{M_{\rm DM}}{20}-\frac{M_{\rm DM}}{30}$ is the standard freeze-out temperature while $T_0$ is the present time temperature of the Universe. The experimentally determined value of the DM relic density \cite{Planck:2018vyg}:
\begin{equation}
    \label{eq:DMOmega}
\Omega_{\rm DM} h^2 = 0.1199 \pm 0.0022\,.
\end{equation}
is achieved for annihilation cross-sections of the order of $10^{-26}\,{\mbox{cm}}^3 {\mbox{s}}^{-1}$.
For the kind of models under scrutiny, $\langle \sigma v \rangle$ includes both DM pair annihilation processes as well as coannihilation processes involving the $t$-channel mediator. Its expression is hence given by \cite{Bai:2013iqa}:
\begin{align}
\langle \sigma v \rangle_{\rm eff}~=~ &\frac{1}{2}\langle \sigma v \rangle_{\rm DM\, DM}\frac{g_{\rm DM}^2}{g_{\rm eff}^2}+\langle \sigma v \rangle_{\rm DM\, M}\frac{g_{\rm DM} g_{\rm M}}{g^2_{\rm eff}}{\left(1+\tilde{\Delta}\right)}^{3/2} \exp\left[-x \tilde{\Delta} \right]+ \\
& \frac{1}{2}\langle \sigma v \rangle_{\rm M^{\dagger}M}\frac{g_{\rm M}^2}{g_{\rm eff}^2}{\left(1+\tilde{\Delta}\right)}^3 \exp\left[-2 x \tilde{\Delta} \right]\,,
\end{align}
in the case of complex scalar or dirac fermionic DM. In the case of real scalar of majorana fermion we have a slightly different expression:
\begin{align}
\langle \sigma v \rangle_{\rm eff} 
~=~ & \langle \sigma v \rangle_{\rm DM \,DM}\frac{g_{\rm DM}^2}{g_{\rm eff}^2}+\langle \sigma v \rangle_{\rm DM\, M}\frac{g_{\rm DM} g_{\rm M}}{g^2_{\rm eff}}{\left(1+\tilde{\Delta}\right)}^{3/2} \exp\left[-x \tilde{\Delta} \right] +
\nonumber \\
& \left(\langle \sigma v \rangle_{\rm M^{\dagger}M}+\langle \sigma v \rangle_{\rm M\,M}\right)\frac{g_{\rm M}^2}{g_{\rm eff}^2}{\left(1+\tilde{\Delta}\right)}^3 \exp\left[-2 x \tilde{\Delta} \right]\,.
\end{align}
In the above equations $\tilde{\Delta}={(M_{\rm M}-M_{\rm DM})}/{M_{\rm DM}}$ is the relative DM/mediator mass splitting while:
\begin{equation}
g_{\rm eff}=g_{\rm DM}+g_{\rm M}{\left(1+\tilde{\Delta}\right)}^{3/2}\exp\left[-x \tilde{\Delta} \right]\,,
\end{equation}
where $g_{\rm M}$ and $g_{\rm DM}$ are the internal degrees of freedom of the mediator and of the DM.
Despite the results presented below have been obtained through the package micrOMEGAs \cite{Belanger:2015nma}.\footnote{As recently shown, e.g.~in~\cite{Biondini:2018ovz,Biondini:2019int,Becker:2022iso}, in case of coannihilation between the DM and a coloured field, additional effects, like thermal corrections, Sommerfeld enhancement and bound state formation might be relevant. We postpone a discussion of these effects to future studies.}, it is helpful, for the comparison with Direct Detection limit, to show approximate expressions, obtained by taking the leading order contributions, in the so-called velocity expansion, of the DM annihilation cross-section into pairs of SM fermions, taking also the limit of massless final states \cite{Bai:2013iqa,Bai:2014osa,Giacchino:2015hvk,Arcadi:2017kky}:
\eqa{
\langle \sigma v \rangle_{{\rm DM}\, {\rm DM}}^{\text{Complex}}
&=&
\sum_f N_c^f\frac{|\Gamma^f_{L,R}|^4 M_{\Phi_{\rm DM}}^2 v^2}{48 \pi {\left(M_{\Phi_{\rm DM}}^2+M_{\Psi_f}^2\right)}^2}\ ,
\\
\langle \sigma v \rangle_{{\rm DM}\, {\rm DM}}^{\text{Dirac}}
&=&
\sum_f N_c^f\frac{|\Gamma^f_{L,R}|^4 M_{\Psi_{\rm DM}}^2}{32 \pi {\left(M_{\Psi_{\rm DM}}^2+M_{\Phi_f}^2\right)}^2}\ ,
\\
\langle \sigma v \rangle_{{\rm DM}\, {\rm DM}}^{\text{Real}}
&=&
\sum_f N_c ^f\frac{|\Gamma^f_{L,R}|^4 M_{\Phi_{\rm DM}}^6 v^4}{60 \pi {\left(M_{\Phi_{\rm DM}}^2+M_{\Psi_f}^2\right)}^4}\ ,
\\
\langle \sigma v \rangle_{{\rm DM}\,{\rm DM}}^{\text{Majorana}}
&=&
\sum_f N_c^f\frac{|\Gamma^f_{L,R}|^4 M_{\Psi_{\rm DM}}^2 \left(M_{\Psi_{\rm DM}}^4+M_{\Phi_f}^4\right) v^2}{48 \pi {\left(M_{\Psi_{\rm DM}}^2+M_{\Phi_f}^2\right)}^4}\ .
}
Notice that the limit $m_f \rightarrow 0$ might not be valid in the case of annihilations into top pairs. In the regime in which the mass of the DM is higher, but not too far from the mass of the top quark, the velocity suppression of the annihilation cross-sections for scalar (both real and complex) and majorana DM is lifted. In such regime the latter cross-sections get s-wave contributions of the form~\footnote{
The cross-section of real scalar DM receives as well an additional p-wave contribution which we do not report explicitly here being a rather long expression. For more details on these analytical approximations we refer e.g. to \cite{Berlin:2014tja}}
:
\begin{align}
& \langle \sigma v \rangle_{\rm DM DM,s-wave}^{\rm complex}=\frac{3 |\Gamma^t_{L,R}|^4 m_t^2}{2 \pi {\left(M_{\Psi_t}^2+M_{\Phi_{\rm DM}}^2-m_t^2\right)}^2} \left(1-\frac{m_t^2}{M_{\Phi_{\rm DM}}^2}\right)\ ,
\nonumber\\
& \langle \sigma v \rangle_{\rm DM DM,s-wave}^{\rm real}=\frac{12 |\Gamma^t_{L,R}|^4 }{\pi}{\left(1-\frac{m_t^2}{M_{\Phi_{\rm DM}}^2}\right)}^{3/2} \frac{m_t^2}{{\left(M_{\Phi_{\rm DM}}^2+M_{\Psi_t}^2-m_t^2\right)}^2}\ ,
\nonumber\\
& \langle \sigma v \rangle_{\rm DM DM,s-wave}^{majo}=\frac{3 |\Gamma_{L,R}^t|^4}{2\pi}\frac{m_t^2}{{\left(M_{\Phi_t}^2+M_{\Psi_{\rm DM}}^2-m_t^2\right)}^2}\sqrt{1-\frac{m_t^2}{M_{\Psi_{\rm DM}}^2}}\ .
\end{align}

\begin{figure}[t]
    \centering
    \subfloat{\includegraphics[width=0.33\linewidth]{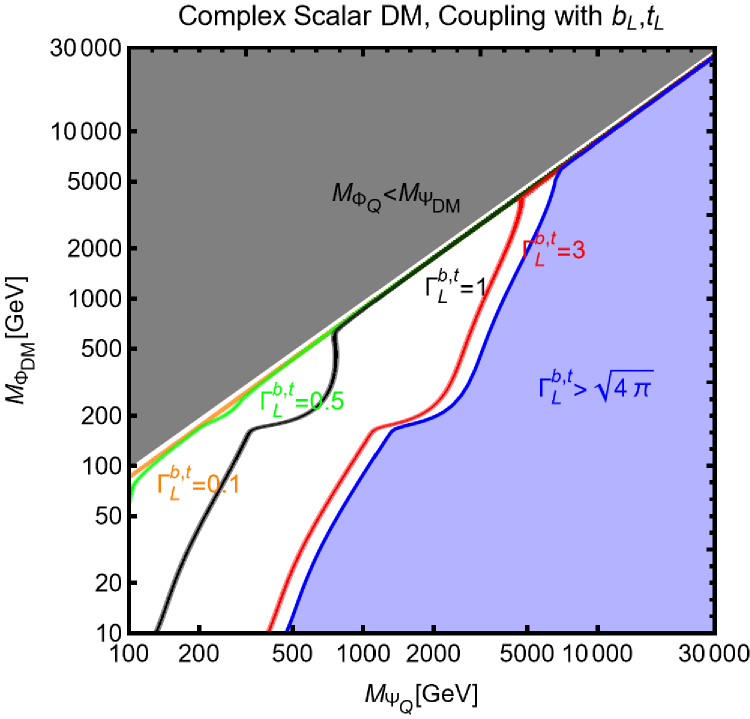}}
    \subfloat{\includegraphics[width=0.33\linewidth]{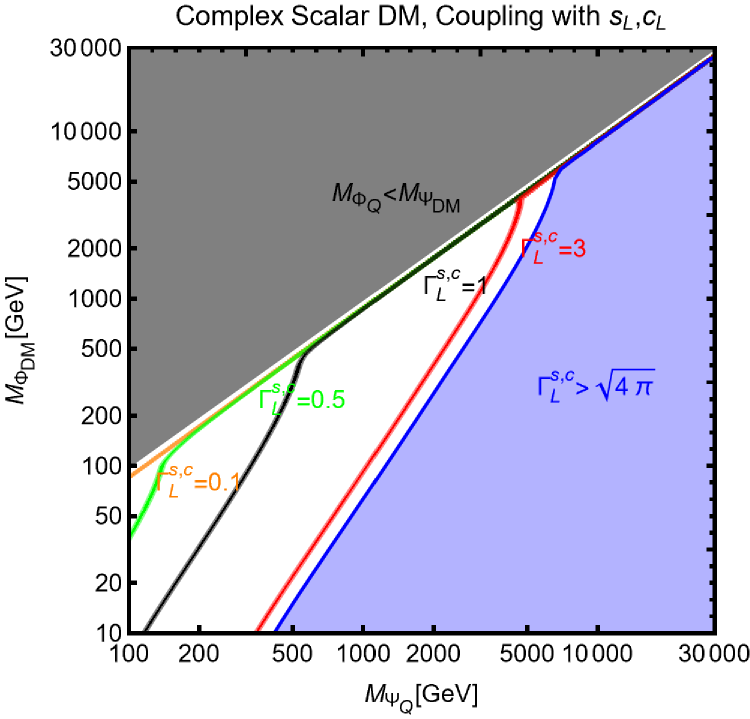}}
     \subfloat{\includegraphics[width=0.33\linewidth]{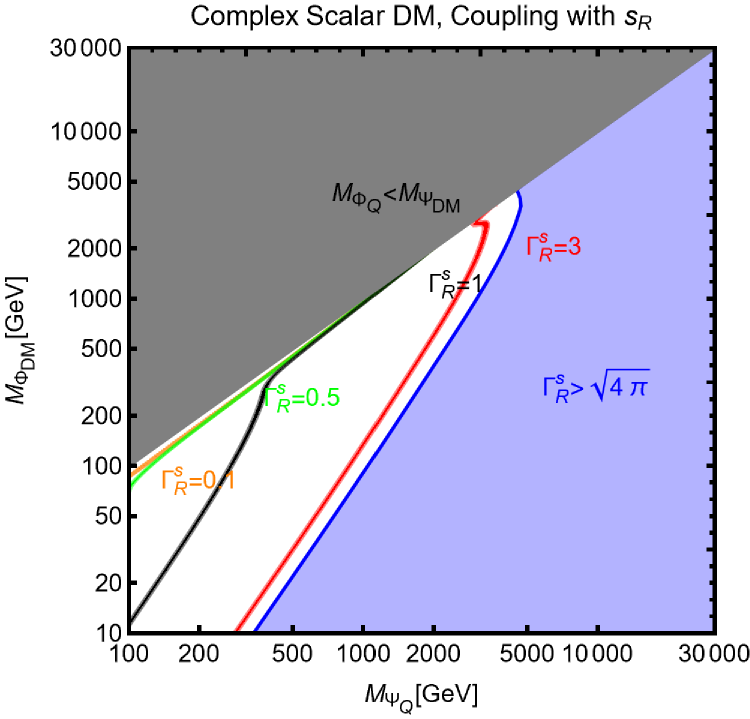}}\\
     \subfloat{\includegraphics[width=0.33\linewidth]{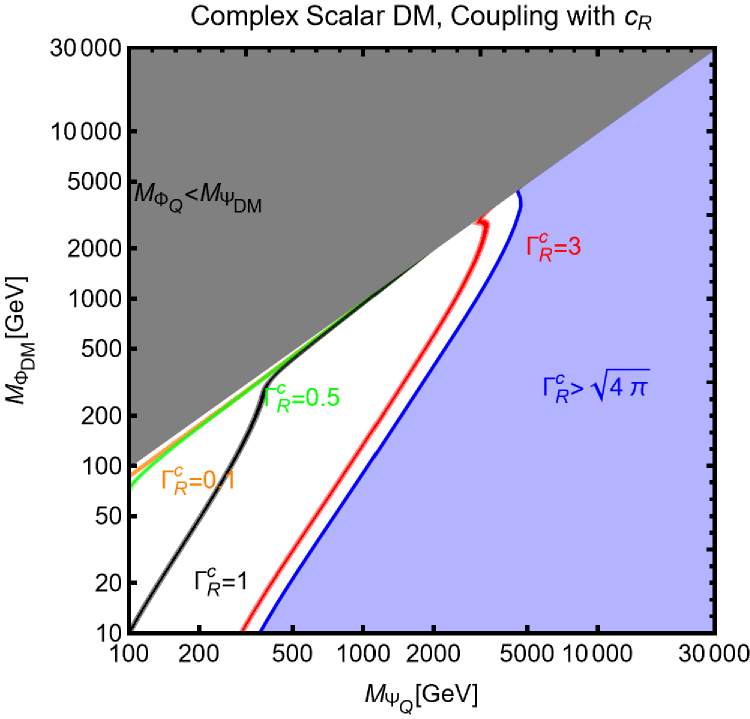}}
    \subfloat{\includegraphics[width=0.33\linewidth]{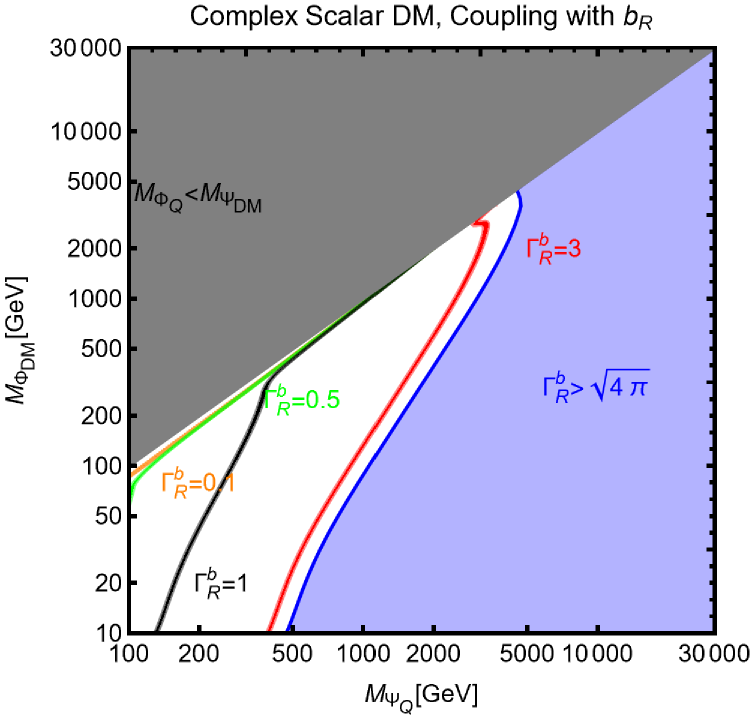}}
     \subfloat{\includegraphics[width=0.33\linewidth]{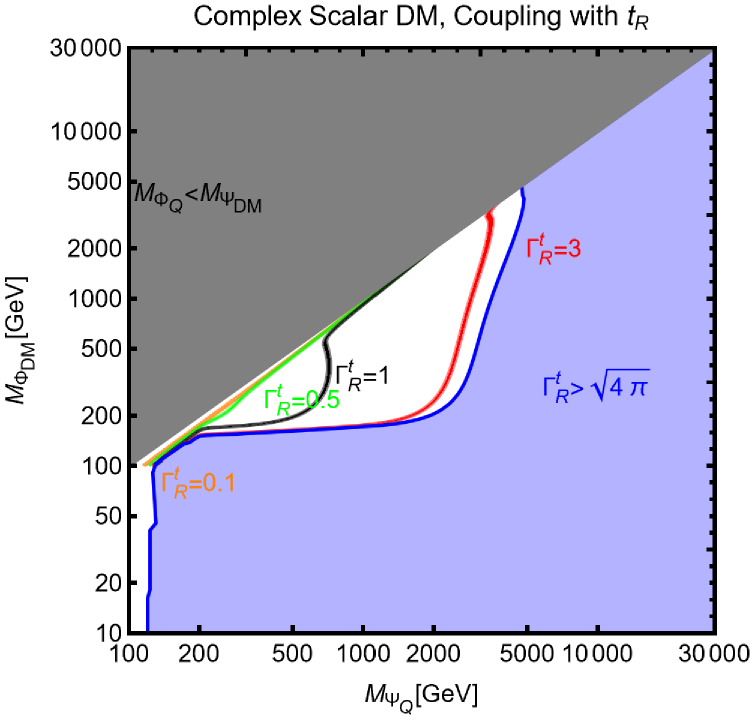}}
    \caption{\it Isocontours of the correct DM relic density, in the $(M_{\Psi_Q},M_{\Phi_{\rm DM}})$ bidimensional plane for complex scalar DM. The different colored solid lines correspond to different assignations of the couplings reported in the panels. The regions in which $\Phi_{\rm DM}$ is not the DM candidate have been marked in gray. The blue regions correspond, instead, to the cases in which the thermally favoured value of the DM annihilation cross-section can be reached only via non-perturbative values of the couplings.}
    \label{fig:prelic_complex}
\end{figure}

\begin{figure}
    \centering
    \subfloat{\includegraphics[width=0.33\linewidth]{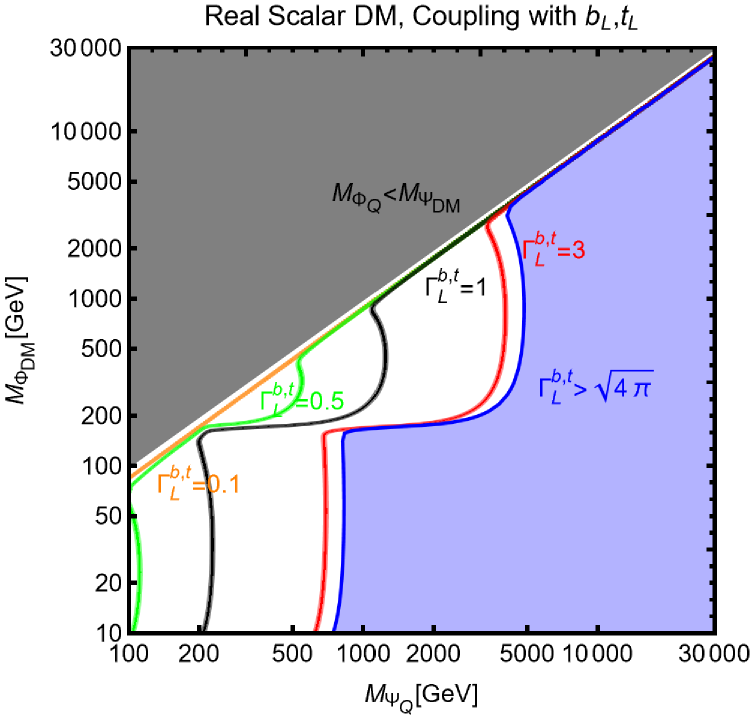}}
    \subfloat{\includegraphics[width=0.33\linewidth]{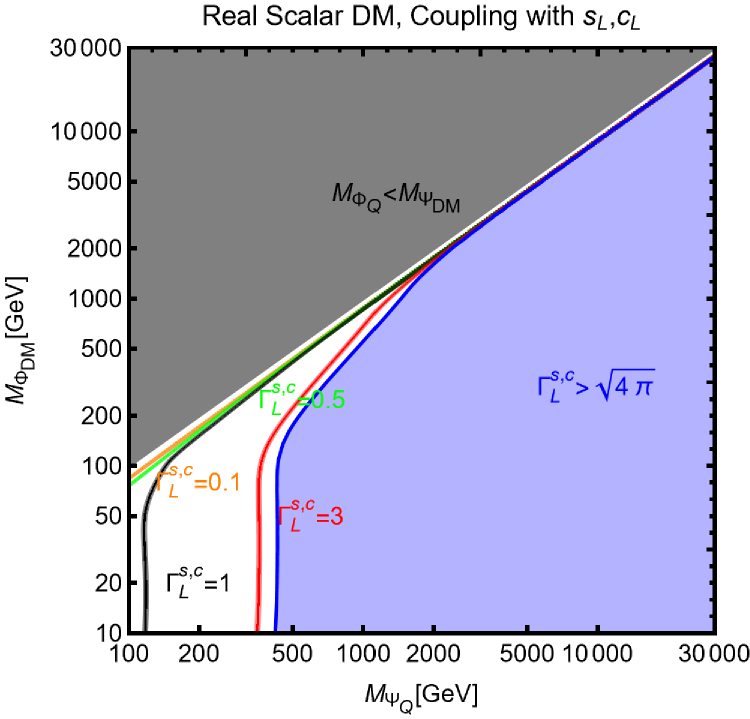}}
     \subfloat{\includegraphics[width=0.33\linewidth]{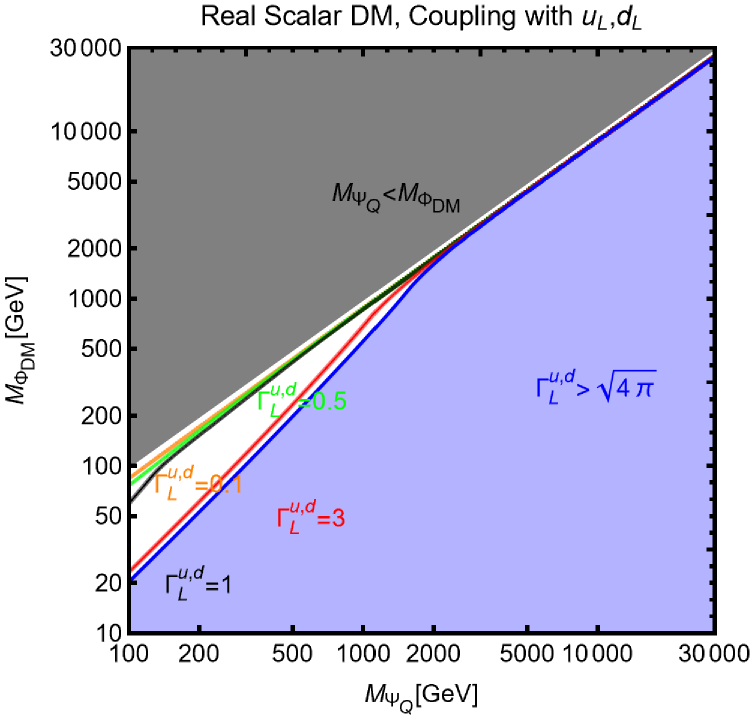}}\\
      \subfloat{\includegraphics[width=0.33\linewidth]{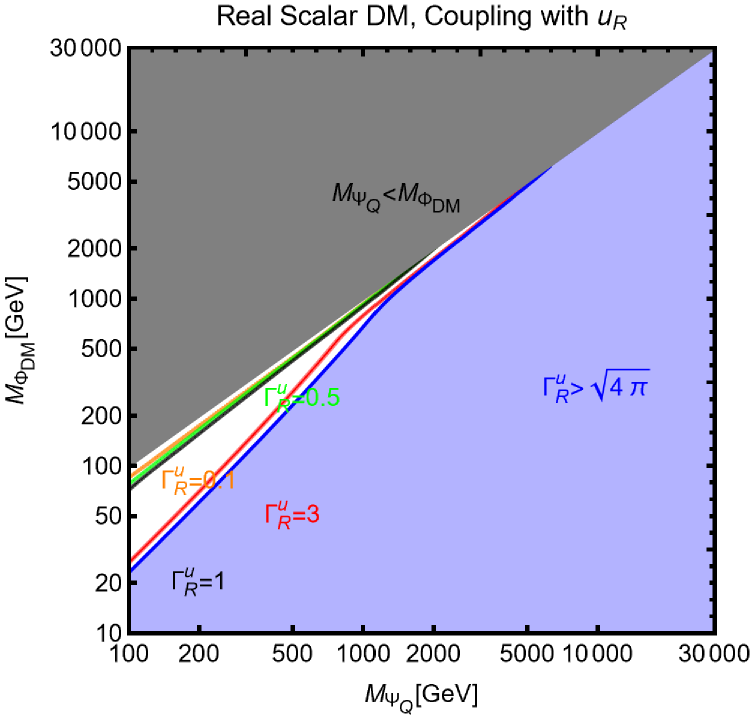}}
    \subfloat{\includegraphics[width=0.33\linewidth]{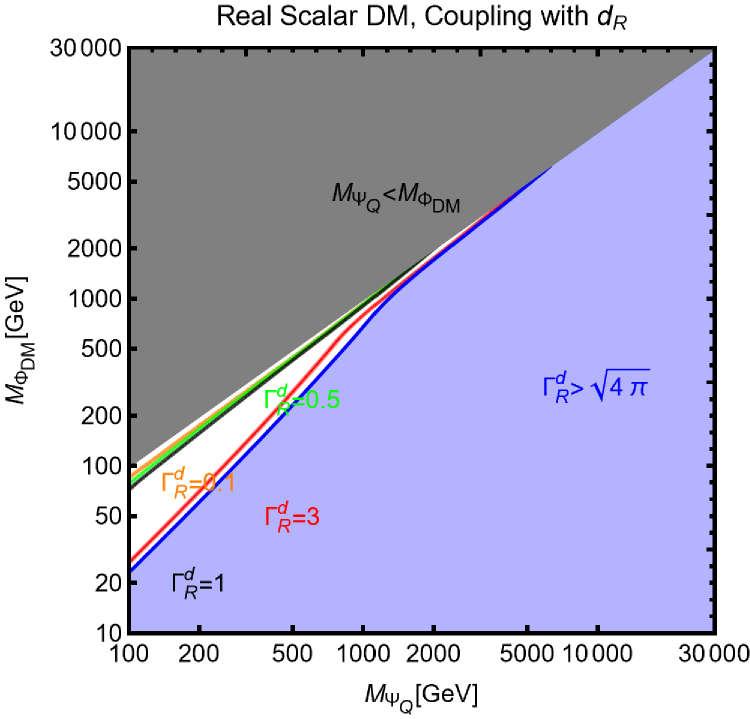}}
     \subfloat{\includegraphics[width=0.33\linewidth]{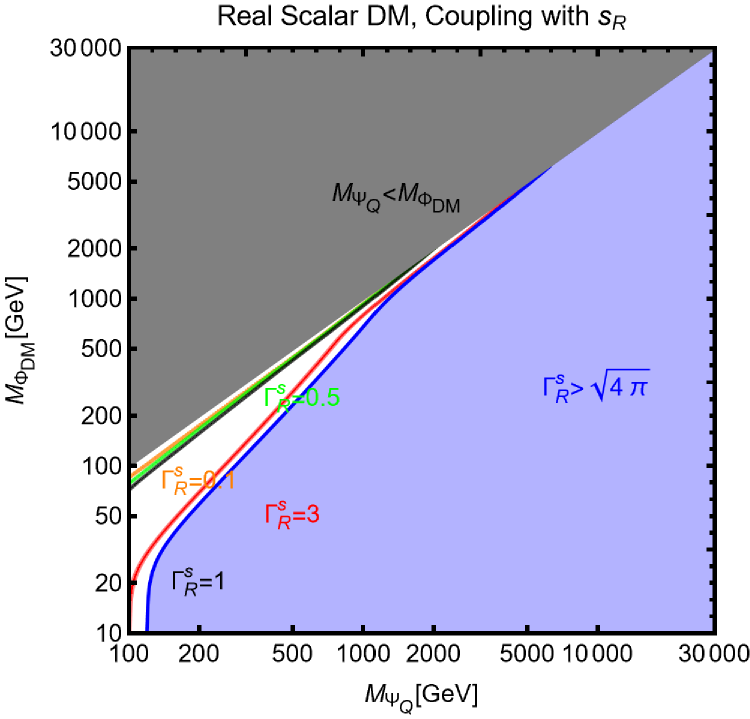}}\\
      \subfloat{\includegraphics[width=0.33\linewidth]{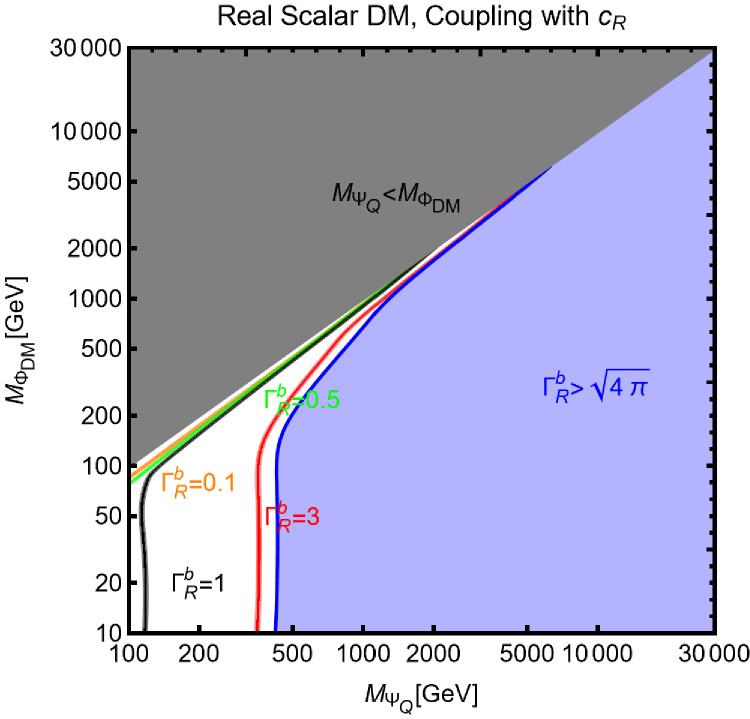}}
    \subfloat{\includegraphics[width=0.33\linewidth]{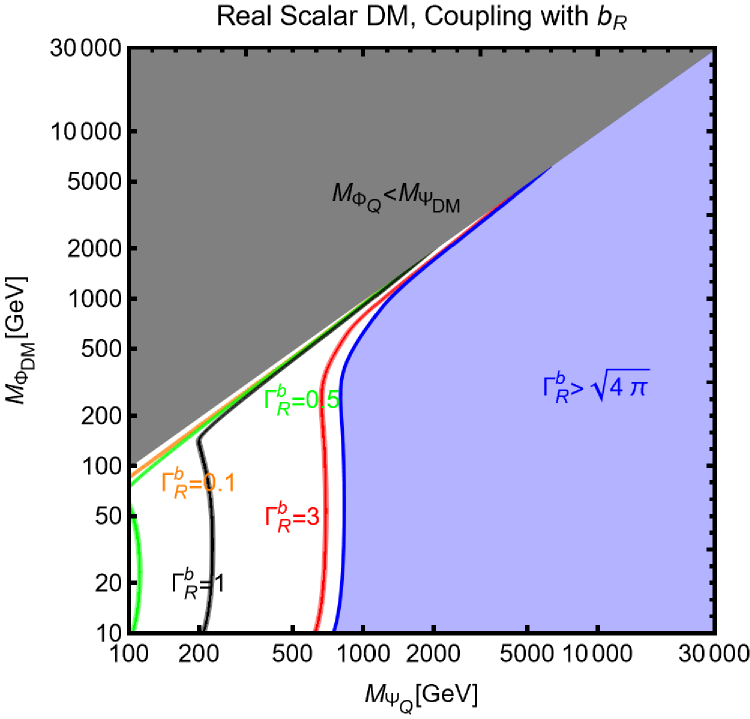}}
     \subfloat{\includegraphics[width=0.33\linewidth]{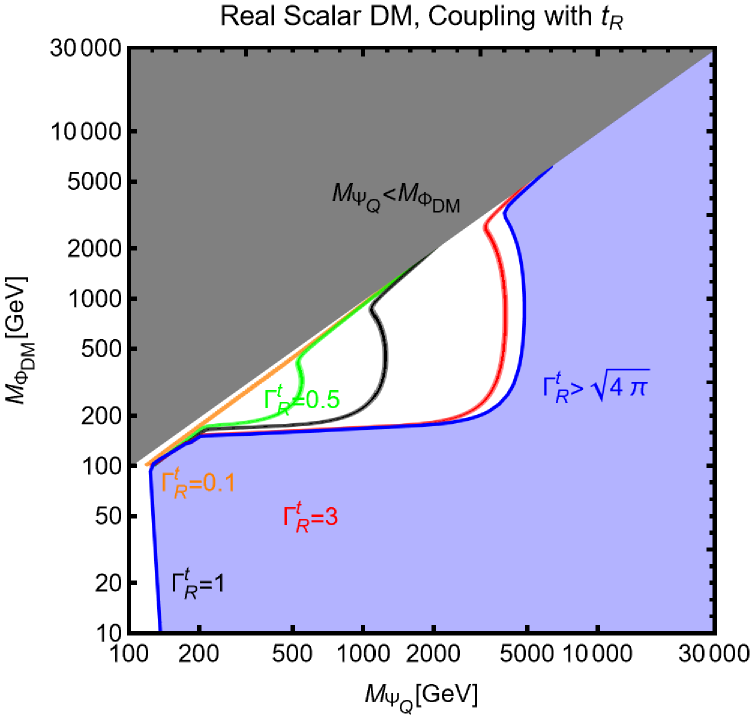}}
    \caption{\it Same as~\Fig{fig:prelic_complex} but for Real Scalar DM}
    \label{fig:prelic_real}
\end{figure}

\begin{figure}[t]
    \centering
    \subfloat{\includegraphics[width=0.33\linewidth]{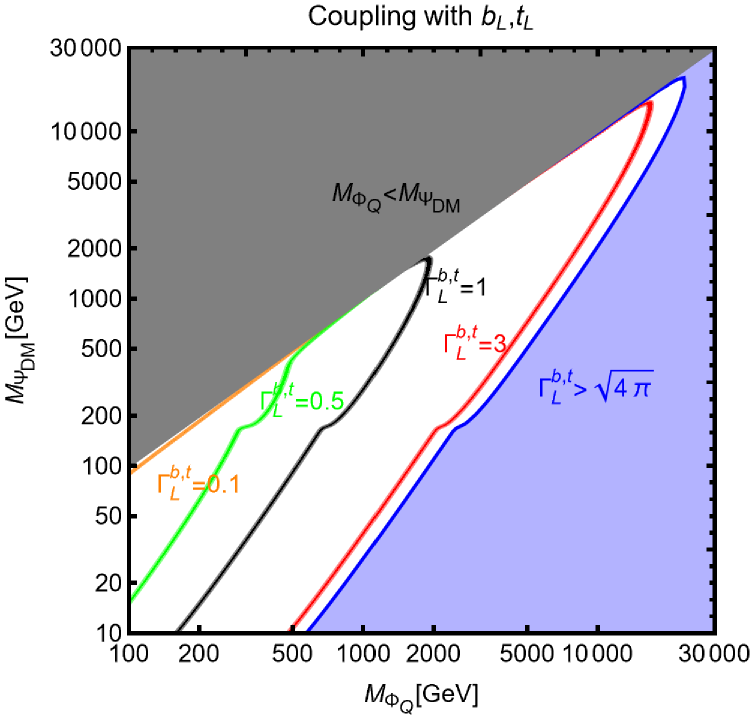}}
    \subfloat{\includegraphics[width=0.33\linewidth]{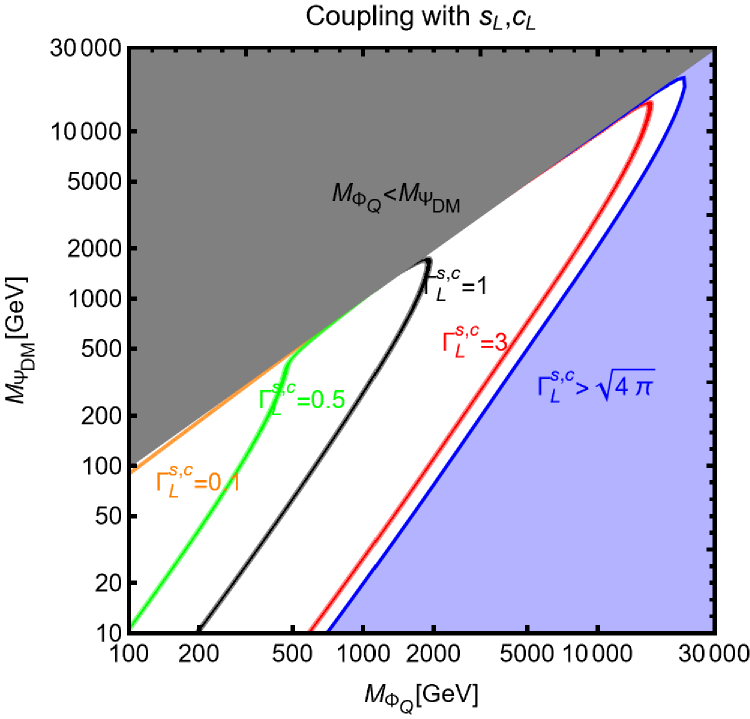}}
    \subfloat{\includegraphics[width=0.33\linewidth]{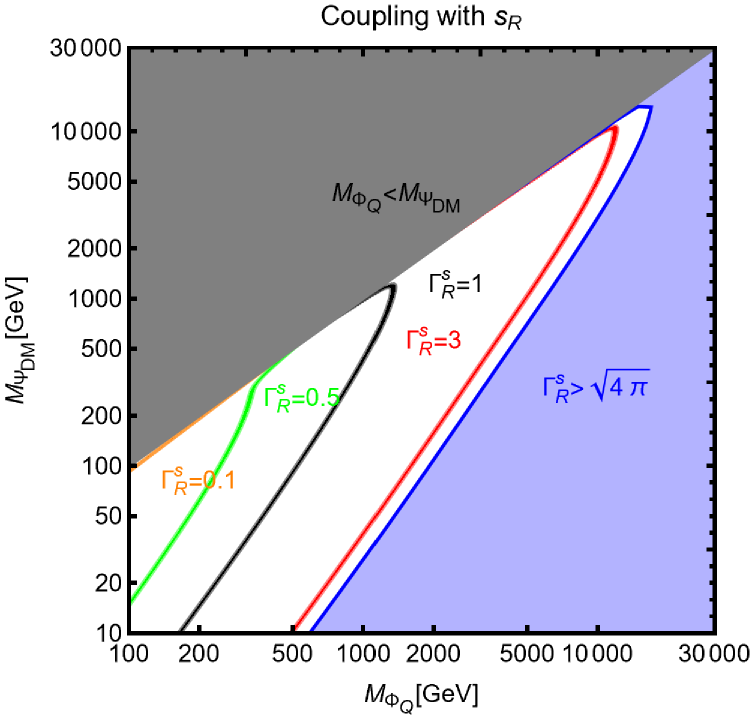}}\\
     \subfloat{\includegraphics[width=0.33\linewidth]{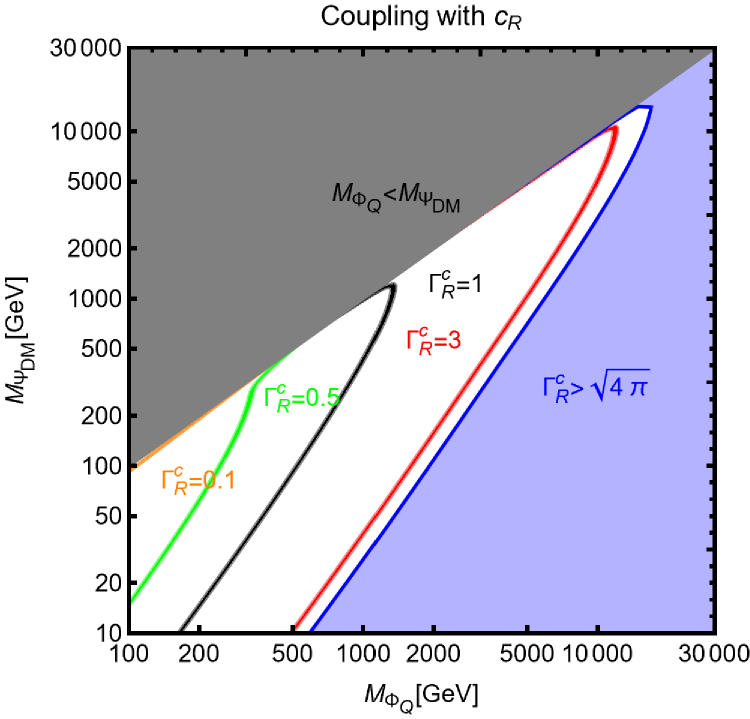}}
    \subfloat{\includegraphics[width=0.33\linewidth]{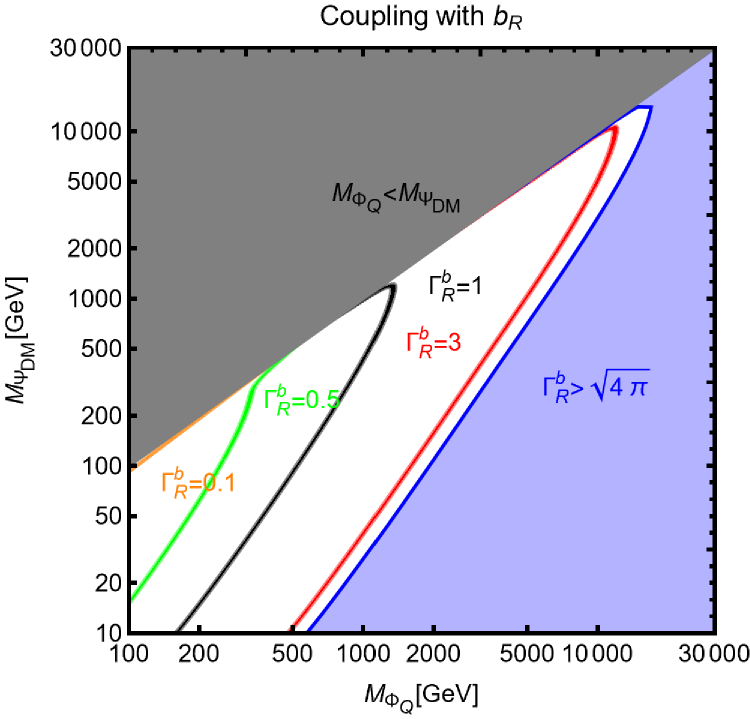}}
    \subfloat{\includegraphics[width=0.33\linewidth]{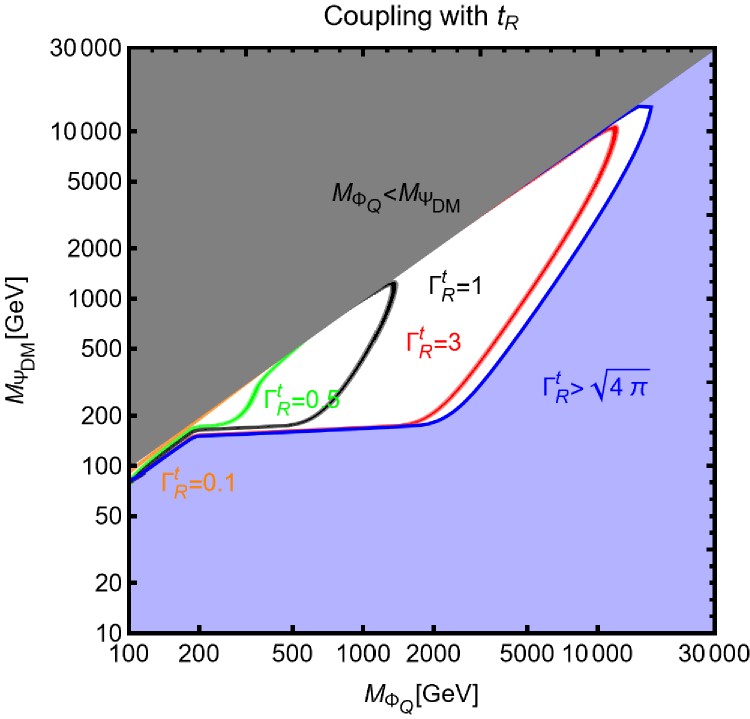}}
    \caption{\it Isocontours of the correct DM relic density, in the $(M_{\Phi_Q},M_{\Psi_{\rm DM}})$ bidimensional plane. The different colored solid lines correspond to different assignations of the the couplings reported in the panels. The region in which $\Psi_{\rm DM}$ is not the DM candidate has been marked in gray. The blue region corresponds, instead, to the case in which the thermally favoured value of the DM annihilation cross-section can be reached only via non-perturbative values of the couplings.}
    \label{fig:prelic_dirac}
\end{figure}

\begin{figure}
    \centering
    \subfloat{\includegraphics[width=0.33\linewidth]{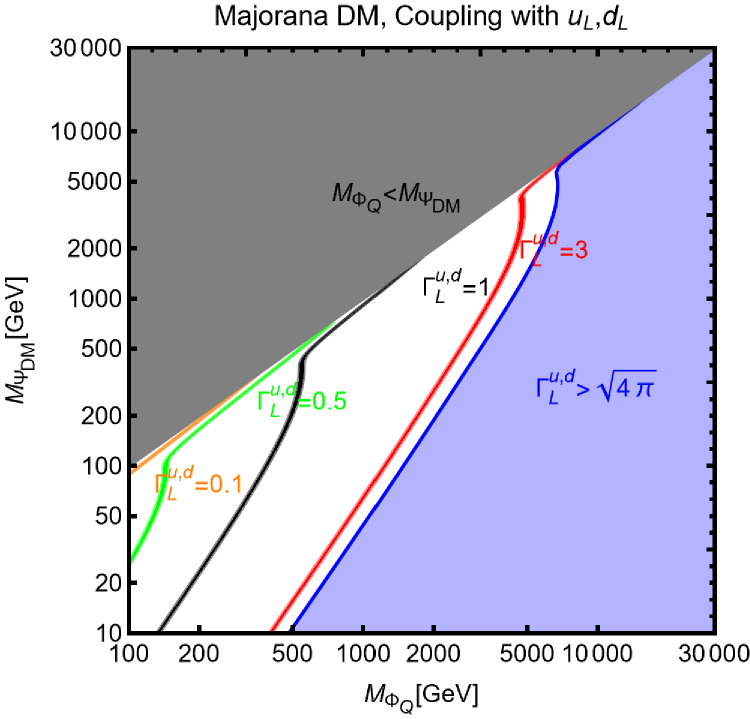}}
    \subfloat{\includegraphics[width=0.33\linewidth]{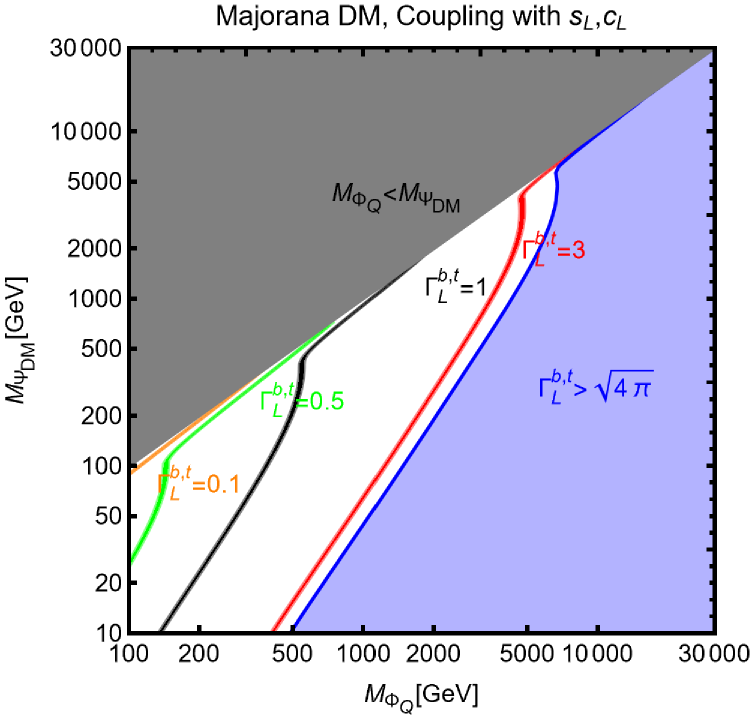}}
    \subfloat{\includegraphics[width=0.33\linewidth]{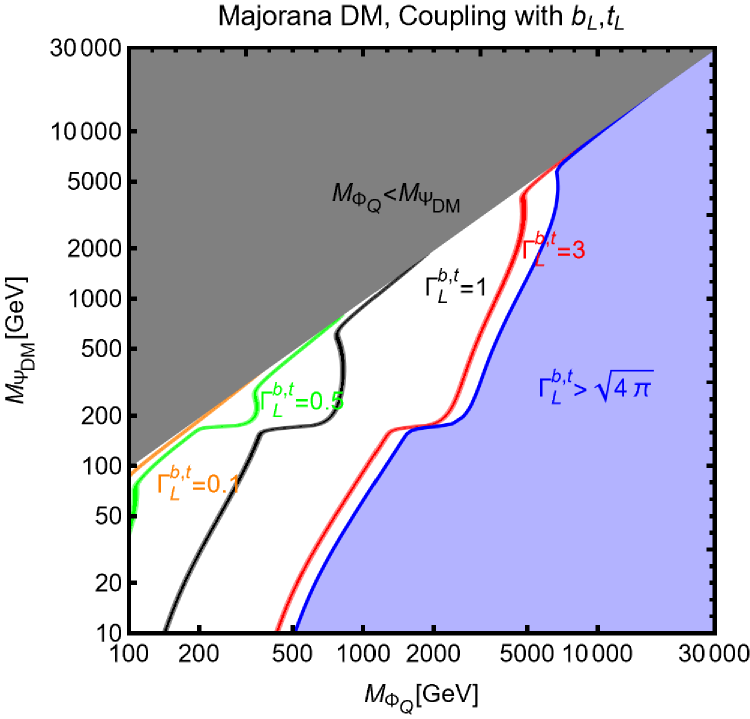}}\\
     \subfloat{\includegraphics[width=0.33\linewidth]{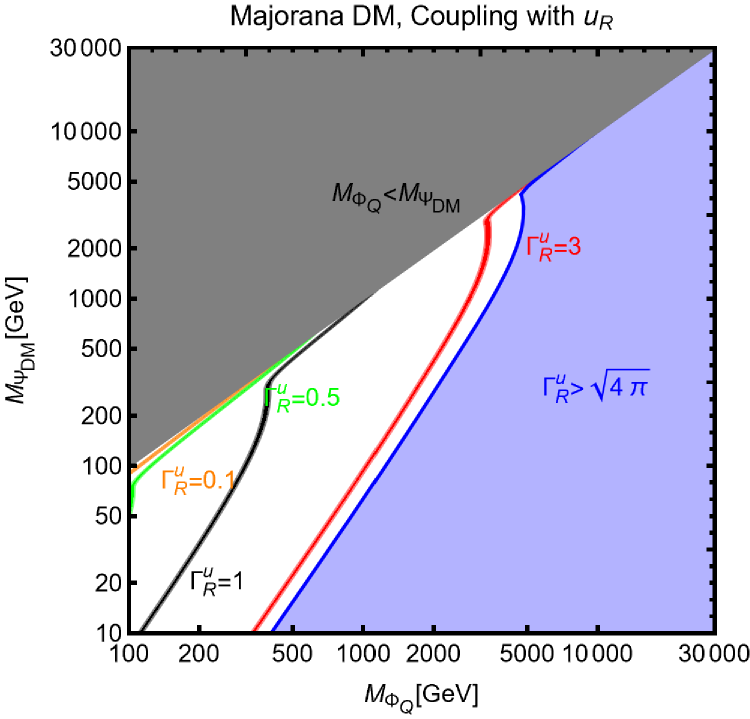}}
    \subfloat{\includegraphics[width=0.33\linewidth]{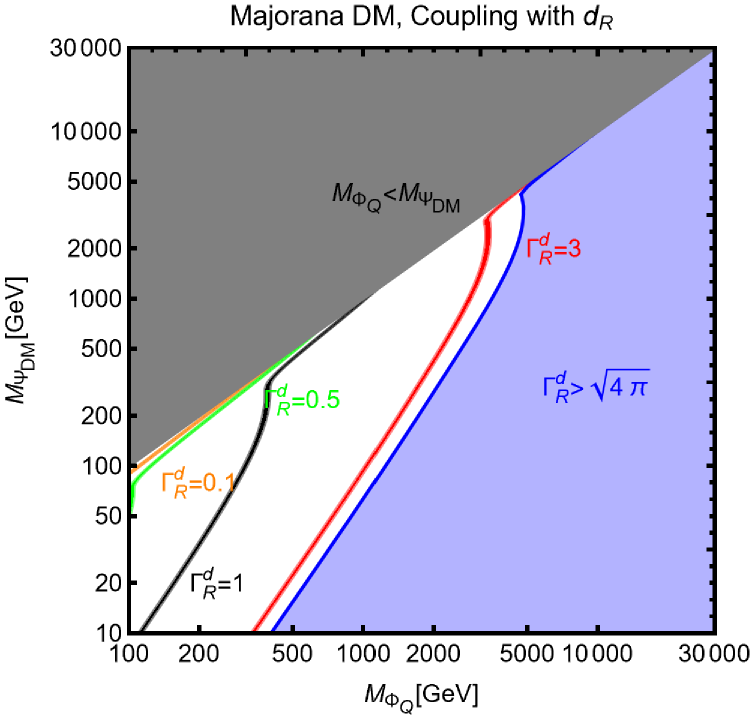}}
    \subfloat{\includegraphics[width=0.33\linewidth]{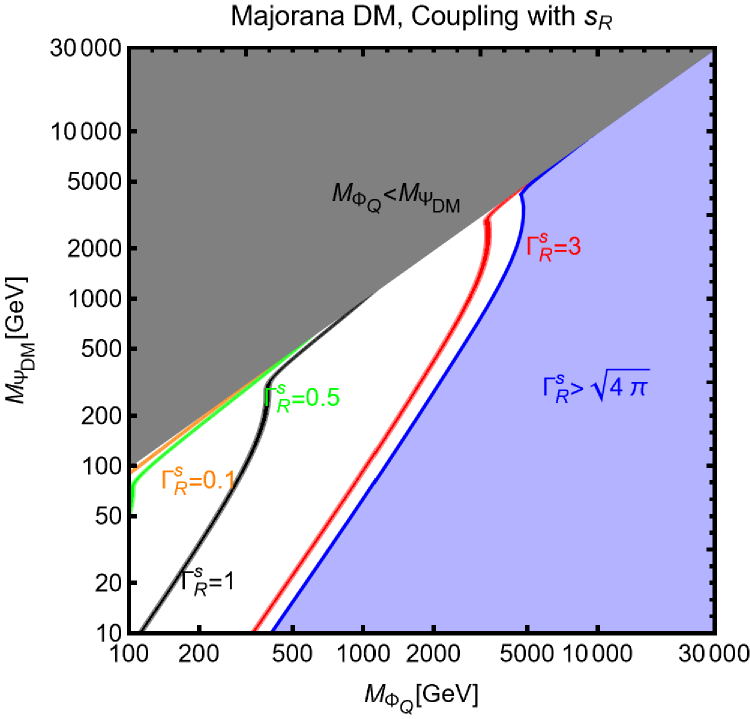}}\\
    \subfloat{\includegraphics[width=0.33\linewidth]{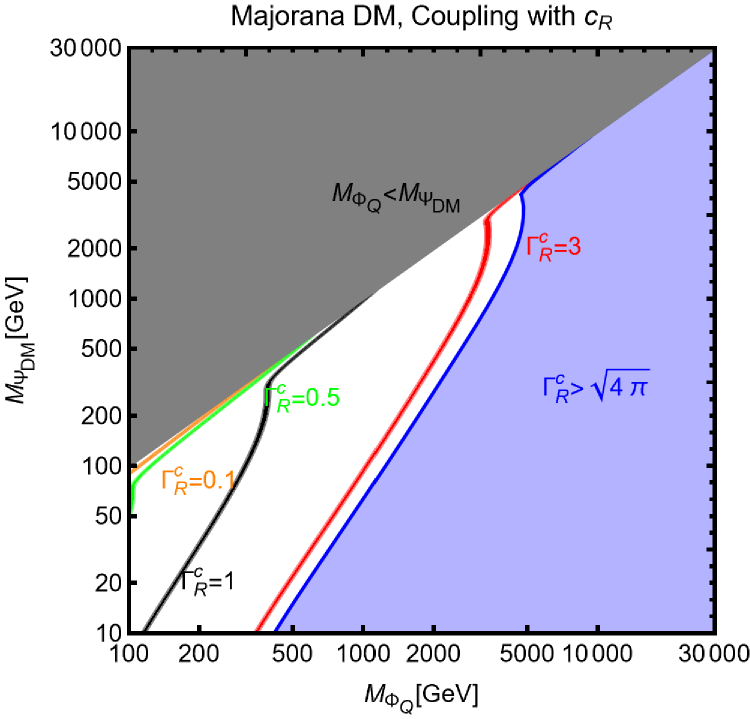}}
    \subfloat{\includegraphics[width=0.33\linewidth]{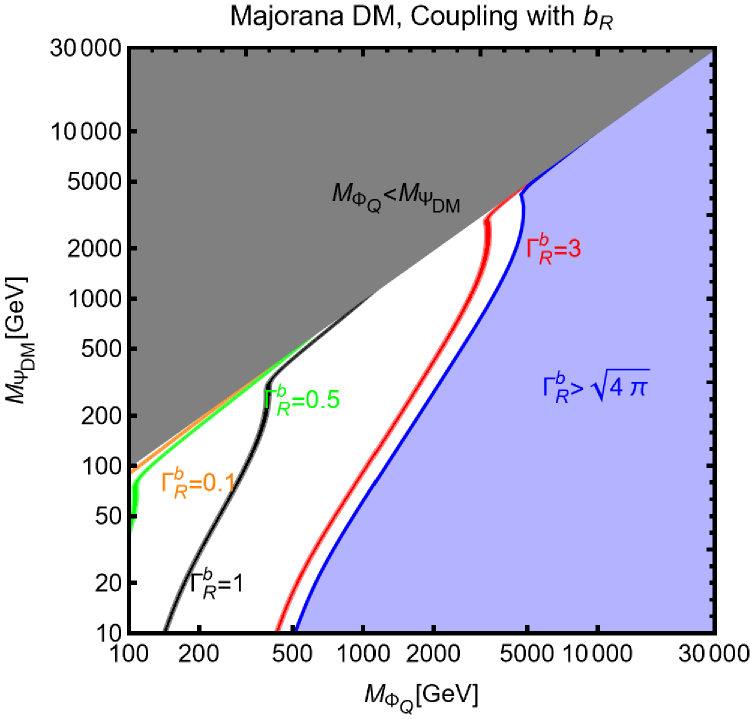}}
    \subfloat{\includegraphics[width=0.33\linewidth]{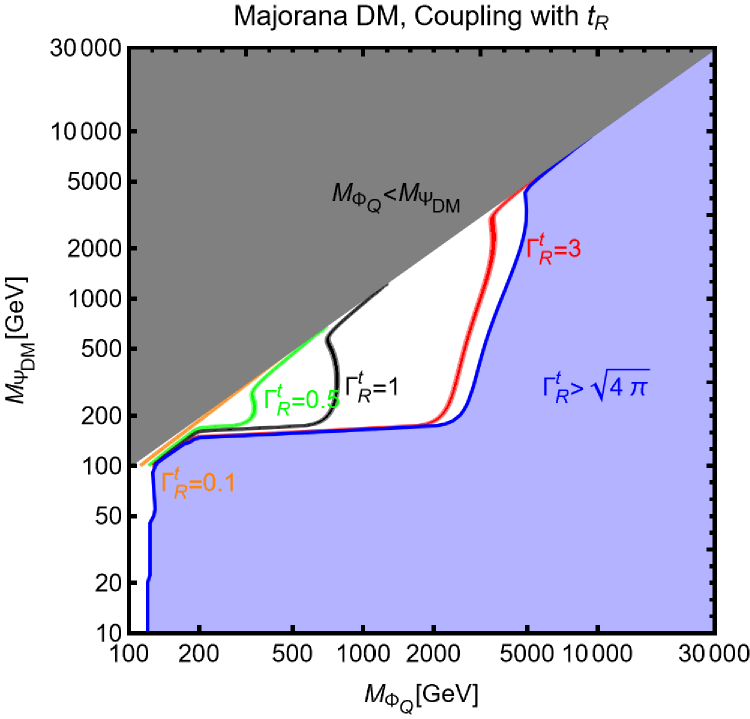}}
    \caption{\it Same as~\Fig{fig:prelic_dirac} but for Majorana DM.}
    \label{fig:prelic_majo}
\end{figure}

The impact of relic density constraints is illustrated in Figures~\ref{fig:prelic_complex}-\ref{fig:prelic_majo}. In each figure, iscontours of the correct relic density, for fixed assignations of the $\Gamma_{L,R}^f$ coupling are shown in the $(M_{\Phi_f},M_{\Psi_{\rm DM}})$ and $(M_{\Psi_f},M_{\Phi_{\rm DM}})$ planes for, respectively, fermionic and scalar DM. Again, we have assumed the coupling of the DM with a single quark species. The different cases correspond to the different panels shown in the figure. We remark again that we have not considered explicitly the cases of couplings of the complex scalar and dirac fermionic DM with the first generation quarks as this scenario corresponds to Direct Detection arising from tree level interactions, not of upmost interest for our study.  In each plot, moreover, the regions in which the DM is not stable (gray) and in which the thermally favoured value of the DM annihilation cross-section can be achieved only via non-perturbative couplings (blue) have been highlighted. In other words, viable relic density, via the freeze-out paradigm, can be achieved only within the white regions of each panel.

In agreement with the analytical estimates, the case of dirac DM seems the most favorable one if only relic density is accounted for. The s-wave dominated cross-section allows, in fact, to achieve the thermally favored value of the annihilation cross-section for values of the couplings down to $0.1$.
Having both a p-wave dominated cross-section, the relic density constraints impact in a very similar way the models with majorana fermion and complex scalar DM. In this cases, the correct relic density
can be obtained away from the coannihilation regime, namely $M_{\Phi_{\rm DM},\Psi_{\rm DM}}\simeq M_{\Psi_f,\Phi_f}$, only for couplings of order 1.
Finally, real scalar DM seems to be disfavored by the relic density constraints.  With the exception of couplings with the top, the viable region of parameter space are very narrow and, in the cases of couplings with $u,d,s$ quarks, essentially limited to the coannihilation regime.
Together with relic density, the other potentially relevant constraint is represented by Indirect Detection, as it is as well a probe of the DM annihilation cross-section. On the other hand, Indirect Detection is sensitive to annihilation processes occurring at present time whose annihilation cross-section might not coincide with the one at freeze-out, as the thermally averaged cross-section is in general a time/temperature dependent function. In the velocity expansion approximation such time dependence is accounted for by the value of the velocity, being $v^2 \simeq 0.1$ at the thermal freeze-out and $v^2 \simeq 0$ at present times. From this it is evident that Indirect Detection probes the parameter regions corresponding to the viable relic density only in the case of s-wave dominated cross-sections, only occurring for dirac fermion DM in our framework. We have, however, verified that the corresponding constraints are always subdominant with respect to the ones from Direct Detection, for some explicit example see also e.g. \cite{Arcadi:2021glq}. Consequently. Indirect Detection constraints are not evidenced in any of the figures shown in this work.

\subsection{Combined Constraints}
\label{sec:combined_constraints}

In this section we combine in a more systematic way the constraints from DM Direct Detection and relic density. To this purpose, we will illustrate the outcome of some scans over the parameter space of the models under consideration. First of all we will consider again the minimal scenarios in which the DM candidate couples with a single quark species via $t$-channel exchange of a single mediator field. We have hence performed the following type of scan:
\begin{align}
\label{eq:min_scan}
    & M_{\Phi_{\rm DM},\Psi_{\rm DM}}\in \left[10,10^5\right]\,\mbox{GeV}\nonumber\\
    & M_{\Phi_f,\Psi_f} \in \left[100,10^5\right]\,\mbox{GeV}\nonumber\\
    & \Gamma_{L,R}^{f}\in \left[10^{-3},\sqrt{4\pi}\right]
\end{align}
and retained only the model points, i.e. assignation for the set $(M_{\Psi_{\rm DM},\Phi_{\rm DM}},M_{\Phi_{f},\Psi_{f}},\Gamma_{L,R}^f)$, giving the correct DM relic density. Notice that the lower bound of 10 GeV on the DM mass is motivated by the fact that the main focus of this work is to assess the capability of Direct Detection facilities to probe the models under scrutiny. At the moment, the sensitivity of current experiments to low masses is limited by their energy threshold cuts. For this reason we have, conservatively, focused on the $M_{\rm DM} \geq 10\,\mbox{GeV}$ case. The lower limit on the mass of the $t$-channel mediator is just due to the LEP bound on new BSM electrically charged states.

\begin{figure}
\centering
\subfloat{\includegraphics[width=0.48\linewidth]{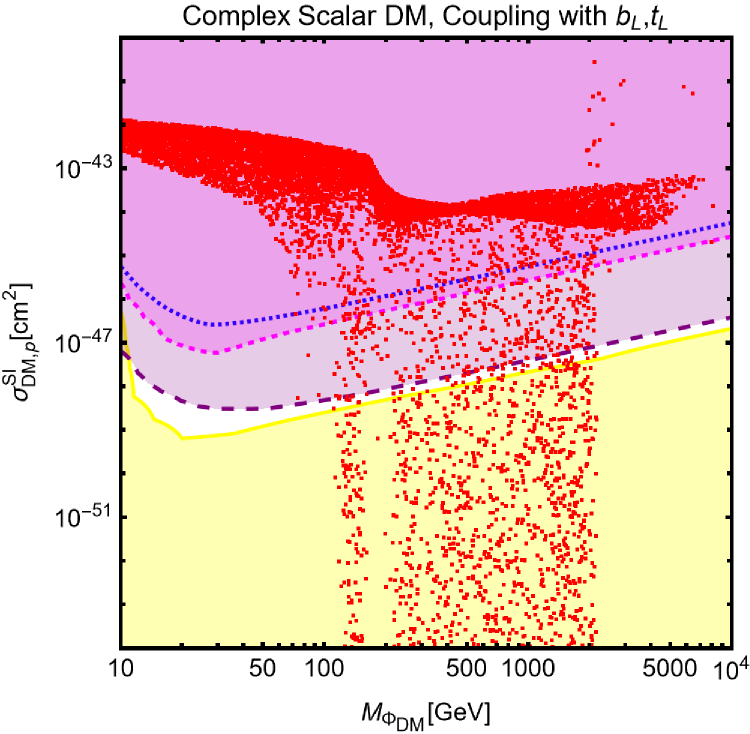}}
\subfloat{\includegraphics[width=0.48\linewidth]{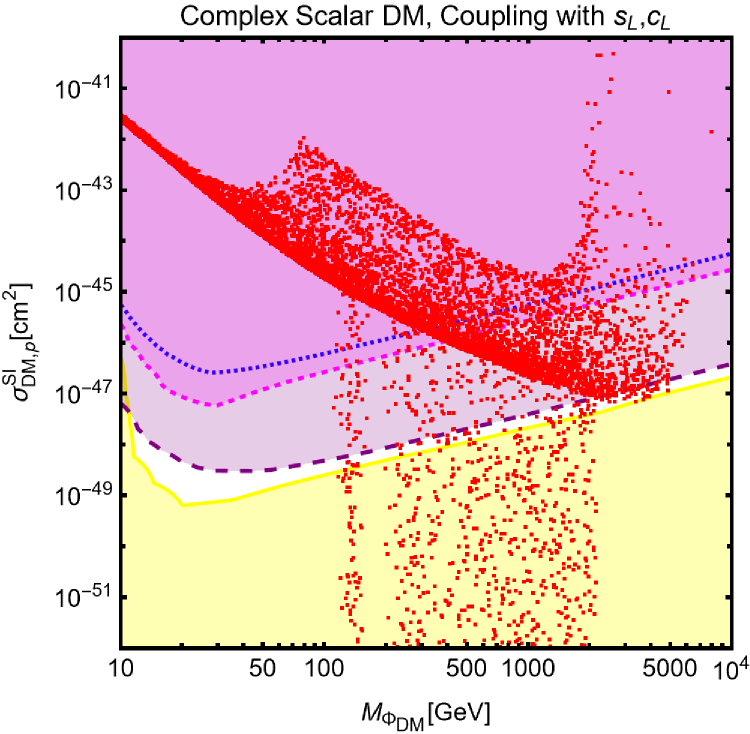}}\\
\subfloat{\includegraphics[width=0.48\linewidth]{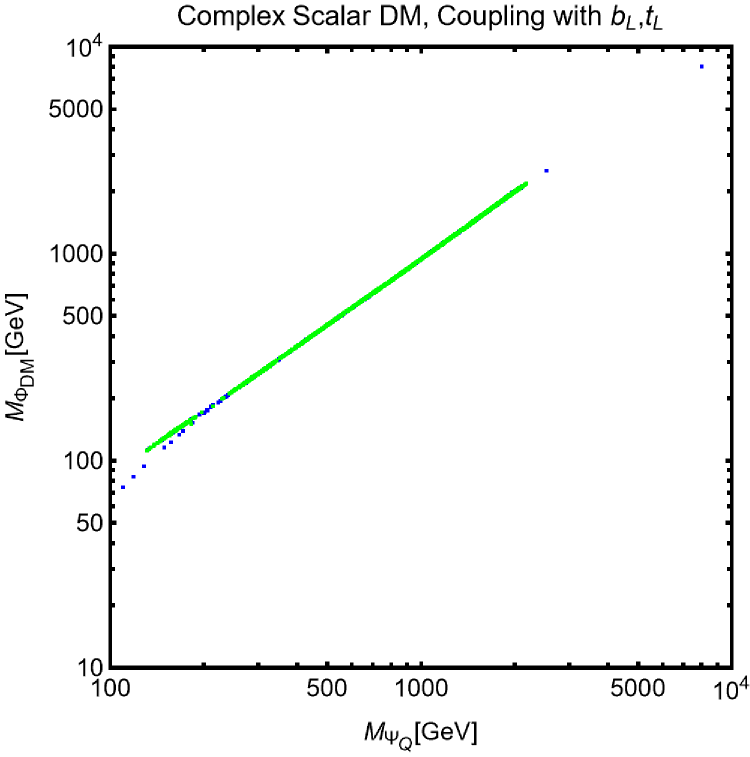}}
\subfloat{\includegraphics[width=0.48\linewidth]{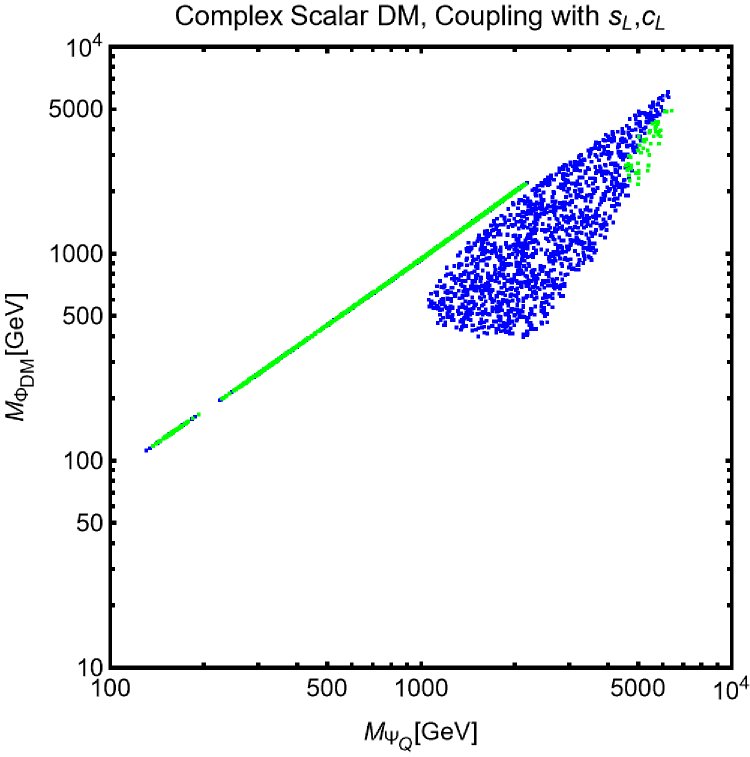}}\\
\caption{\it Model points emerged from the parameter scan, eq. (\ref{eq:min_scan}), for complex scalar DM. The different panels refer to coupling with different quark species. For each case two plots are shown. First all the points complying with the correct relic density are reported in the $(M_{\Phi_{\rm DM}},\sigma_{\Phi_{\rm DM},p}^{\rm SI})$ bidimensional plane. Together with them the exclusion by LZ (magenta region), XENONnT (region above the blue dashed line), the projected sensitivity by DARWIN (purple region) and the region corresponding to the neutrino floor (yellow region) are evidenced. The second panel shows, in the $(M_{\Psi_f},M_{\Phi_{\rm DM}})$ plane, the model point complying with the constraint by LZ (blue points) and the ones which would survive even upon a negative signal at the DARWIN experiment (green points). For definiteness only the cases of couplings of the DM with second and third generation left-handed quarks have been considered.}
\label{fig:scan_complex}
\end{figure}

\begin{figure}
\centering
\subfloat{\includegraphics[width=0.48\linewidth]{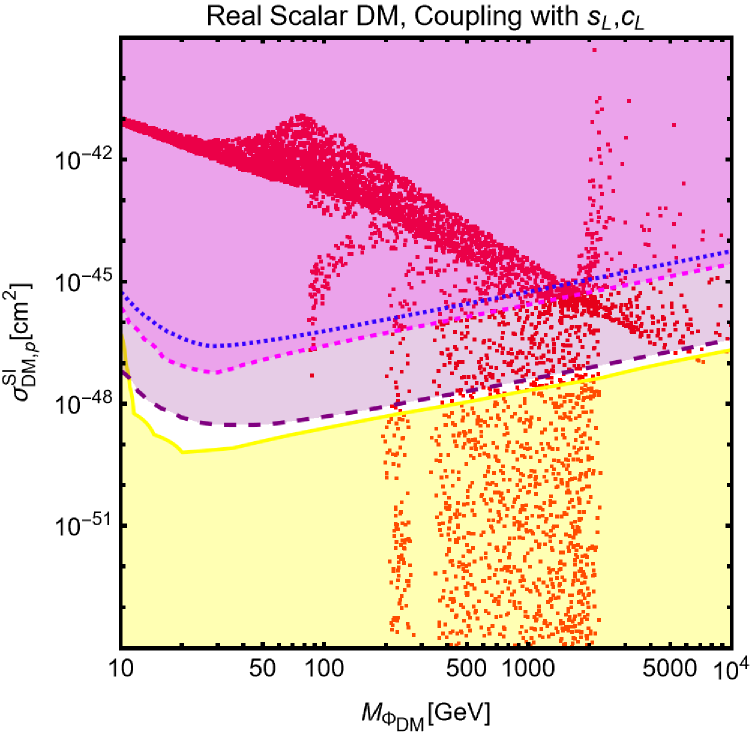}}
\subfloat{\includegraphics[width=0.48\linewidth]{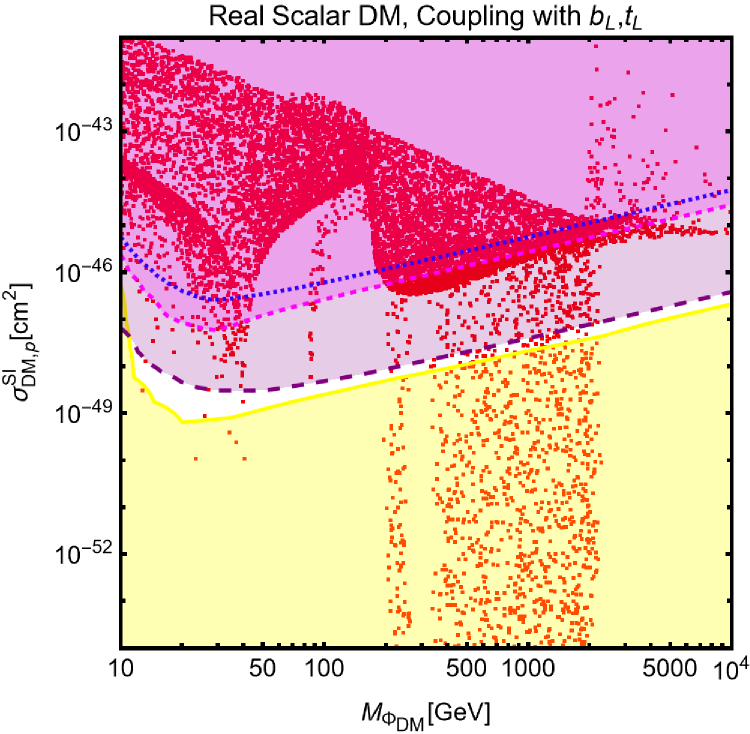}}\\ 
\subfloat{\includegraphics[width=0.48\linewidth]{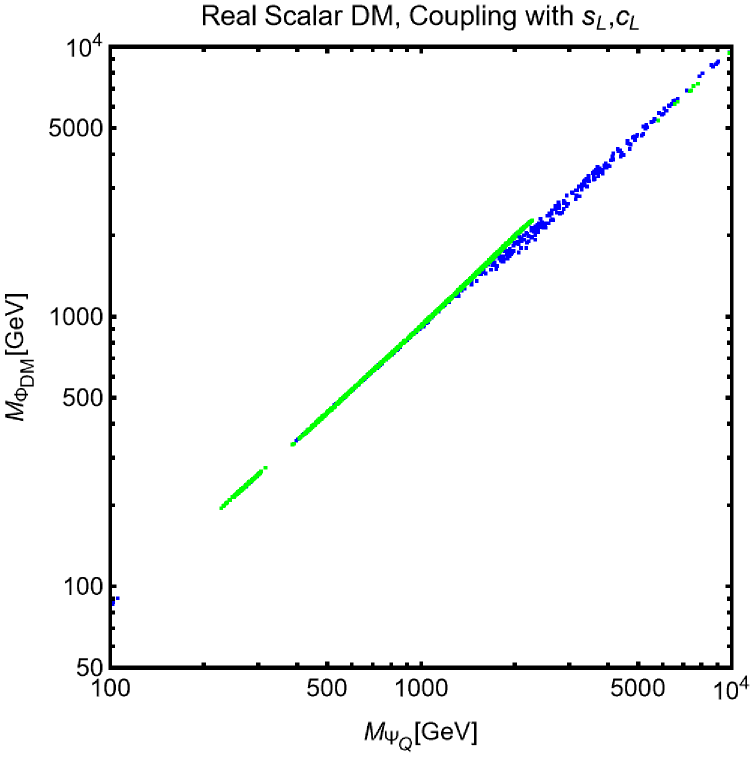}}
\subfloat{\includegraphics[width=0.48\linewidth]{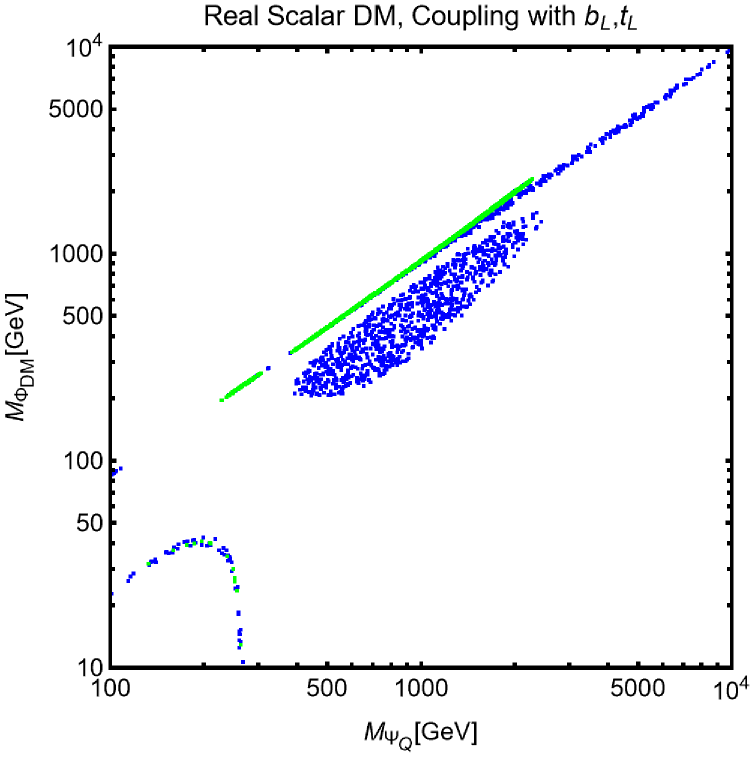}}
\caption{\it The same as~\Fig{fig:scan_complex} but for real scalar DM coupled with left-handed quarks.}
\label{fig:scan_real_L}
\end{figure}

\begin{figure}
\centering
\subfloat{\includegraphics[width=0.48\linewidth]{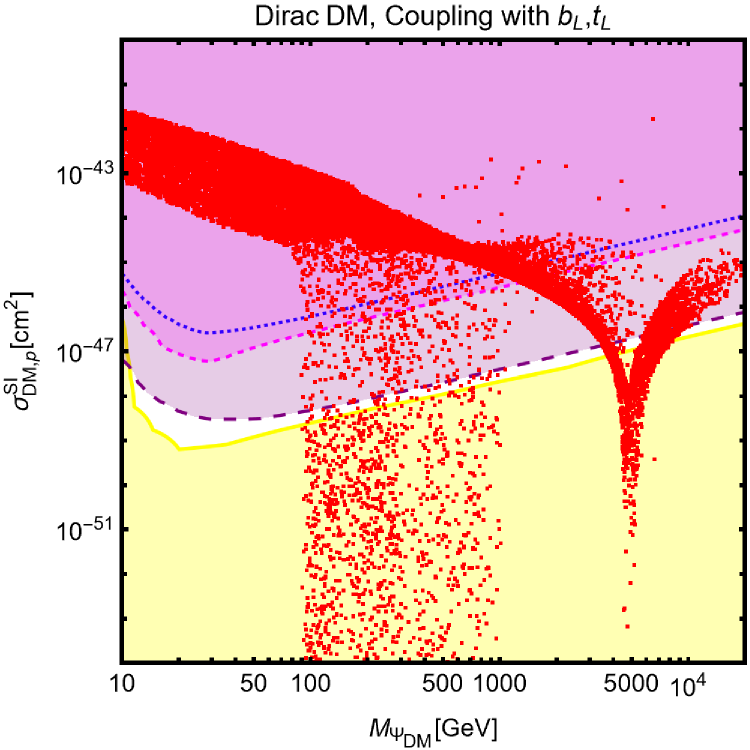}}
\subfloat{\includegraphics[width=0.48\linewidth]{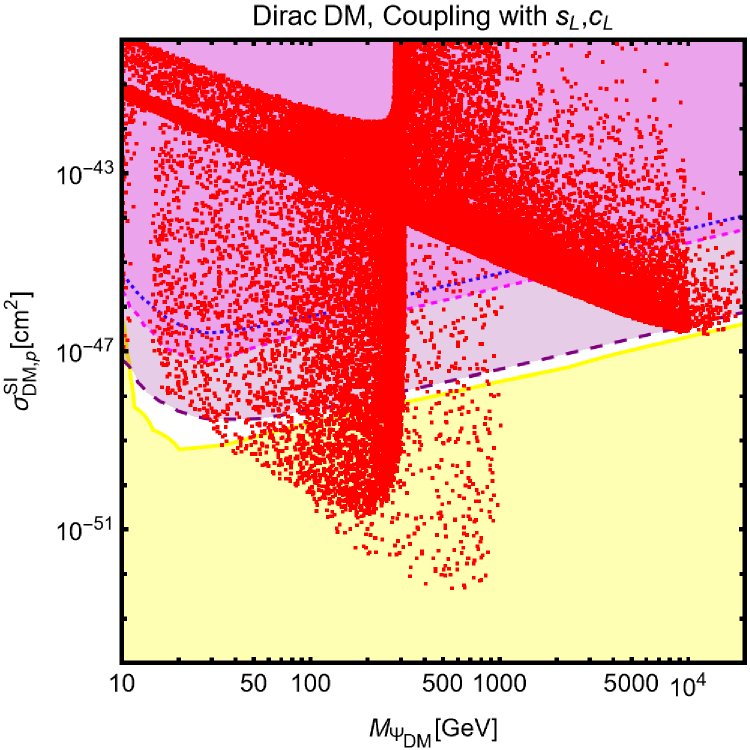}}\\
\subfloat{\includegraphics[width=0.48\linewidth]{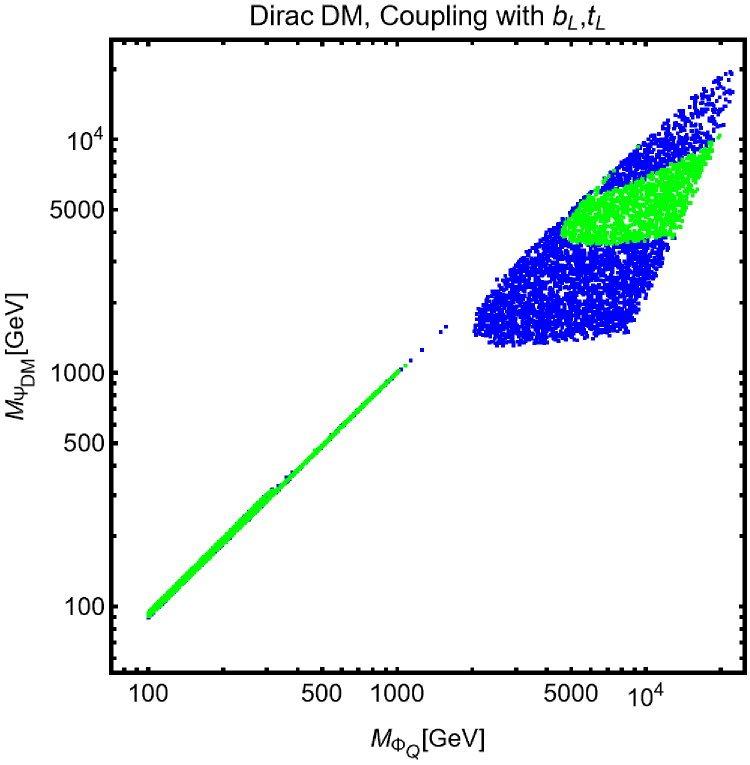}}
\subfloat{\includegraphics[width=0.48\linewidth]{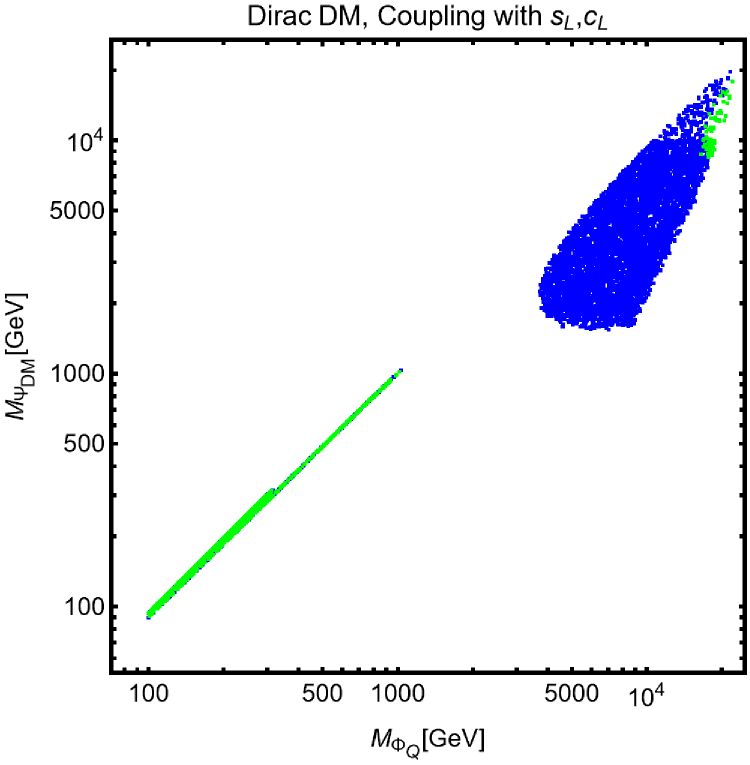}}
\caption{\it Model points emerged from the parameter scan, eq. (\ref{eq:min_scan}), for dirac fermion DM. The different panels refer to coupling with different quark species. For each case two plots are shown. First all the points complying with the correct relic density are reported in the $(M_{\Phi_{\rm DM}},\sigma_{\Phi_{\rm DM},p}^{\rm SI})$ bidimensional plane. Together with them the exclusion by LZ (magenta region), XENONnT (region above the blue dashed line), the projected sensitivity by DARWIN (purple region) and the region corresponding to the neutrino floor (yellow region) are evidenced. The second panel shows, in the $(M_{\Psi_f},M_{\Phi_{\rm DM}})$ plane, the model point complying with the constraint by LZ (blue points) and the ones which would survive even upon a negative signal at the DARWIN experiment (green points).}
    \label{fig:scan_dirac}
\end{figure}

\begin{figure}
\centering
\subfloat{\includegraphics[width=0.48\linewidth]{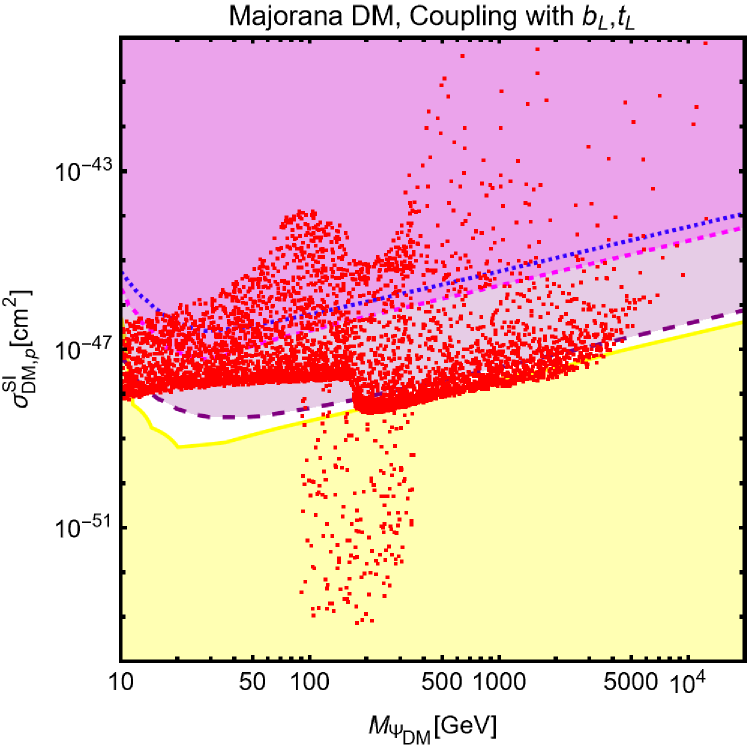}}
\subfloat{\includegraphics[width=0.48\linewidth]{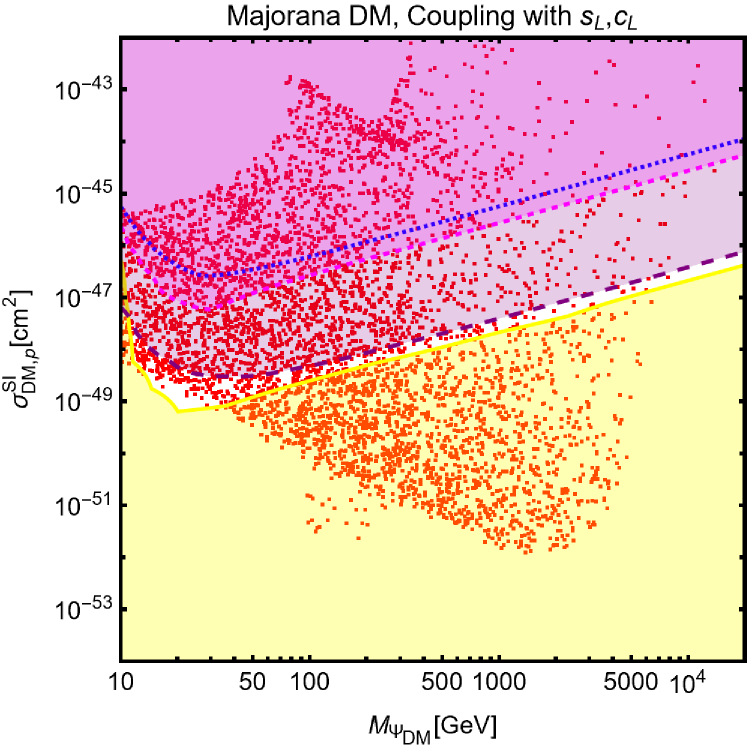}}\\
\subfloat{\includegraphics[width=0.48\linewidth]{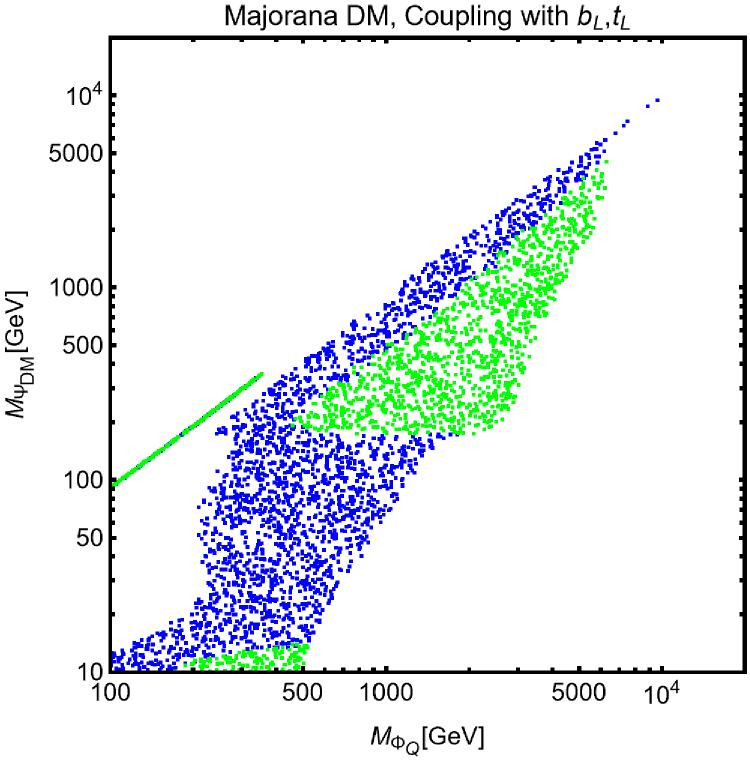}}
\subfloat{\includegraphics[width=0.48\linewidth]{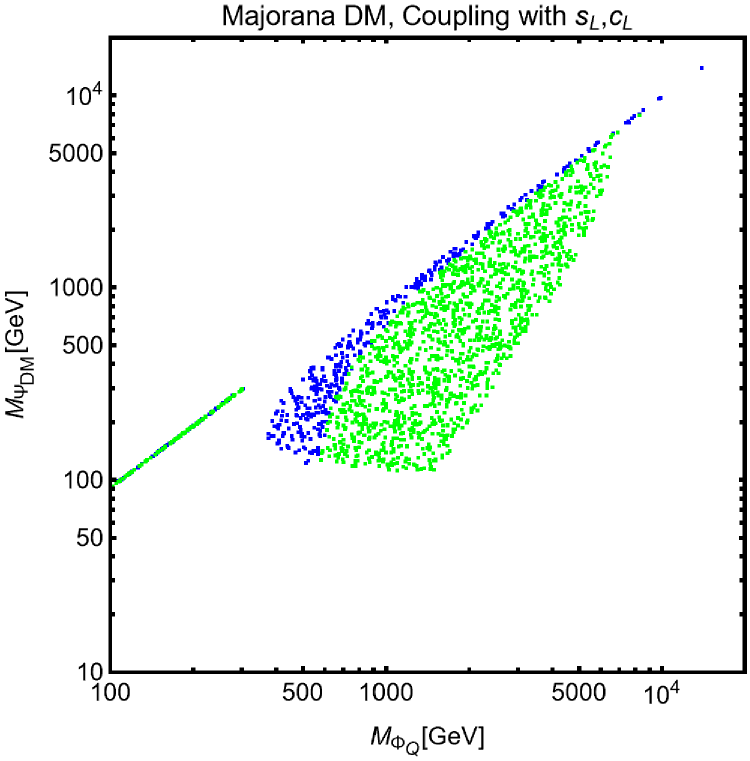}}
\caption{\it The same as~\Fig{fig:scan_dirac} but for majorana DM coupled with left-handed quarks.}
\label{fig:scan_majo_LL}
\end{figure}

The outcomes of the various scans are reported in Figures.~\ref{fig:scan_complex}-\ref{fig:scan_majo_LL}. For each model variant, two plots have been shown. First of all we have reported, in the $(M_{\Psi_{\rm DM},\Phi_{\rm DM}},\sigma_{\Phi_{\rm DM},\Psi_{\rm DM},p}^{\rm SI})$ bidimensional plane all the model points complying with the requirement of the correct DM relic density \footnote{As already pointed out, the scattering cross-section over nucleons does not provide a complete picture of the Direct Detection phenomenology of dirac-fermion DM. Nevertheless it is useful to provide a qualitative understanding and for comparison with the other kind of DM candidates considered in this work.}. Together with them we have highlighted in magenta the region excluded by LZ (experimental limit has been shown as dashed magenta line) and, in purple, the region which would be ruled out in case of negative signal from DARWIN. The plot also shows, in yellow, the region conventionally dubbed as neutrino floor, namely the values of the DM scattering cross-section below which the sensitivity of the Direct Detection experiment, at least within the currently considered design, is limited by the irreducible background represented by the coherent scattering of SM neutrinos over nucleons mediated by the $Z$-boson. The second kind of plot which is shown for each model features, in the $(M_{\Psi_f},M_{\Phi_{\rm DM}})$ or $(M_{\Phi_f},M_{\Psi_{\rm DM}})$ plane, the model points (in blue) complying with the present Direct Detection bound from LZ and the model points (in green) which would survive an exclusion bound by the DARWIN experiment. 

For each DM candidate, the scans have been repeated individually for the couplings with each quark flavour. For simplicity we have reported, in this section, only few representative cases, corresponding to the coupling with the second and third generation left-handed quarks. The full set of plot are nevertheless shown in appendix A.

In analogous way with what already done in the previous subsection let's consider in more detail the individual scenarios starting with the case of complex scalar DM shown in fig. \ref{fig:scan_complex}. Regardless the quark species with which the DM is coupled, one deals in general with very constrained scenarios. Indeed, most of the model points have scattering cross-section above the present bounds. Outside the coannihilation regime, some residual viable regions, in the case that no couplings with the top quark are present, exist for DM and mediator masses at the TeV scale, where the DM scattering cross-section is just suppressed by the heavy scale of the NP states. Such regions will be nevertheless ruled out in case of absence of signals at the DARWIN experiment. This latter ultimate detector will hence completely probe the parameter space of the model with the sole exception of very narrow and fine-tuned coannihilation strips. In the case that the DM is coupled with top quarks, direct detection and relic density constraints can be encompassed at the same time only in a narrow coannihilation strip. This is due to the enhancement, previously discussed, of the contribution from Higgs penguin diagram. Because of the latter, one is forced to rely on very tiny yukawa couplings for the DM. Consequently, the correct relic density can be achieved only via coannihilations.

Scenarios with real scalar DM appear to be very strongly constrained as well. Despite for equivalent values of the DM/mediator masses and of the couplings the DM scattering cross-section is typically smaller due to the absence of photon and $Z$-penguins, the DM annihilation cross-section is comparatively even more suppressed as consequence of the $v^4$ dependence. One would then need higher values of the coupling $\Gamma_{L,R}^f$ or strongly rely on coannihilations to achieve the correct DM relic density. As shown by the figure, the former option is disfavored by direct detection. The only exception to this picture is represented by the case of coupling of the DM with third generation quarks. As already pointed, when the DM mass is not far above the one of the final state quark, the velocity dependence can be lifted so that one can obtain viable model points outside the coannihilation regions. These points, however lie above the sensitivity of the DARWIN experiment and hence will be potentially ruled out in case of negative results from the latter experiment.

While behaving in very similar way, for what Direct Detection is concerned, to complex scalar DM, the scenario of dirac DM appears to be more favorable, once also DM relic density is accounted for. As discussed in the previous section, dirac DM features a s-wave dominated annihilation cross-section. Consequently, the correct relic density is achieved, for same values of the DM and mediator masses, for lower values of the couplings with respect to the case of complex scalar DM. This helps to overcome direct detection constraints, which remain still rather strong though.

Considering, finally, Majorana DM, we notice that it represents the scenario with the largest regions of allowed parameter space outside the coannihilation regime and even surviving a negative detection by the next future DARWIN experiment. We nevertheless notice a lower bound of around $M_{\Psi_{\rm DM}} \gtrsim 100\,\mbox{GeV}$ on the DM mass in the case one considers coupling with $u,d,s$ quarks. This is due to the exclusion bound from Spin Dependent interactions.

\subsection{DM coupled with leptons}

As already pointed out, complex and dirac fermionic DM can have a loop induced scattering cross-section even if they have a yukawa interaction only with leptons.

\begin{figure}
\centering
\subfloat{\includegraphics[width=0.48\linewidth]{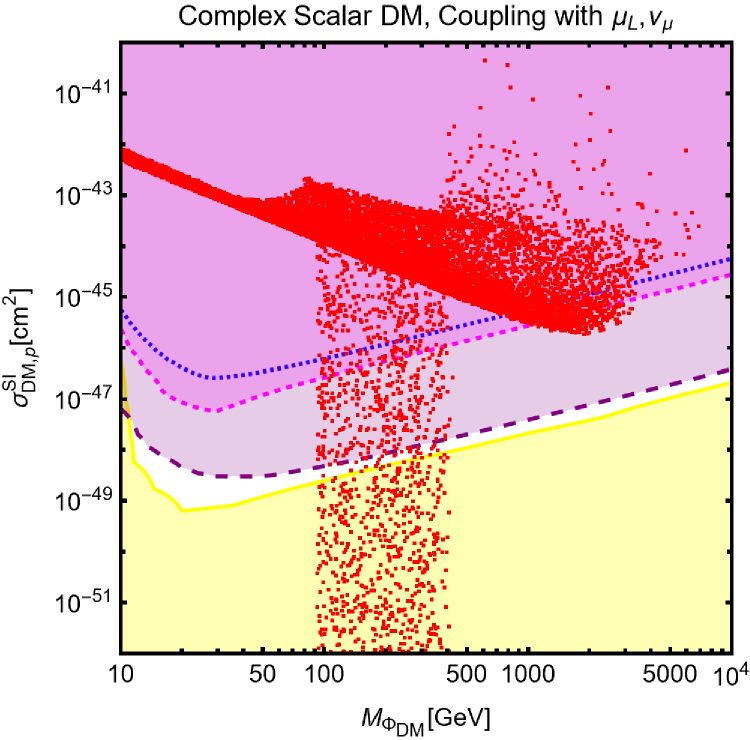}}
\subfloat{\includegraphics[width=0.48\linewidth]{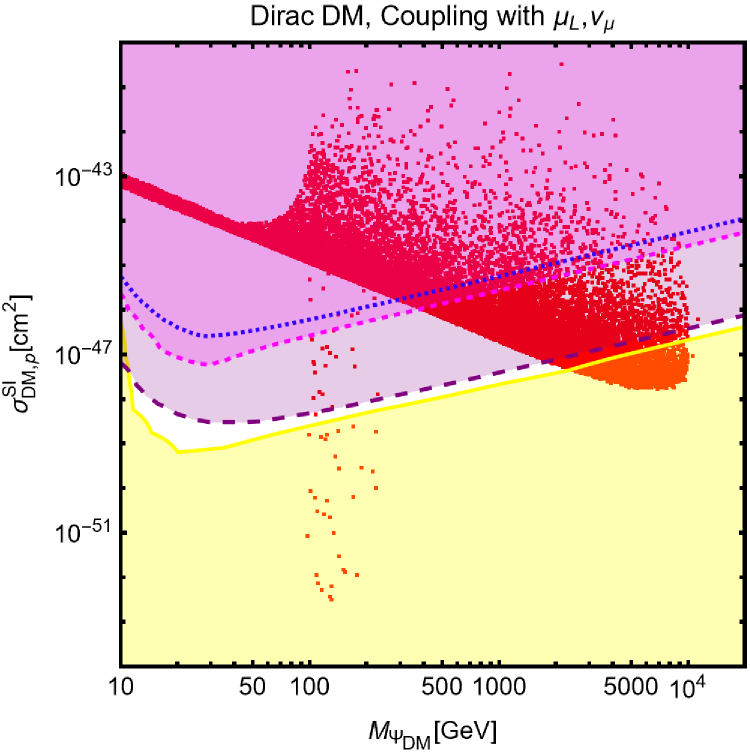}}\\
\subfloat{\includegraphics[width=0.48\linewidth]{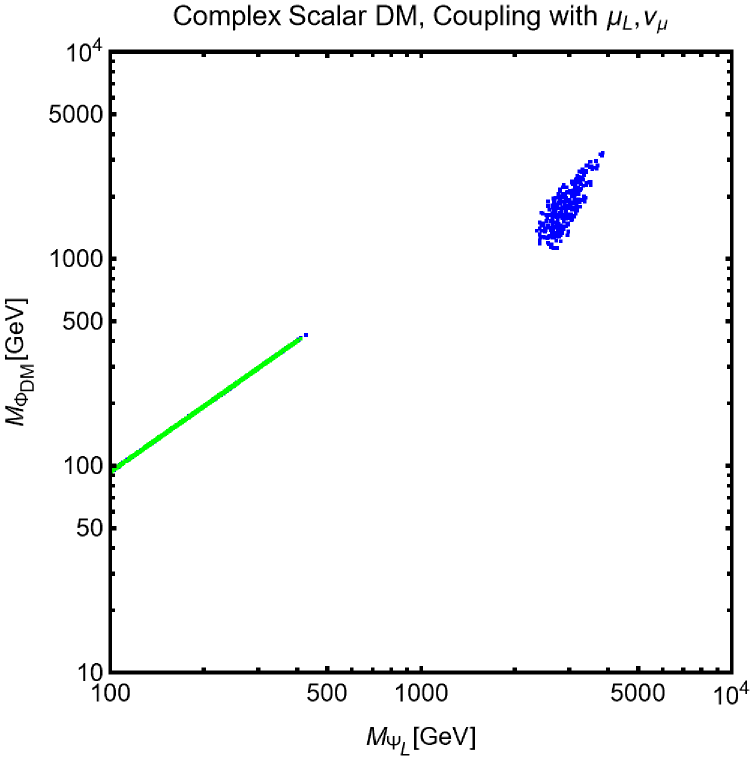}}
\subfloat{\includegraphics[width=0.48\linewidth]{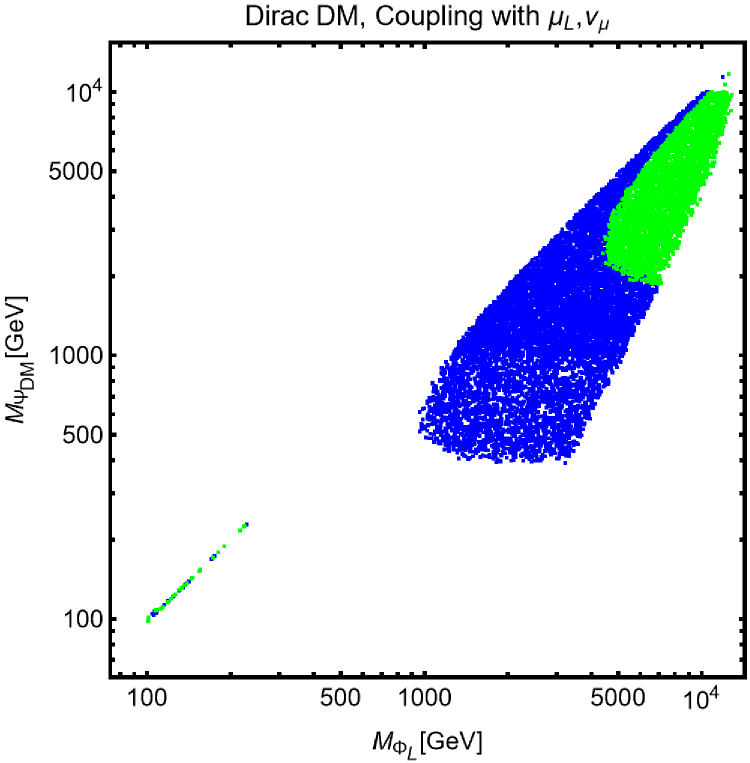}}
\caption{\it Combined constraints for complex scalar and dirac fermion DM coupled with the second generation of left-handed leptons.}
\label{fig:plot_lmu}
\end{figure}

We hence show in~\Fig{fig:plot_lmu}  the outcome a combined DD/relic density analysis in the case of effective coupling of the DM with the second generation of left-handed leptons (we do not consider the case of other generation has there are no noticeable differences). As evident, the results are very similar to the cases of couplings with the $s,c$ quark flavors. This can be understood with the fact that in all these cases the dominant contributions to the DM scattering interactions come from photon penguin diagrams.

\subsection{Scalar DM with portal coupling}

As already pointed out, one of the main results of our study is that in the case of scalar DM, the presence of a radiatively induced coupling of a DM pair with the Higgs boson is unavoidable. Until know, we have adopted $\lambda (M \equiv M_{\Psi_f})=0$ as initial condition for the RG evolution. We will relax here this hyphothesis and perform again a parameter scan by also varying the coupling $\lambda$ in the $[10^{-3},1]$ range. For simplicity we have considered only the scenarios of coupling with $s_L,c_L$ and $b_L,t_L$.

\begin{figure}
\centering    
\subfloat{\includegraphics[width=0.25\linewidth]{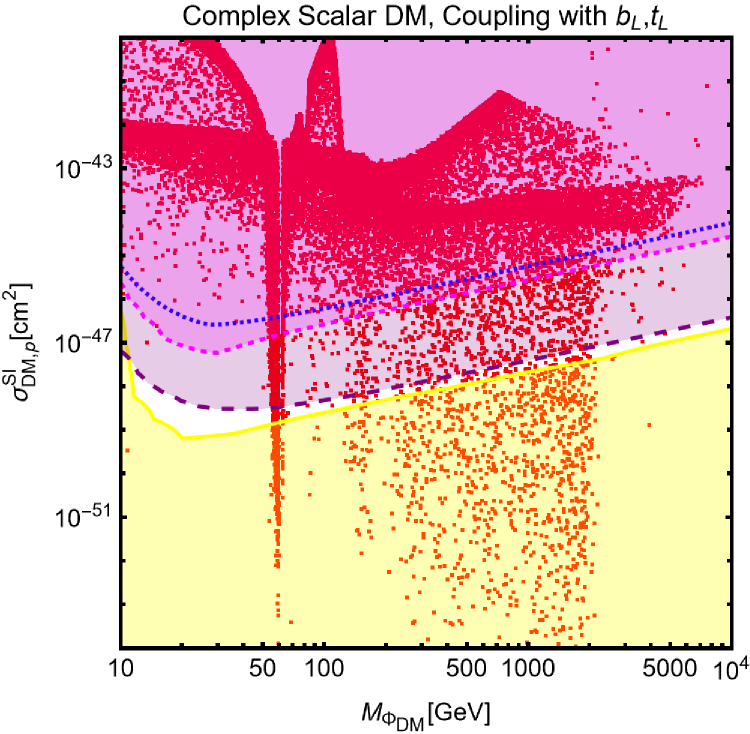}}
\subfloat{\includegraphics[width=0.25\linewidth]{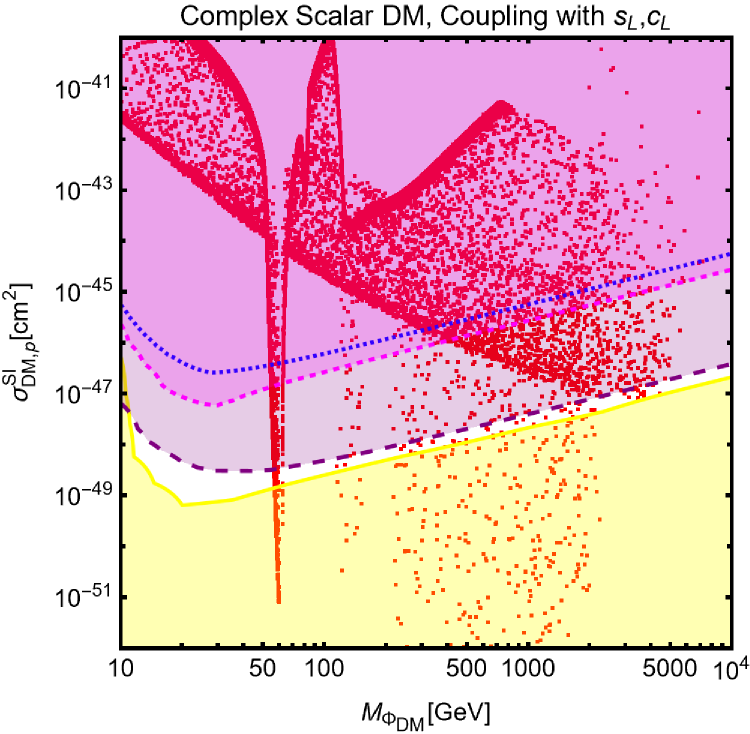}}
\subfloat{\includegraphics[width=0.25\linewidth]{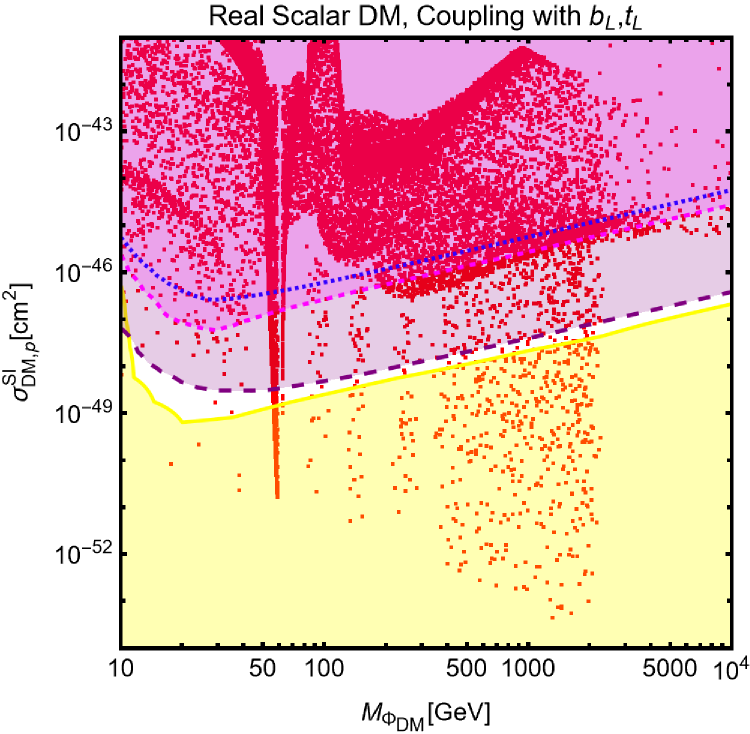}}
\subfloat{\includegraphics[width=0.25\linewidth]{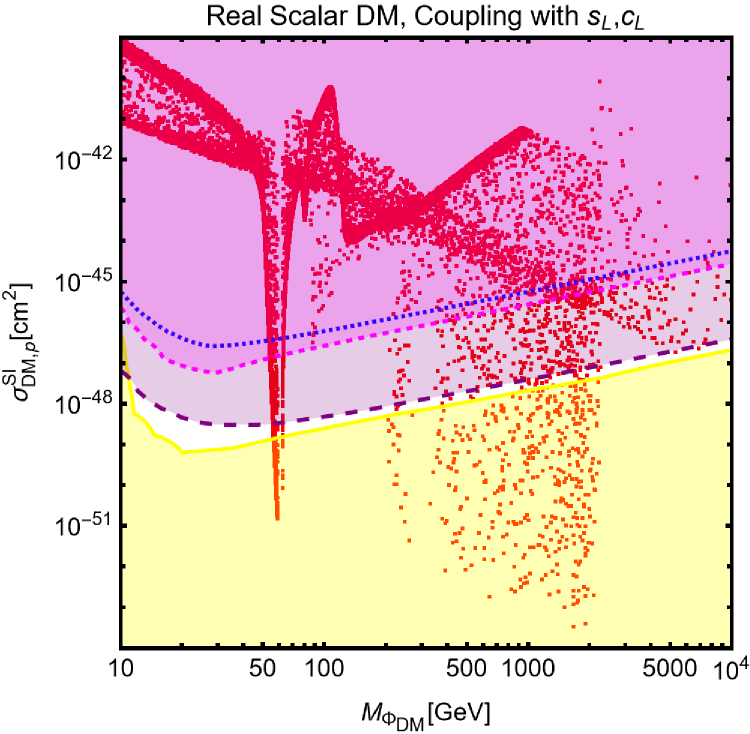}}\\
\subfloat{\includegraphics[width=0.25\linewidth]{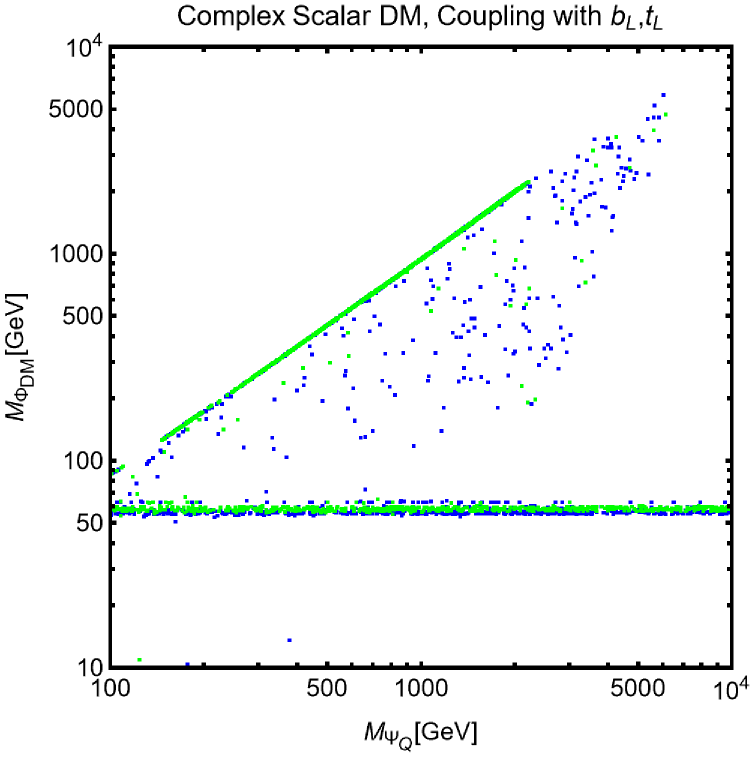}}
\subfloat{\includegraphics[width=0.25\linewidth]{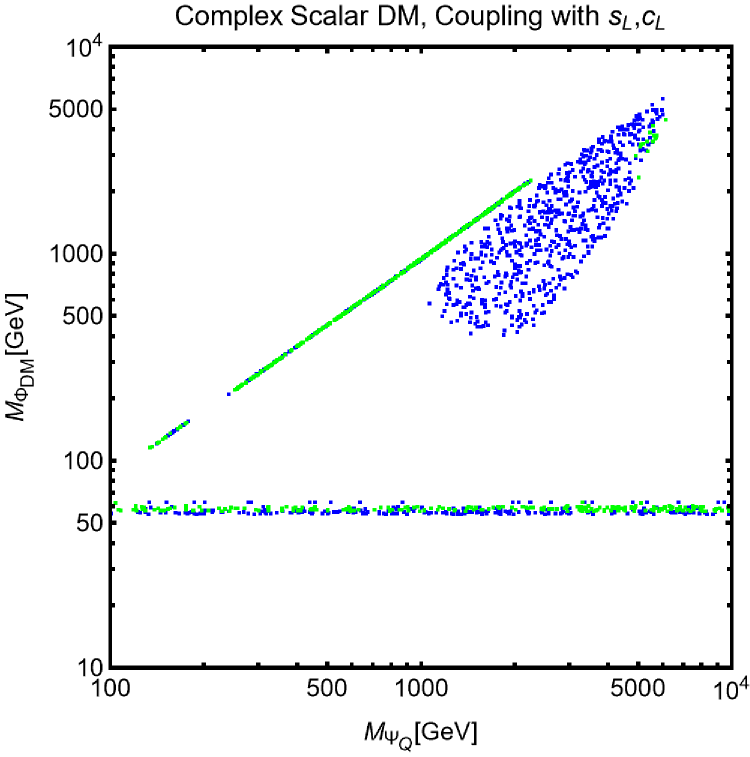}}
\subfloat{\includegraphics[width=0.25\linewidth]{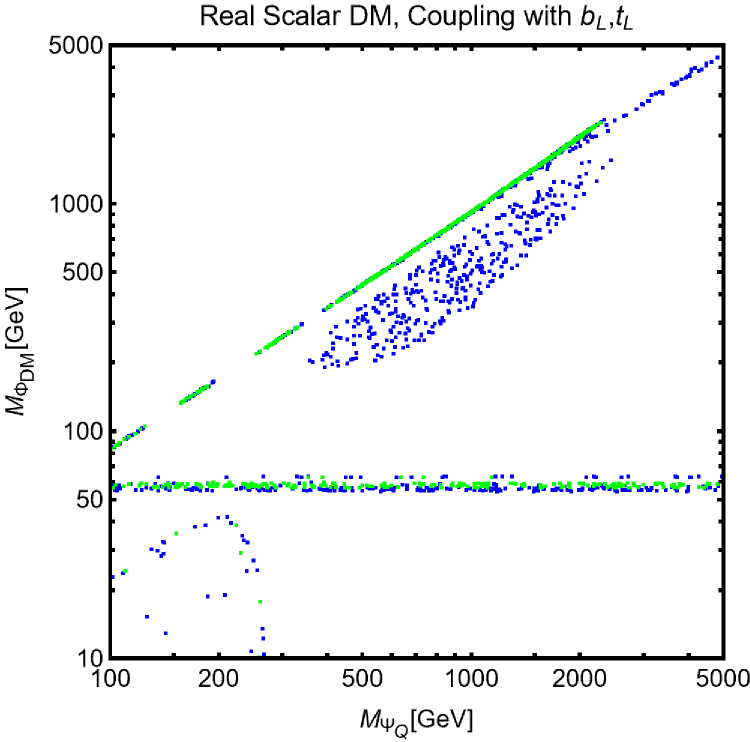}}
\subfloat{\includegraphics[width=0.25\linewidth]{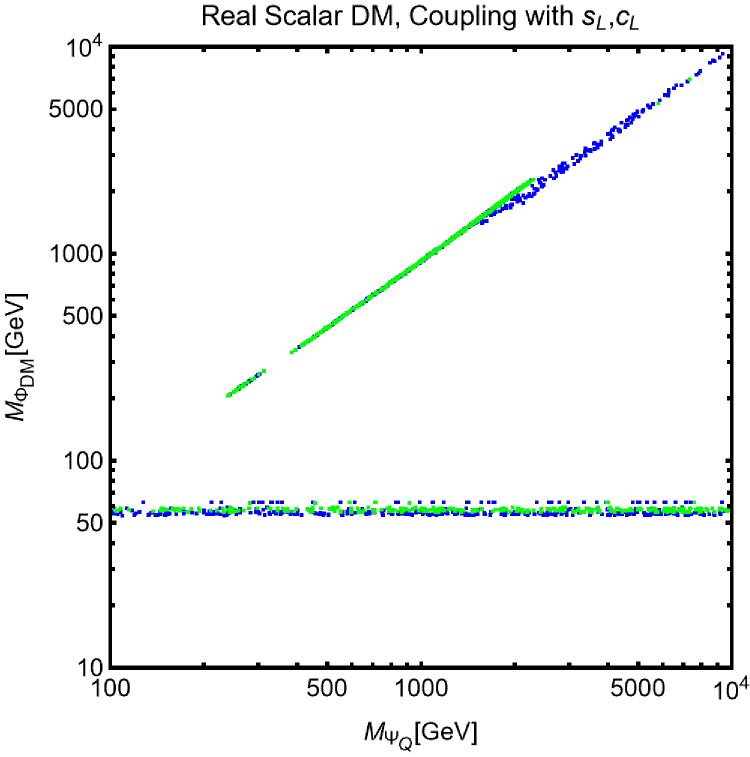}}
\caption{\it Combined constraints for scalar DM with a non-zero portal coupling.}
\label{fig:scan_lqlmu}
\end{figure}

The results are displayed, in the customary form, in~\Fig{fig:scan_lqlmu}. The presence of a coupling with the Higgs boson does not alter substantially the Direct Detection prospects of the model as it is responsible of an enhancement of the SI interactions of the DM. A new, fine-tuned region appears, for $M_{\Phi_{\rm DM}} \simeq m_h/2$. This region corresponds to the s-channel enhancement of the annihilation cross-section of the DM via Higgs exchange.

\section{Conclusions}
\label{sec:conclusions}

We have provided a reappraisal of the DM phenomenology, with strong focus on Direct Detection, of a class of DM models dubbed $t$-channel portals. We have considered the case of scalar (both complex and real) DM with a Yukawa-like coupling with SM fermions and a BSM fermion state as well as the complementary scenario of a fermionic (both Dirac and Majorana) DM candidate coupled with SM fermions and a BSM scalar. With the exception of the complex scalar and dirac fermionic DM coupled with light quark flavors, to properly assess the Direct Detection prospects of this class of models, it is necessary to evaluated the coefficients of the relevant effective operators which arise from loop induced interactions of the $t$-channel mediators, the DM and the SM states. To our best knowledge, we have provided the first complete computations for both scalars and fermionic DM candidates. The outcome of our computation have been compared by the most up-to-date experimental limits, as given by the LZ collaboration, as well as with the expected sensitivity of next generation detectors, represented by the DARWIN experiment. The obtained results have been further complemented by the requirement of the correct DM relic density assuming the standard freeze-out paradigm. Considering the simplest realizations of the $t$-channel portals, namely DM coupled with a single fermion flavor, we see that the different assignations of the DM spin and representation under the Lorentz group lead to rather different outcomes. The case of complex scalar DM appears to be very constrained; this is due to the combination of sizable one-loop induced SI cross-section and a p-wave annihilation cross-section. With the exception of the very fine-tuned coannihilation region, which would evade also the future bound form the DARWIN experiment, current experiments set an approximate lower bound of 500 GeV on the DM mass in the case the latter is coupled to s,c and/or b quarks. Negative signals from the DARWIN experiment would, however, rule-out the parameter space of the model ad exception of the $M_{\Phi_{\rm DM}}\simeq M_{\Psi_f}$ case. Much more constrained is the case of coupling between the DM and the top. This is due to the fact that theoretical consistency requires the radiative generation of a portal coupling between the DM and the Higgs boson. In the case of real scalar DM, the case of couplings with $u,d,s,c$ quarks is substantially ruled out, again besides coannihilations. This is mostly due to the extreme suppression of the DM annihilation cross-section. Wider viable parameter regions appear in the case of couplings with the $b$ and $t$ quarks. The latter will be ruled out in absence of signals by DARWIN. Moving to dirac DM, direct detection sets an analogous lower bound on the DM mass as for complex scalar DM. Thanks to the s-wave dominated annihilation cross-section, we have a wider viable parameter space, above such lower bound. The case of majorana DM results the most favored among the ones considered in this work and the only allowing for viable masses of order or below 100 GeV. This occurs, however, only in the case in which the DM is coupled with heavy quark flavors, namely $c,b,t$. For the other quark flavors, light DM is, instead, ruled-out by limits from SD interactions.

\section*{Acknowledgments} 
The authors thank Robert Ziegler for the fruitful discussions.
FM is supported by the State Agency for Research of the Spanish Ministry of Science and Innovation through the Unit of Excellence María de Maeztu 2020-2023 award to the Institute of Cosmos Sciences (CEX2019-000918-M) and from PID2019-105614GB-C21, 2017-SGR-929 and 2021-SGR-249 grants.
J.V.~acknowledges funding from the Spanish MINECO through the ``Ram\'on y Cajal'' program RYC-2017-21870,
the ``Unit of Excellence María de Maeztu 2020-2023'' award to the Institute of Cosmos Sciences (CEX2019-000918-M)
and from the grants PID2019-105614GB-C21 and 2017-SGR-92, 2021-SGR-249 (Generalitat de Catalunya).

\newpage

\appendix

\section{Complete results of the parameter scan}

In this appendix we show the full set of plots describing the outcome of the parameters scan illustrated in~\Sec{sec:combined_constraints}.

\bigskip
\bigskip

\begin{figure}[h]
    \centering
    \subfloat{\includegraphics[width=0.33\linewidth]{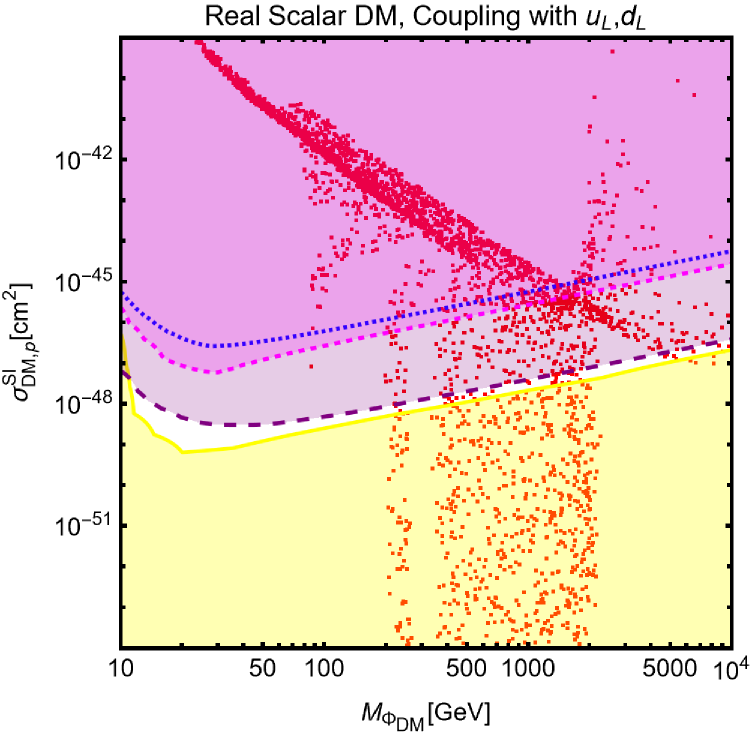}}
    \subfloat{\includegraphics[width=0.33\linewidth]{Figs_scalars/preal_SIs.pdf}}
    \subfloat{\includegraphics[width=0.33\linewidth]{Figs_scalars/preal_SIt.pdf}}\\
    \subfloat{\includegraphics[width=0.33\linewidth]{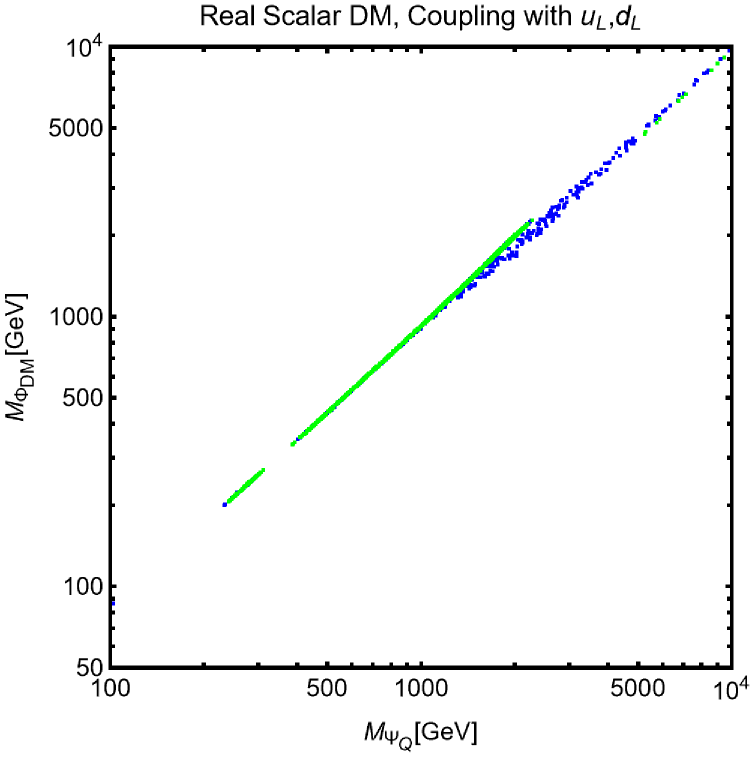}}
    \subfloat{\includegraphics[width=0.33\linewidth]{Figs_scalars/preal_mms.pdf}}
    \subfloat{\includegraphics[width=0.33\linewidth]{Figs_scalars/preal_mmt.pdf}}
    \caption{\it Full results of the parameter scan,~\Eq{eq:min_scan}, for real scalar DM coupled with left-handed quarks.}
    \label{fig:scan_real_L_complete}
\end{figure}

\begin{figure}
    \centering
     \subfloat{\includegraphics[width=0.3\linewidth]{Figs_scalars/pcomplex_Sib.pdf}}
     \subfloat{\includegraphics[width=0.3\linewidth]{Figs_scalars/pcomplex_SIs.pdf}}
    \subfloat{\includegraphics[width=0.3\linewidth]{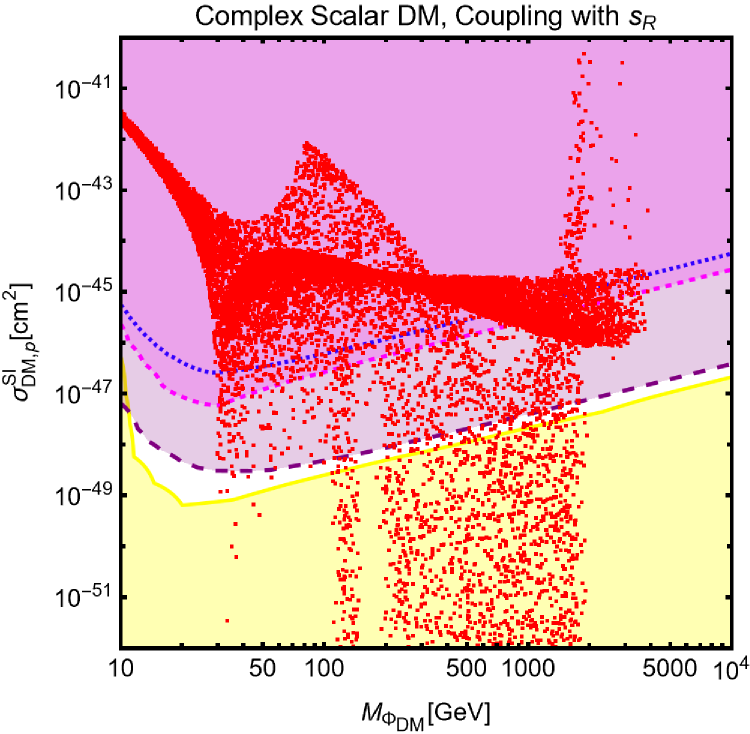}}\\
    \subfloat{\includegraphics[width=0.3\linewidth]{Figs_scalars/pcomplex_mmb.pdf}}
     \subfloat{\includegraphics[width=0.3\linewidth]{Figs_scalars/pcomplex_mms.pdf}}
    \subfloat{\includegraphics[width=0.3\linewidth]{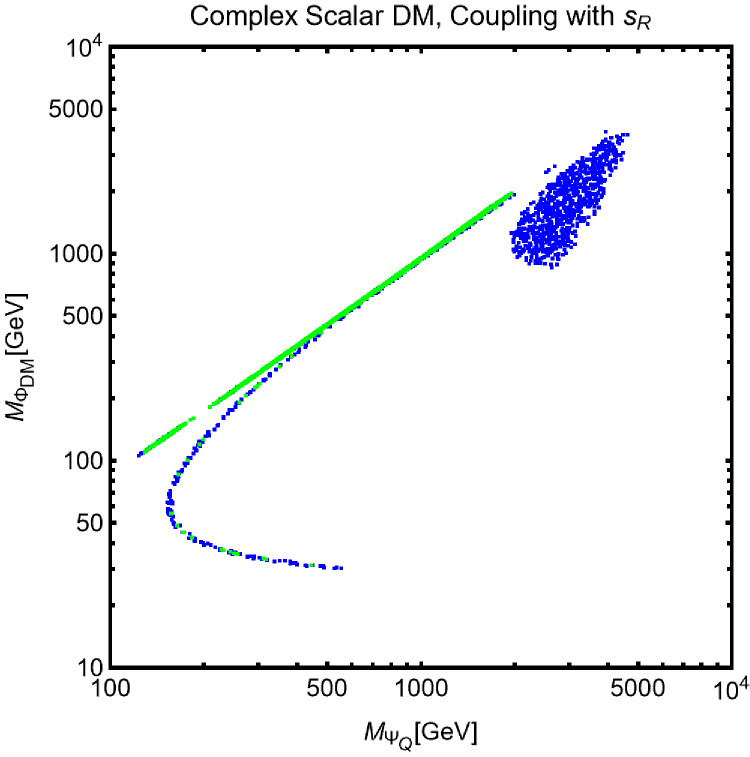}}\\
     \subfloat{\includegraphics[width=0.3\linewidth]{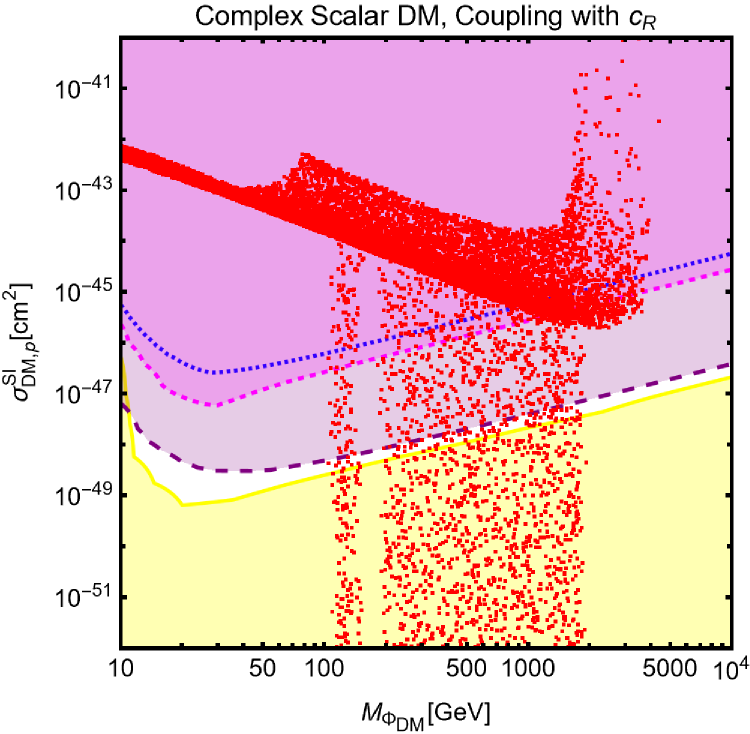}}
     \subfloat{\includegraphics[width=0.3\linewidth]{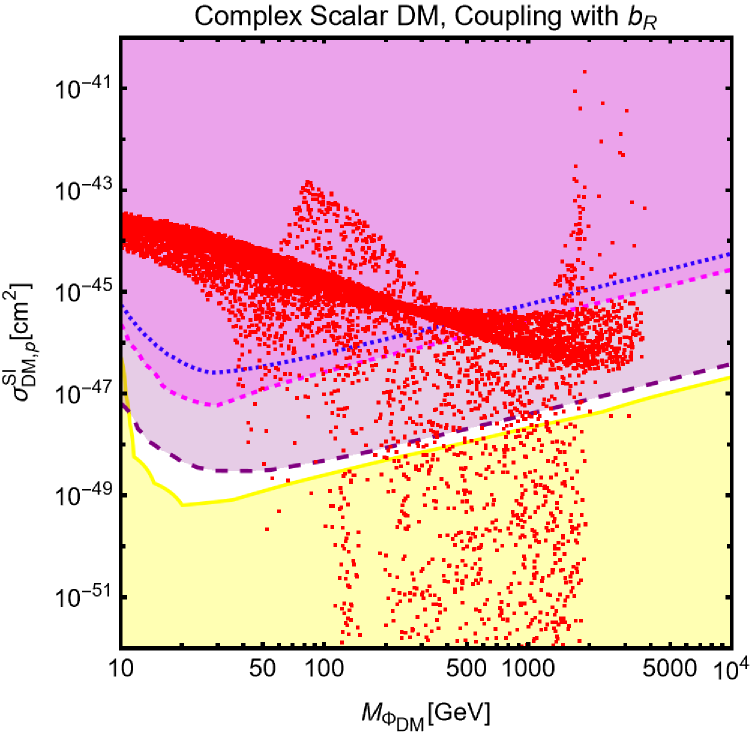}}
    \subfloat{\includegraphics[width=0.3\linewidth]{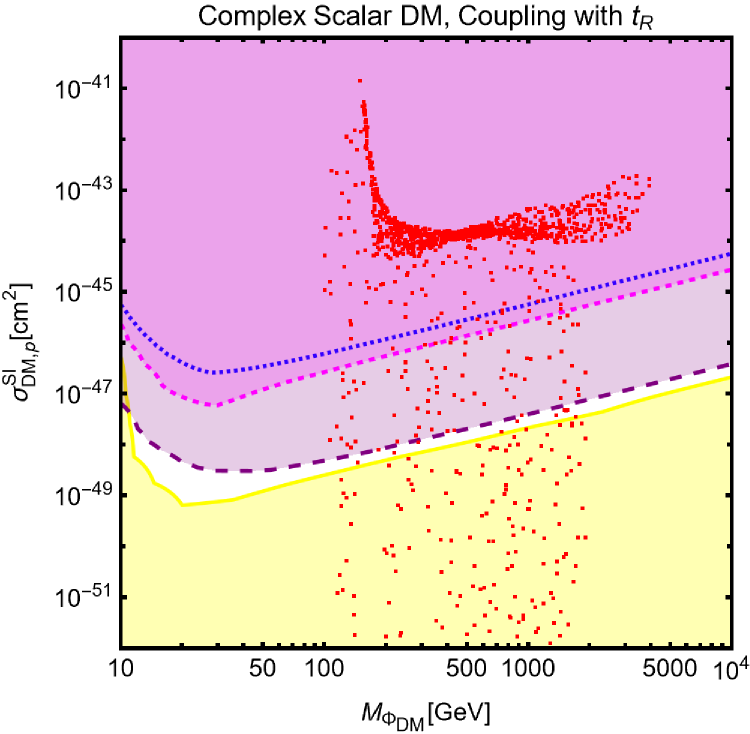}}\\
     \subfloat{\includegraphics[width=0.3\linewidth]{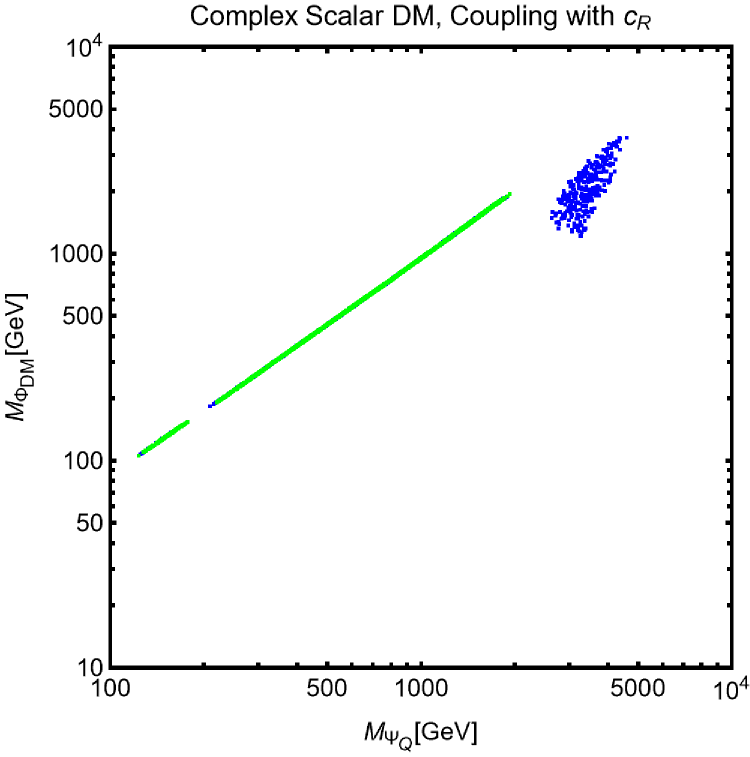}}
      \subfloat{\includegraphics[width=0.3\linewidth]{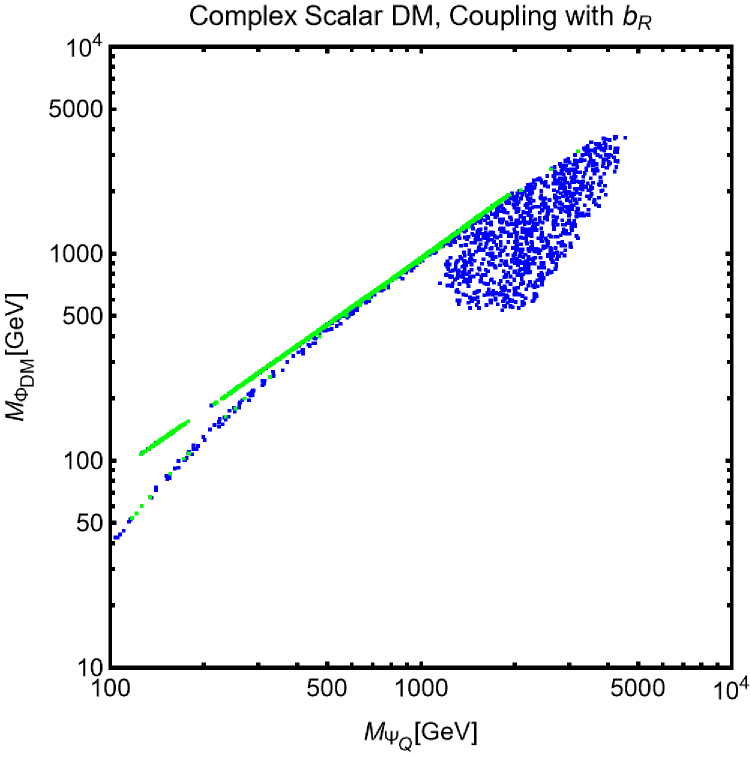}}
      \subfloat{\includegraphics[width=0.3\linewidth]{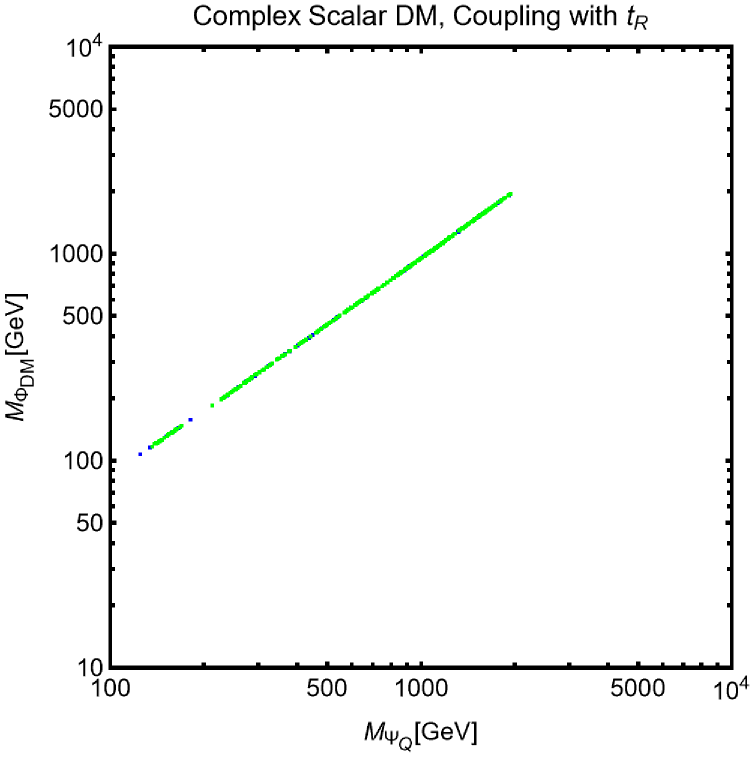}}
    \caption{\it Full results of the parameter scan,~\Eq{eq:min_scan}, for complex scalar DM.}
    \label{fig:scan_complex_complete}
\end{figure}

\begin{figure}
    \centering
    \subfloat{\includegraphics[width=0.3\linewidth]{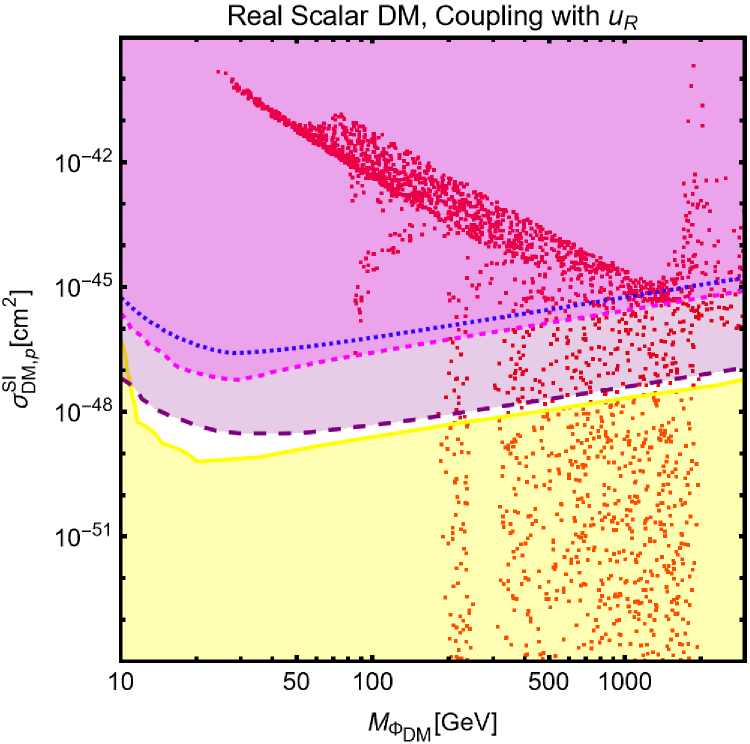}}
    \subfloat{\includegraphics[width=0.3\linewidth]{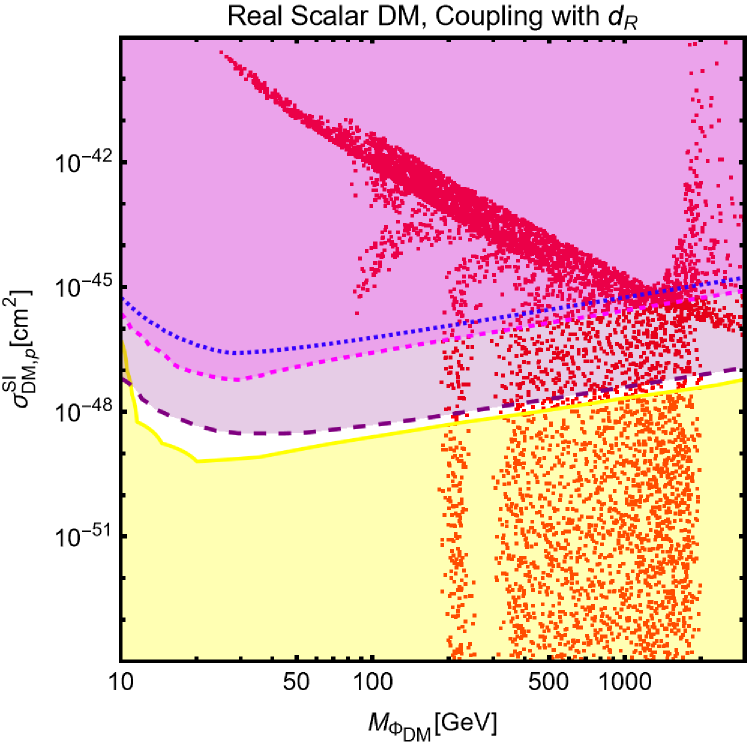}}
    \subfloat{\includegraphics[width=0.3\linewidth]{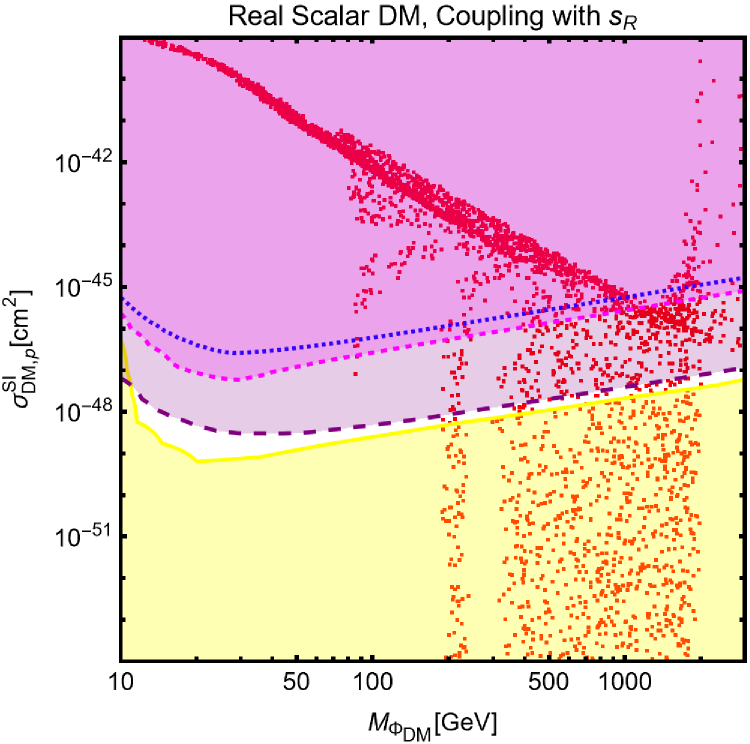}}\\
     \subfloat{\includegraphics[width=0.3\linewidth]{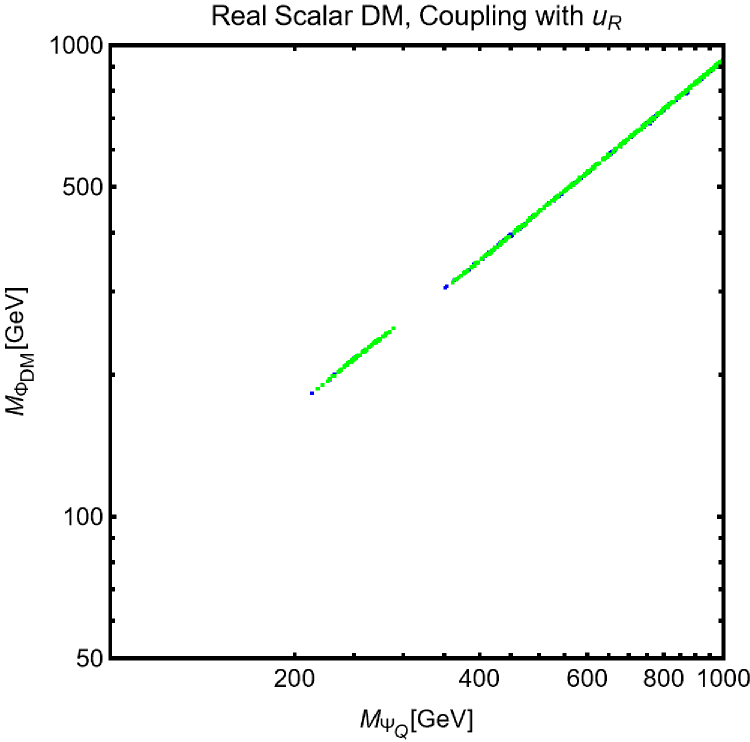}}
    \subfloat{\includegraphics[width=0.3\linewidth]{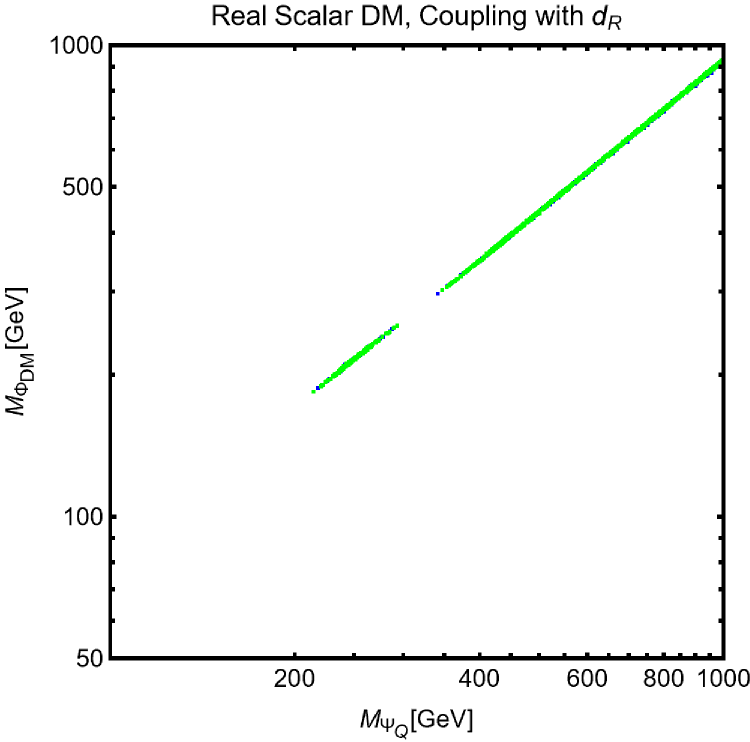}}
    \subfloat{\includegraphics[width=0.3\linewidth]{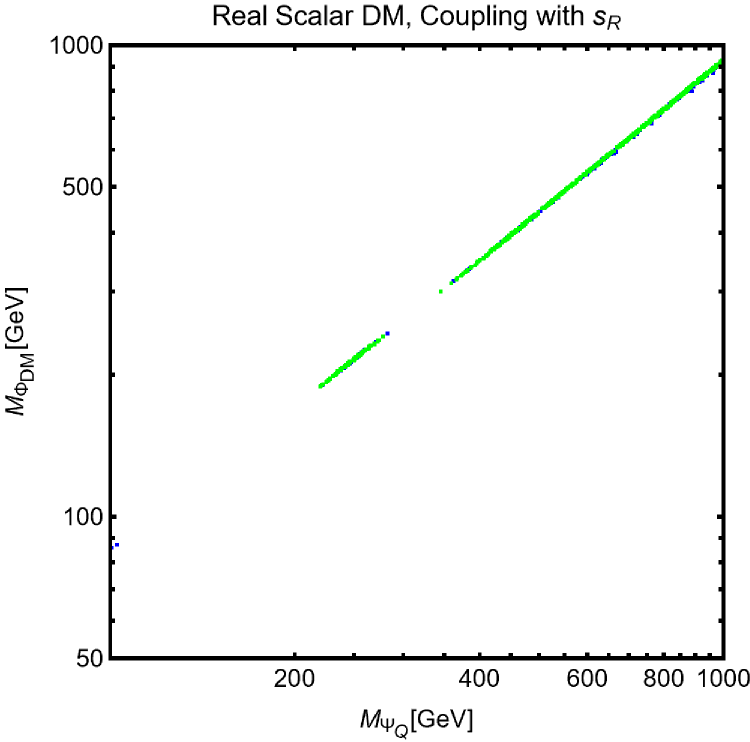}}\\
     \subfloat{\includegraphics[width=0.3\linewidth]{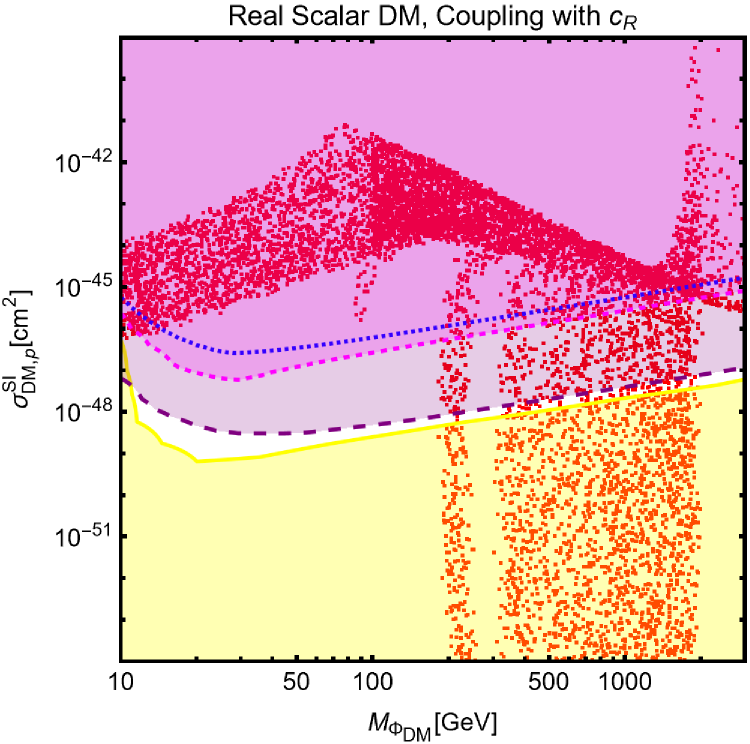}}
    \subfloat{\includegraphics[width=0.3\linewidth]{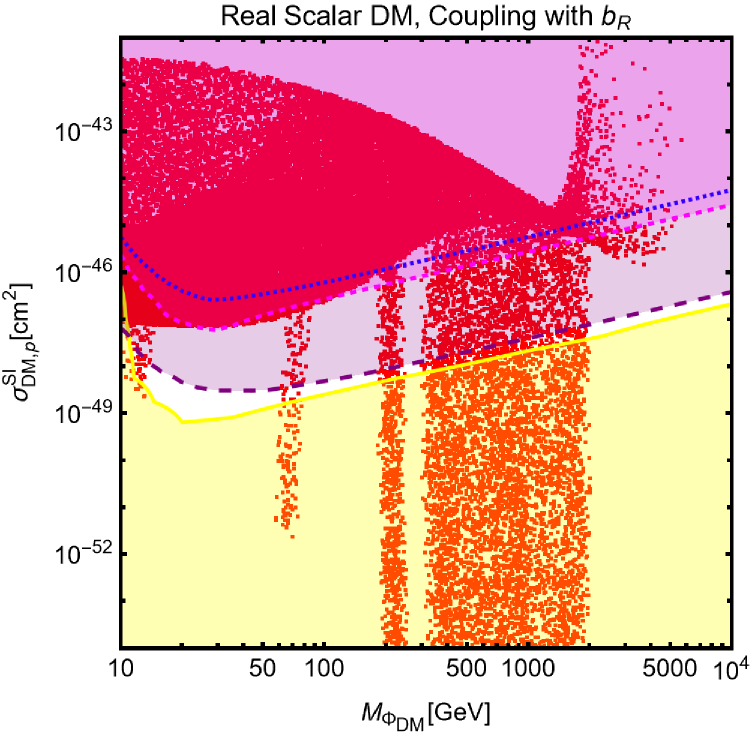}}
    \subfloat{\includegraphics[width=0.3\linewidth]{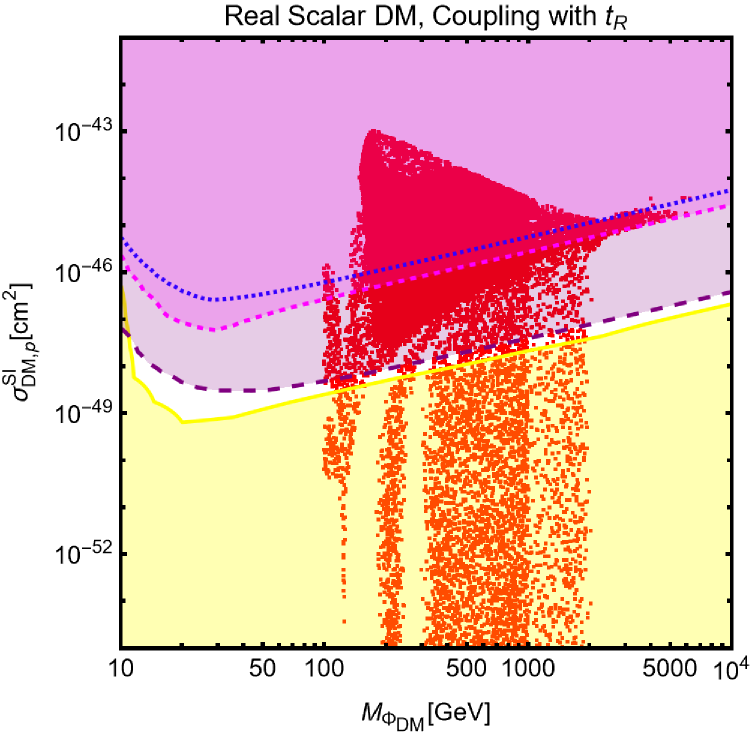}}\\
     \subfloat{\includegraphics[width=0.3\linewidth]{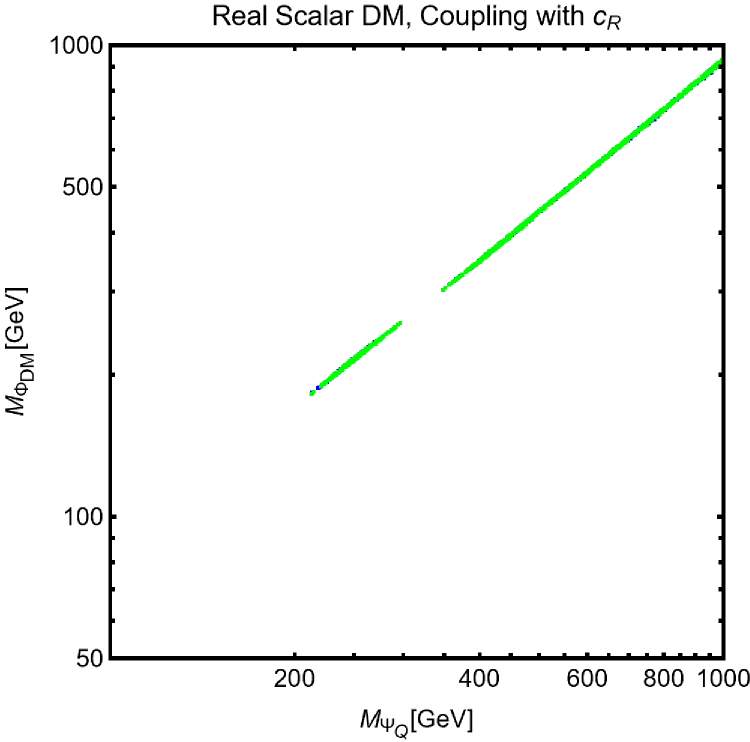}}
    \subfloat{\includegraphics[width=0.3\linewidth]{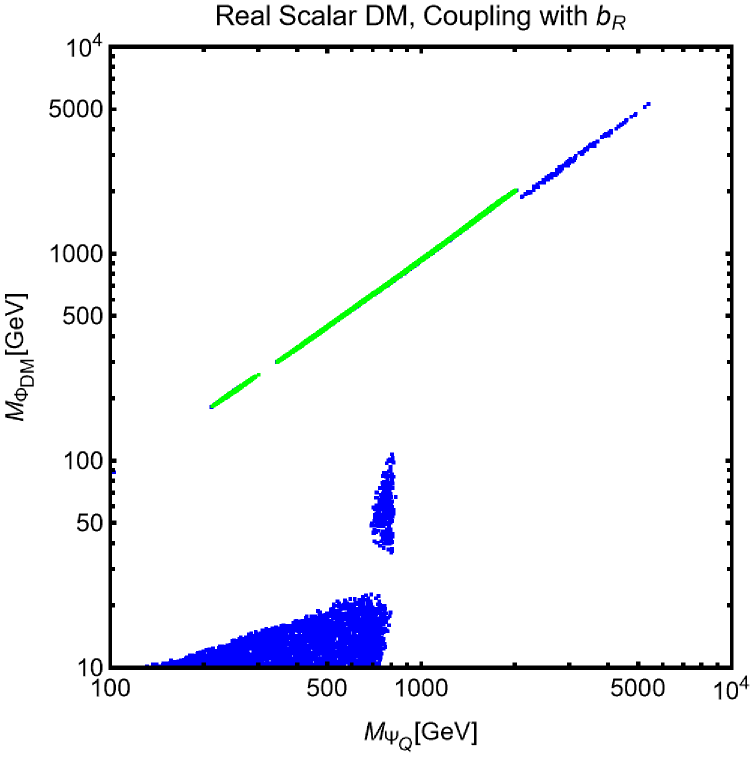}}
    \subfloat{\includegraphics[width=0.3\linewidth]{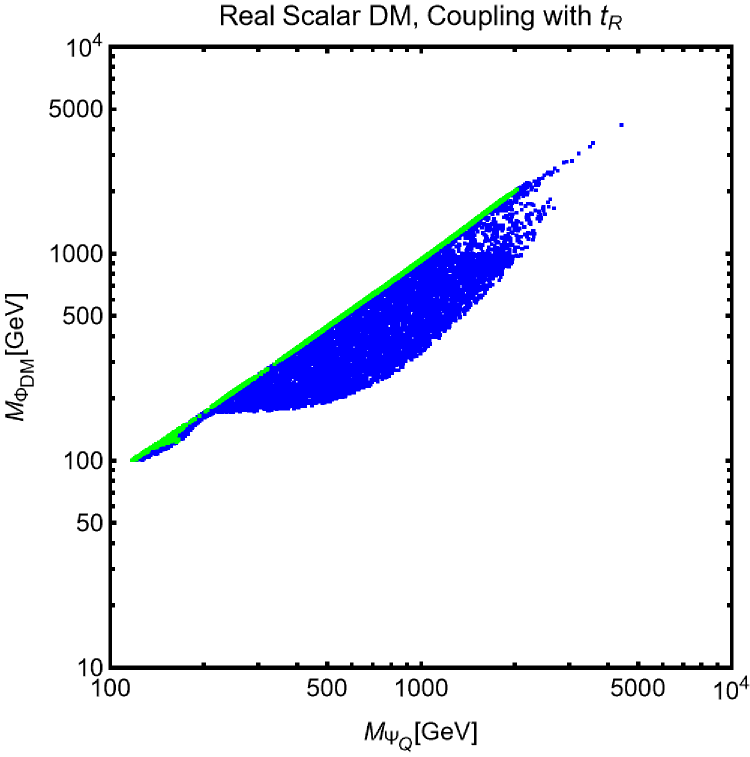}}
    \caption{\it The same as~\Fig{fig:scan_complex_complete} but for real scalar DM coupled with left-handed quarks.}
    \label{fig:scan_real_R_complete}
\end{figure}

\begin{figure}
    \centering
    \subfloat{\includegraphics[width=0.3\linewidth]{Figs_fermions/pdirac_SIt.pdf}}
    \subfloat{\includegraphics[width=0.3\linewidth]{Figs_fermions/pdirac_SIs.pdf}}
    \subfloat{\includegraphics[width=0.3\linewidth]{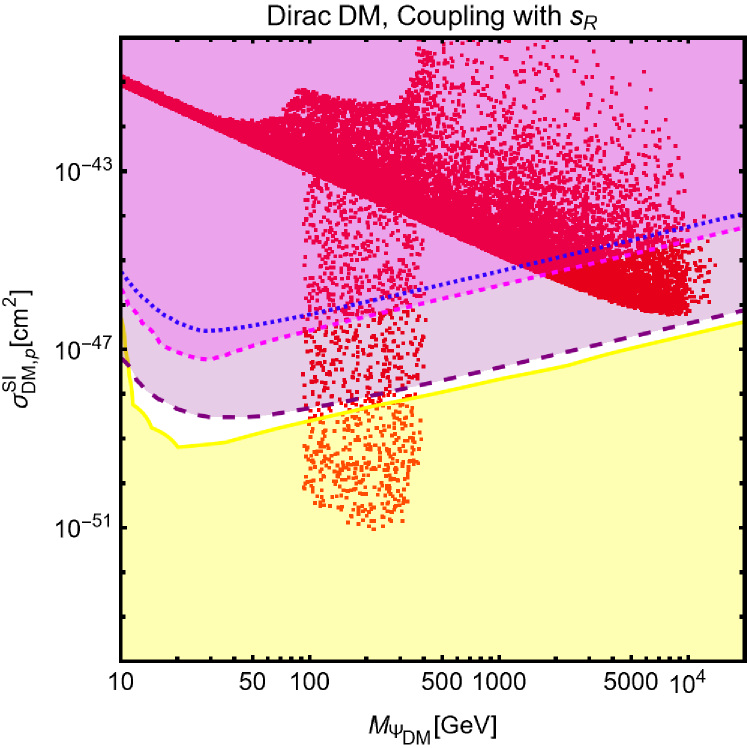}}\\
    \subfloat{\includegraphics[width=0.3\linewidth]{Figs_fermions/pdirac_mmt.pdf}}
     \subfloat{\includegraphics[width=0.3\linewidth]{Figs_fermions/pdirac_mms.pdf}}
    \subfloat{\includegraphics[width=0.3\linewidth]{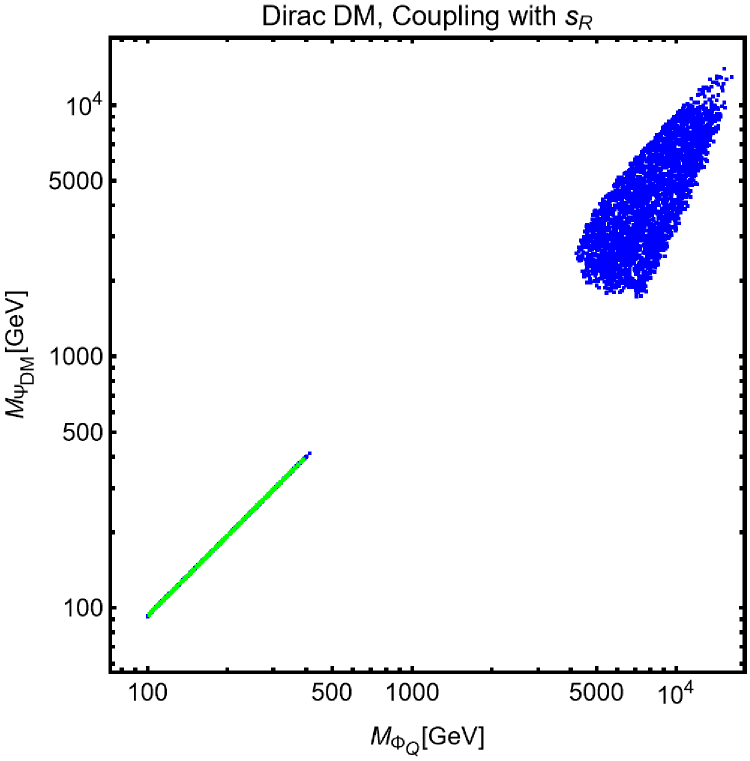}}\\
    \subfloat{\includegraphics[width=0.3\linewidth]{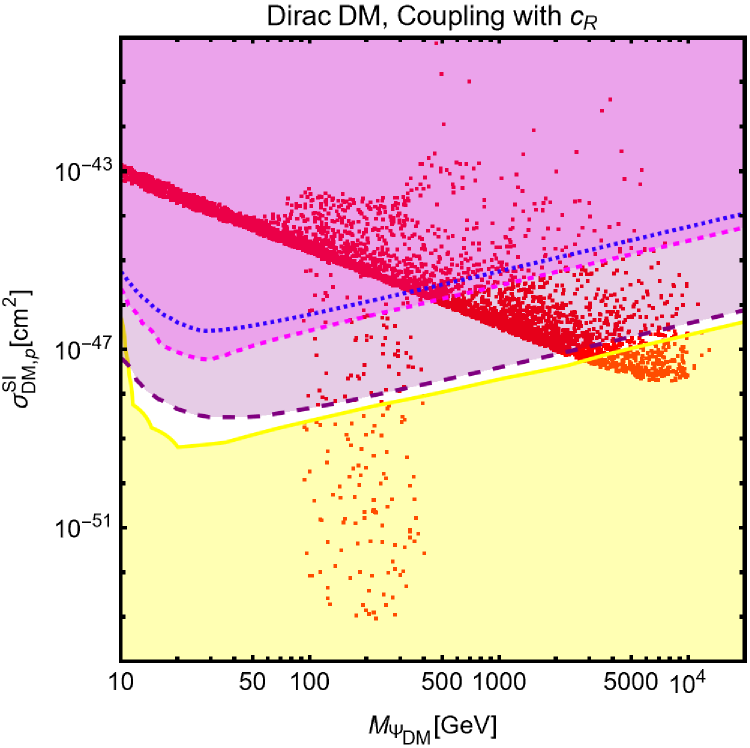}}
    \subfloat{\includegraphics[width=0.3\linewidth]{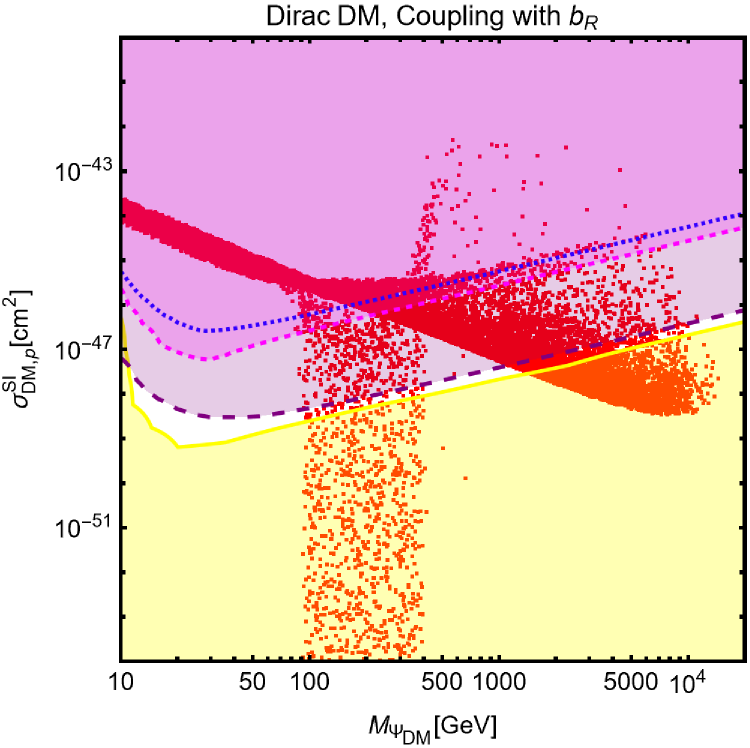}}
     \subfloat{\includegraphics[width=0.3\linewidth]{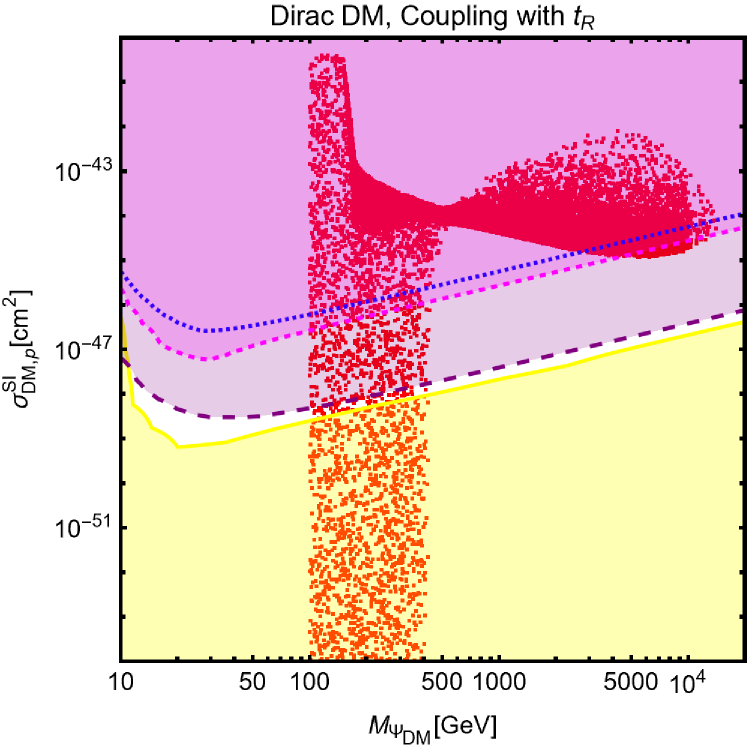}}\\
      \subfloat{\includegraphics[width=0.3\linewidth]{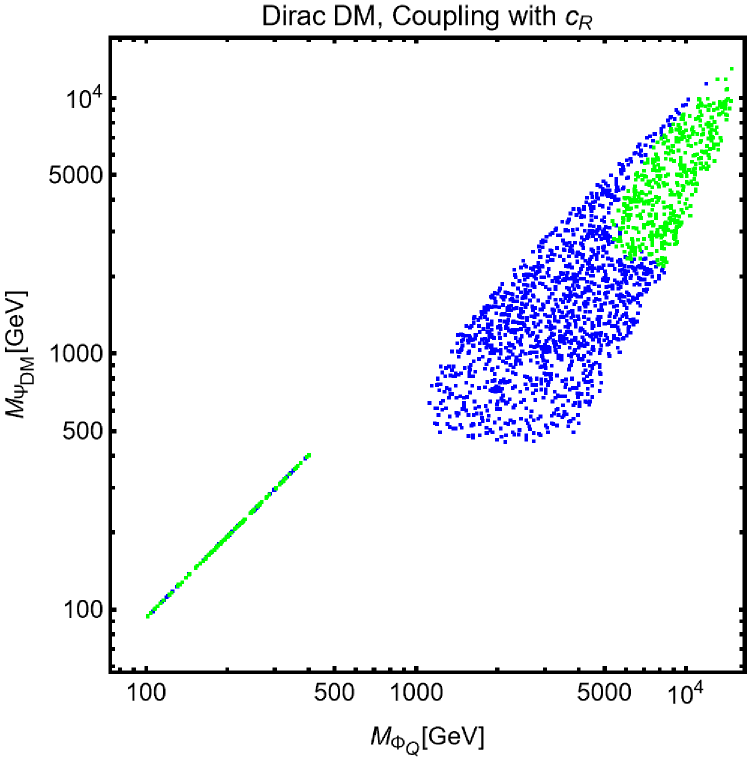}}
      \subfloat{\includegraphics[width=0.3\linewidth]{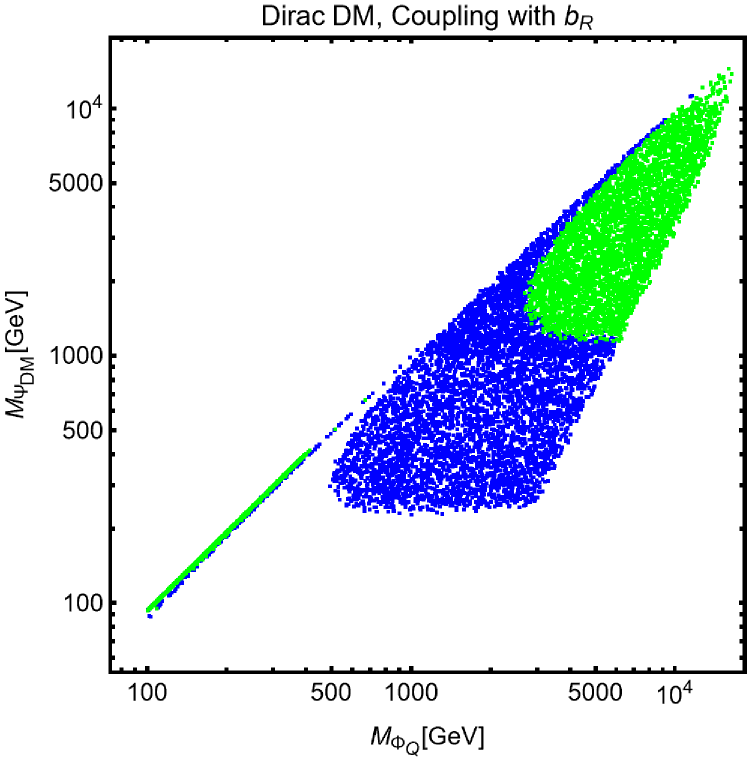}}
      \subfloat{\includegraphics[width=0.3\linewidth]{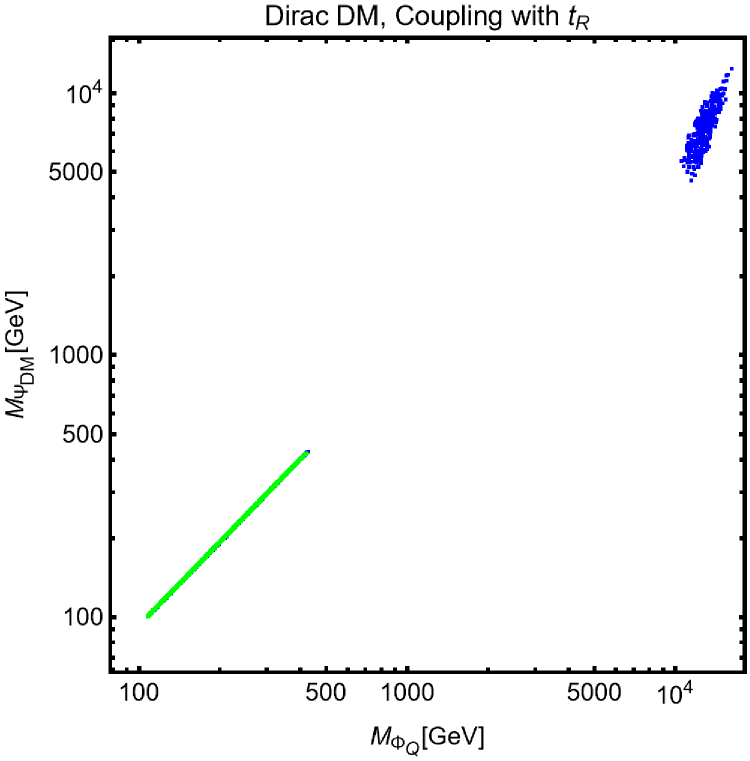}}
    \caption{\it Full results of the parameter scan,~\Eq{eq:min_scan}, for dirac fermion DM.}
    \label{fig:scan_dirac_complete}
\end{figure}

\begin{figure}
    \centering
    \subfloat{\includegraphics[width=0.3\linewidth]{Figs_fermions/pmajo_SIt.pdf}}
    \subfloat{\includegraphics[width=0.3\linewidth]{Figs_fermions/pmajo_SIs.pdf}}
    \subfloat{\includegraphics[width=0.3\linewidth]{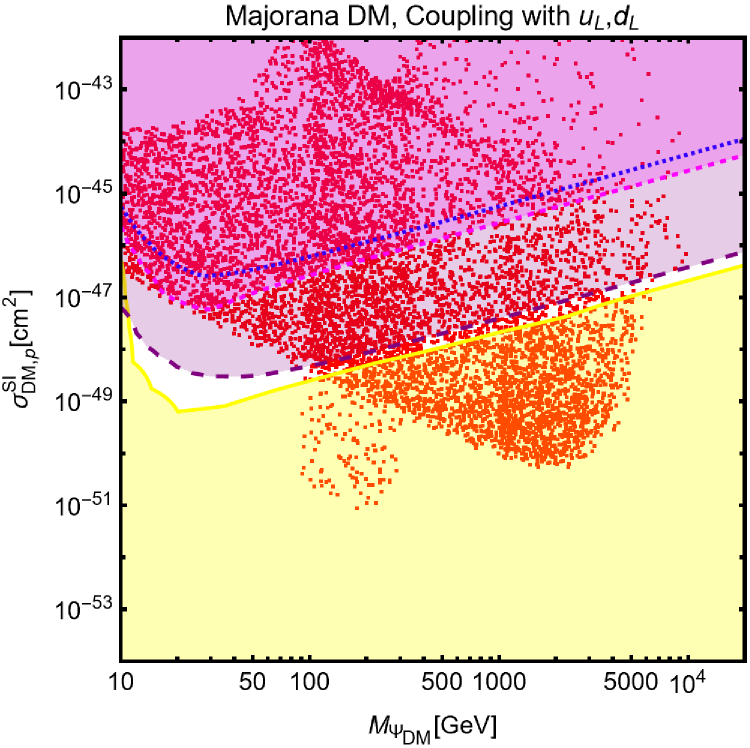}}\\
    \subfloat{\includegraphics[width=0.3\linewidth]{Figs_fermions/pmajo_mmt.pdf}}
     \subfloat{\includegraphics[width=0.3\linewidth]{Figs_fermions/pmajo_mms.pdf}}
      \subfloat{\includegraphics[width=0.3\linewidth]{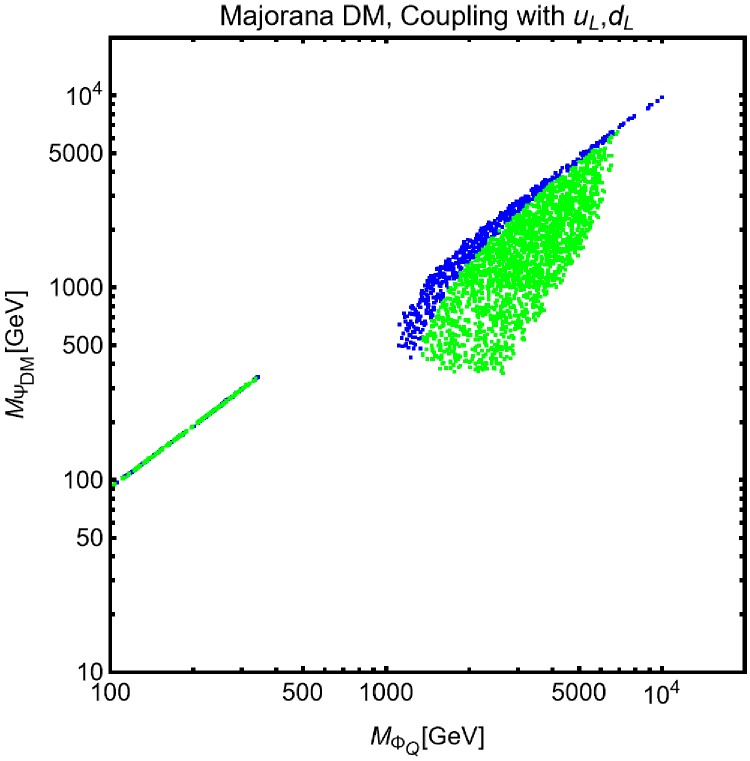}}
    \caption{\it The same as~\Fig{fig:scan_dirac_complete} but for majorana DM coupled with left-handed quarks.}
    \label{fig:scan_majo_LL_complete}
\end{figure}

\begin{figure}
    \centering
    \subfloat{\includegraphics[width=0.3\linewidth]{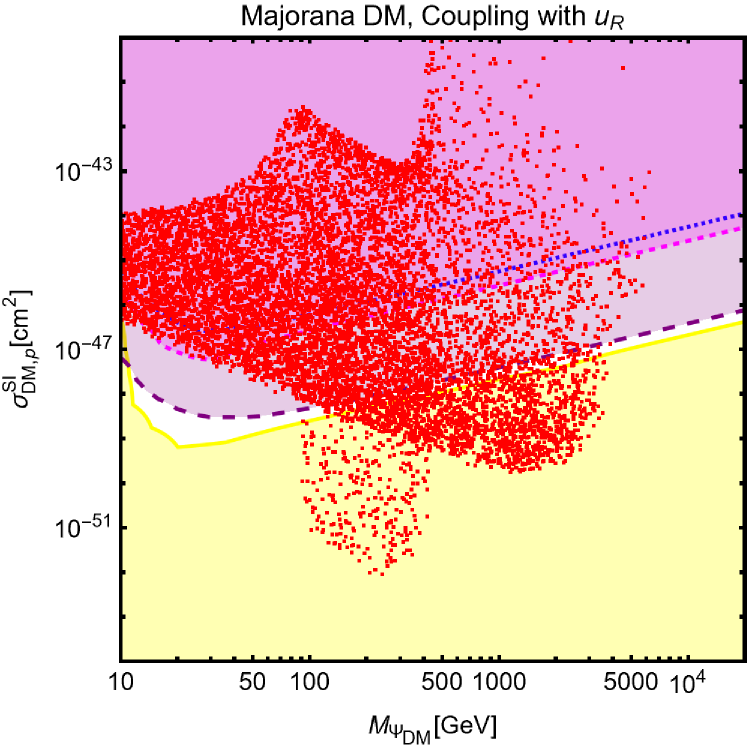}}
    \subfloat{\includegraphics[width=0.3\linewidth]{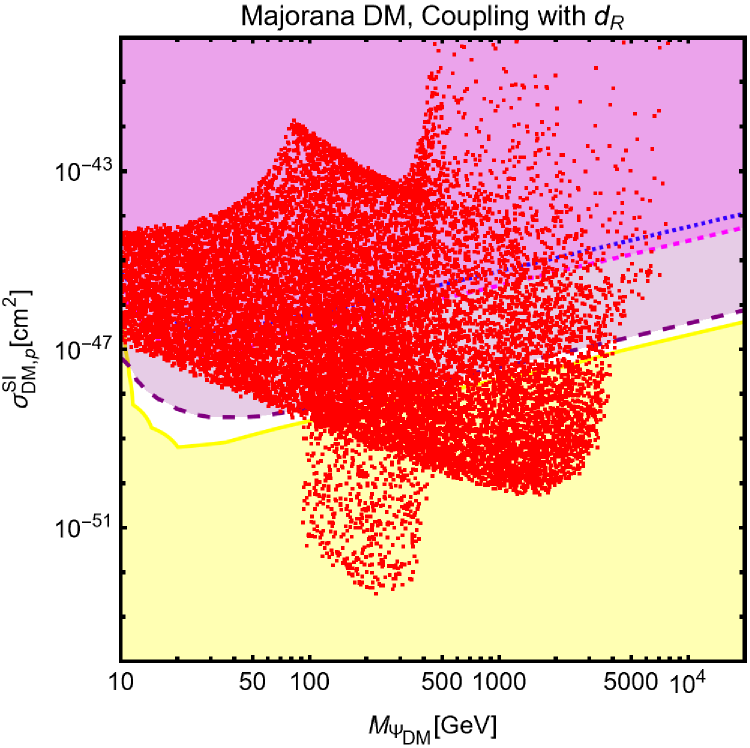}}
    \subfloat{\includegraphics[width=0.3\linewidth]{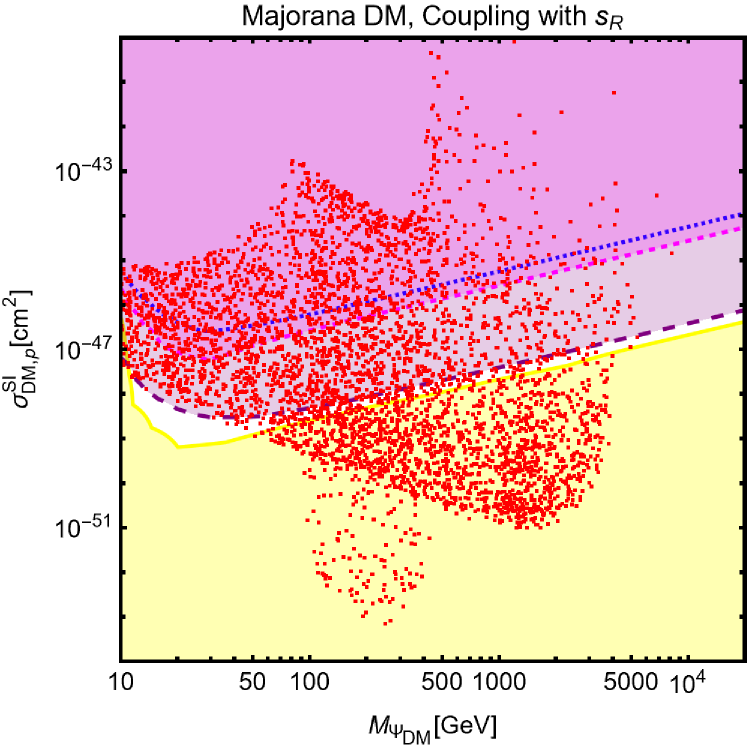}}\\
    \subfloat{\includegraphics[width=0.3\linewidth]{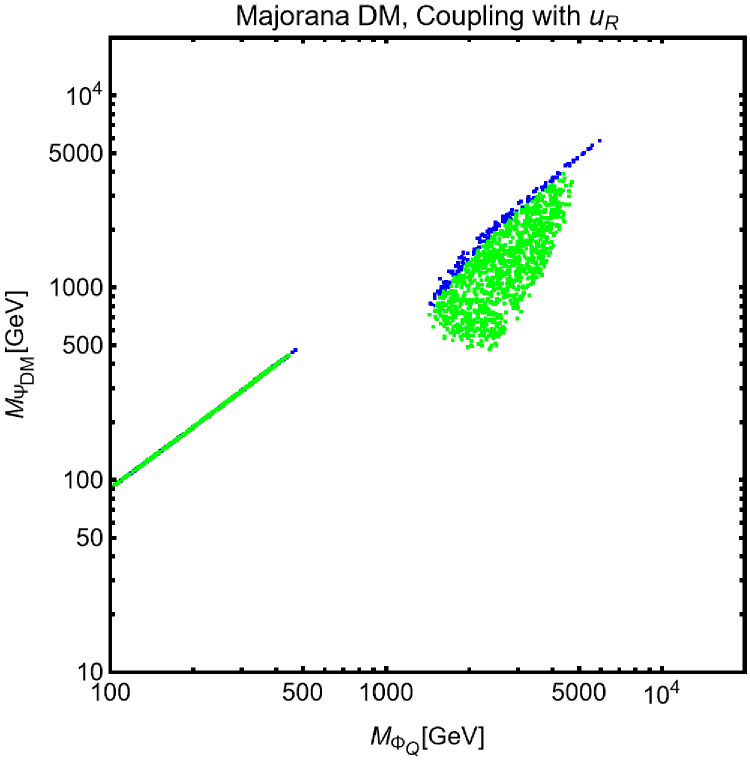}}
     \subfloat{\includegraphics[width=0.3\linewidth]{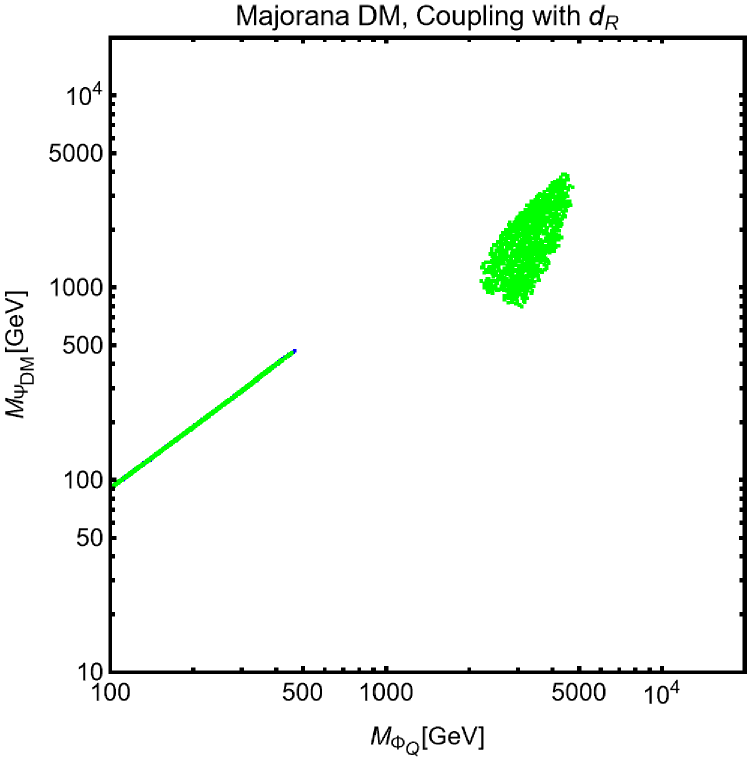}}
      \subfloat{\includegraphics[width=0.3\linewidth]{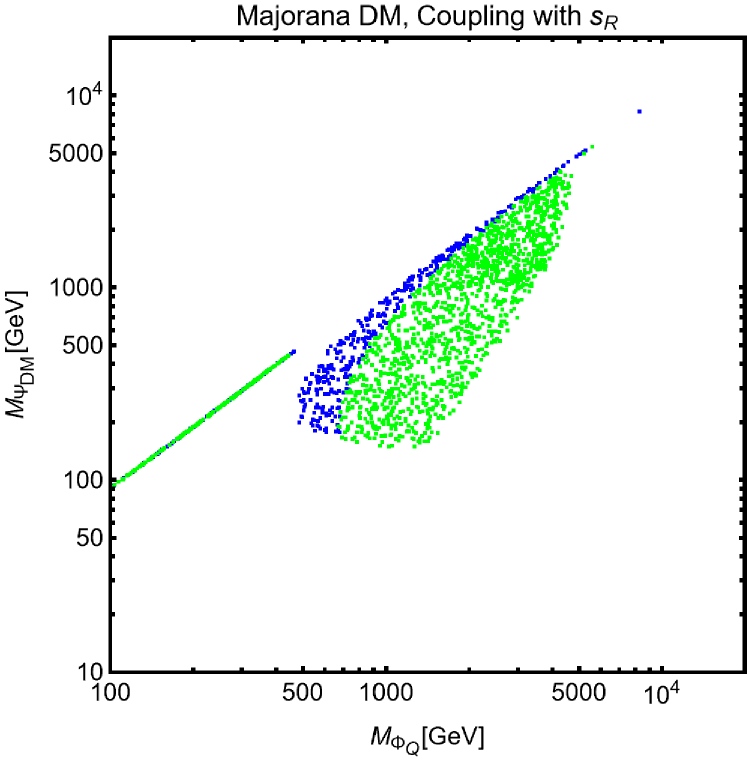}}\\
       \subfloat{\includegraphics[width=0.3\linewidth]{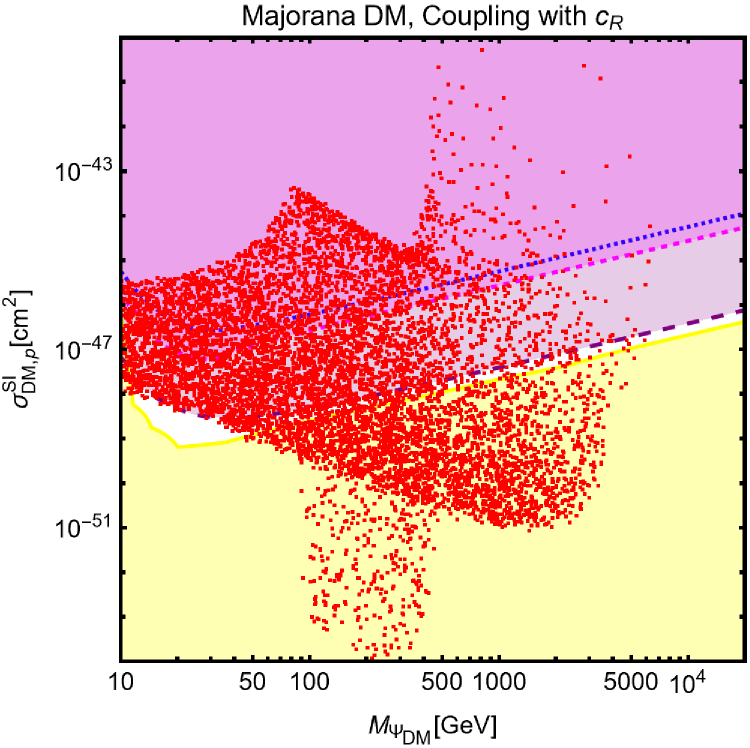}}
    \subfloat{\includegraphics[width=0.3\linewidth]{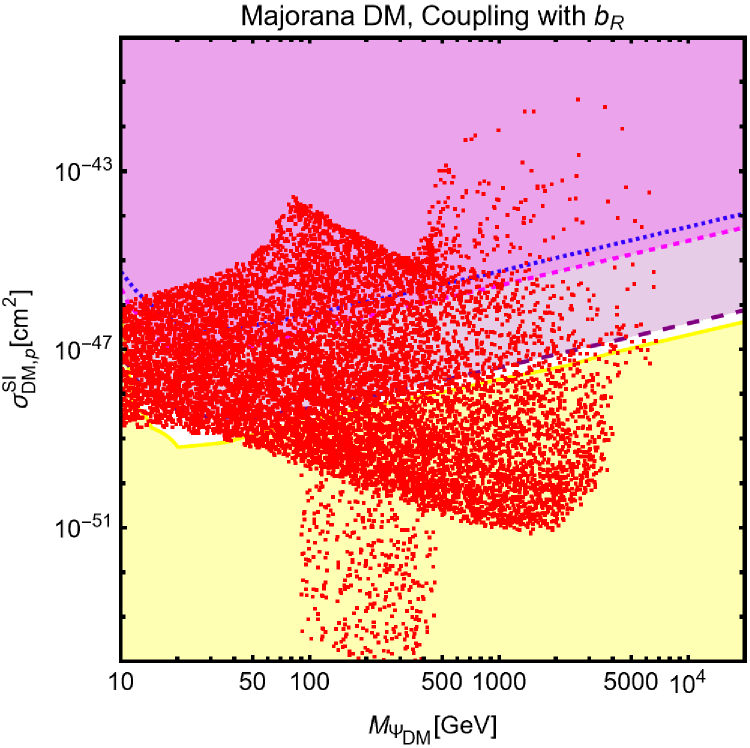}}
    \subfloat{\includegraphics[width=0.3\linewidth]{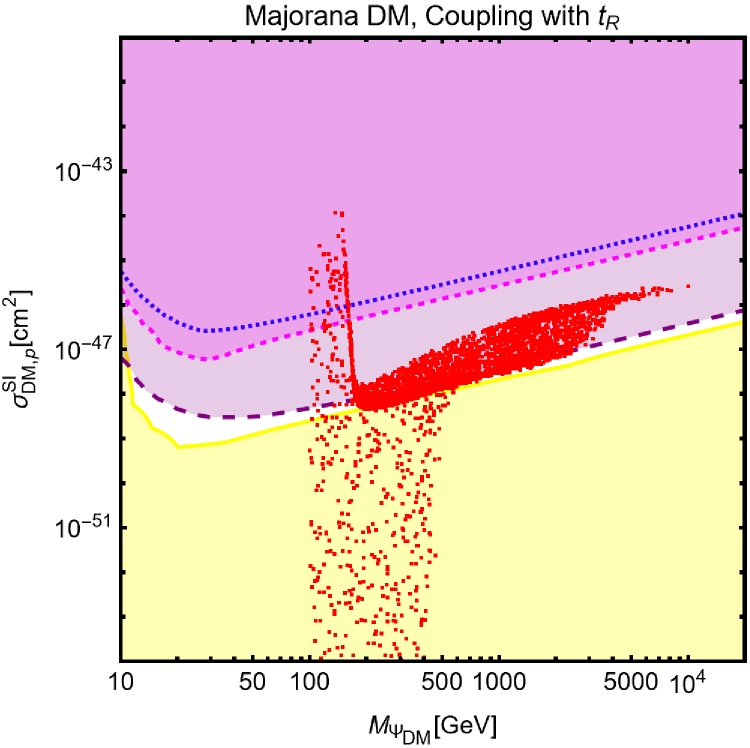}}\\
    \subfloat{\includegraphics[width=0.3\linewidth]{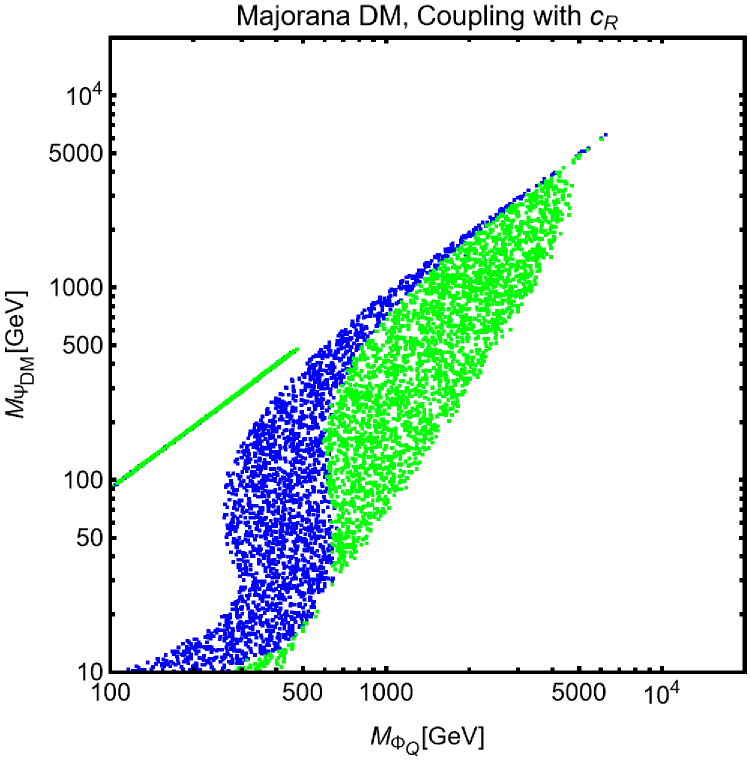}}
     \subfloat{\includegraphics[width=0.3\linewidth]{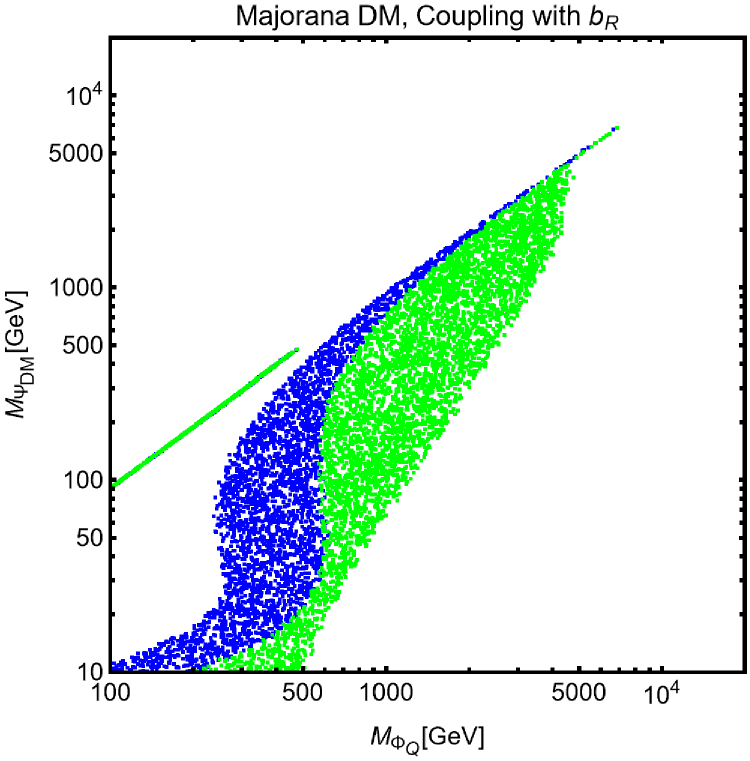}}
      \subfloat{\includegraphics[width=0.3\linewidth]{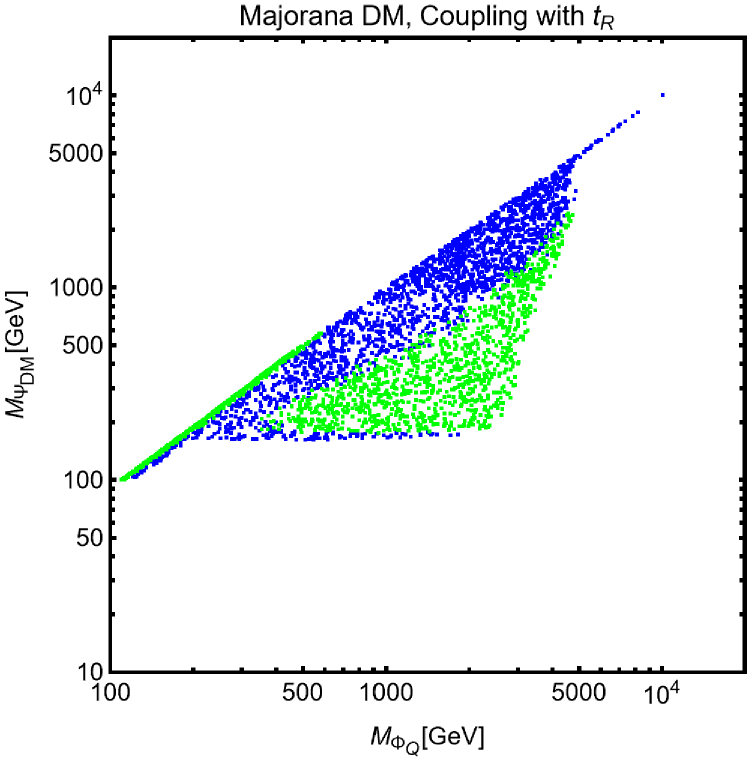}}
    \caption{\it The same as~\Fig{fig:scan_dirac_complete} but for DM coupled with right-handed quarks.}
    \label{fig:scan_majo_RR_complete}
\end{figure}

\end{document}